\documentclass[prb,aps,twocolumn,showpacs,superscriptaddress]{revtex4-1}
\usepackage{graphicx}
\usepackage{amsmath}
\usepackage{amssymb}
\usepackage{bm}
\usepackage{dsfont}
\usepackage[hidelinks]{hyperref}
\usepackage{multirow}
\usepackage{appendix}
\usepackage{color}
\usepackage{placeins}

\begin{document}
\title{The Axion Insulator as a Pump of Fragile Topology}
\author{Benjamin J. Wieder}
\affiliation{Department of Physics,
Princeton University,
Princeton, NJ 08544, USA
            }
\author{B. Andrei Bernevig}
\affiliation{Department of Physics,
Princeton University,
Princeton, NJ 08544, USA
	}
\affiliation{Dahlem Center for Complex Quantum Systems and Fachbereich Physik,
Freie Universit{\"a}t Berlin, Arnimallee 14, 14195 Berlin, Germany
	}
\affiliation{Max Planck Institute of Microstructure Physics, 
06120 Halle, Germany
}

\date{\today}
\begin{abstract}
The axion insulator (AXI) has long been recognized as the simplest example of a 3D magnetic topological insulator (TI).  The most familiar AXI results from magnetically gapping the surface states of a 3D $\mathbb{Z}_{2}$ TI while preserving the bulk gap.  Like the 3D TI, it exhibits a quantized magnetoelectric polarizability of $\theta=\pi$, and can be diagnosed from bulk symmetry eigenvalues when inversion symmetric.  However, whereas a 3D TI is characterized by bulk Wilson loop winding, 2D surface states, and the pumping of the 2D $\mathbb{Z}_{2}$ TI index, we show that an AXI with a large number of bulk bands displays no Wilson loop winding, exhibits chiral hinge states, and does not pump any previously identified quantity.  Crucially, as the AXI exhibits the topological angle $\theta=\pi$, its occupied bands cannot be formed into maximally localized symmetric Wannier functions, \emph{despite its absence of Wilson loop winding.}  In this letter, we revisit the AXI from the perspective of the recently introduced notion of ``fragile'' topology, and discover that it in fact can be generically expressed as the cyclic pumping of a ``trivialized'' fragile phase: a 2D inversion-symmetric insulator with no Wilson loop winding which nevertheless carries a nontrivial topological index, the nested Berry phase $\gamma_{2}$.  We numerically show that the nontrivial value $\gamma_{2}=\pi$ indicates the presence of anomalous 0D corner charges in a 2D insulator, and therefore, that the chiral pumping of $\gamma_{2}$ in a 3D AXI corresponds to the presence of chiral hinge states.  We also briefly generalize our results to time-reversal-symmetric higher-order TIs, and discuss the related appearance of nontrivial $\gamma_{2}$ protected by $C_{2}\times\mathcal{T}$ symmetry in twisted bilayer graphene, and its implications for the presence of 0D corner states. 
\end{abstract}

\maketitle

The discovery of the first topological insulators (TIs)~\cite{AndreiTI,CharlieTI,KaneMeleZ2,FuKaneMele,FuKaneInversion,AndreiInversion,CavaHasanTI,HgTeExp,QHZ} has fueled a decade of rapid discovery in condensed matter physics.  Building on the principles used to predict TI phases in 2D~\cite{AndreiTI,KaneMeleZ2,CharlieTI,HgTeExp} and 3D materials~\cite{FuKaneMele,FuKaneInversion,AndreiInversion,CavaHasanTI,QHZ}, researchers have proposed and identified a tremendous variety of crystalline (TCI)~\cite{LiangTCI,TeoFuKaneTCI,HsiehTCI,FangFuNSandC2,HarukiRotation,HourglassInsulator,Cohomological,DiracInsulator,ChenRotation} and higher-order (HOTI)\cite{multipole,WladTheory,HOTIBernevig,HOTIChen,HigherOrderTIPiet,HigherOrderTIPiet2,FanHOTI,EzawaMagneticHOTI,ZeroBerry,FulgaAnon,TMDHOTI,HarukiLayers,HOTIBismuth,ChenTCI,AshvinIndicators,AshvinTCI,S4Guys,TitusInteractHOTI,VDBHOTI}  variants, as well as underlying theoretical frameworks~\cite{QuantumChemistry,Bandrep1,Bandrep2,Bandrep3,JenFragile1,BarryFragile,RobertCharlieIrreps,AshvinIndicators,ChenTCI} for their large-scale identification in real materials~\cite{AndreiMaterials,ChenMaterials,AshvinMaterials1,AshvinMaterials2,AshvinMaterials3}. 

The defining hallmark of a TI is the inability to form maximally-localized, symmetry Wannier functions from its occupied bands~\cite{QuantumChemistry,Bandrep1,Bandrep2,Bandrep3,JenFragile1,BarryFragile,TKNN,ThoulessWannier,ThoulessPump,AndreiXiZ2,AlexeyVDBWannier,AlexeyVDBTI}.  In the simplest TIs, this is reflected in a momentum-dependent flow of Berry phase~\cite{TKNN,ThoulessWannier,ThoulessPump,ZakPhase,VDBpolarization}, whereas in more complicated TCIs and time-reversal- ($\mathcal{T}$-) symmetric TIs, it is reflected in the winding of the Wilson loop (holonomy) matrix~\cite{AndreiXiZ2,Fidkowski2011,ArisInversion,Cohomological,HourglassInsulator,DiracInsulator,BarryFragile}.  Central to the study of TIs has been an effort to understand the topological consequences of imposing additional crystal symmetries on 2D and 3D insulators.  In symmetry-indicated strong TIs, the eigenvalues of bulk crystal symmetries enable the quick diagnosis of the bulk topology~\cite{FuKaneInversion,RobertCharlieIrreps,AshvinIndicators,ChenTCI}, whereas in TCIs, the presence of \emph{surface} crystal symmetries protects additional surface degeneracies~\cite{LiangTCI,TeoFuKaneTCI,HsiehTCI,FangFuNSandC2,HarukiRotation,ChenRotation} and connectivities~\cite{HourglassInsulator,Cohomological,DiracInsulator}.  For both TCIs and symmetry-indicated TIs (which are not mutually exclusive categories), an outstanding question has been whether their occupied bands become Wannierizable under the relaxation of bulk and surface symmetries.  

\begin{figure}[t]
\includegraphics[width=0.98\columnwidth]{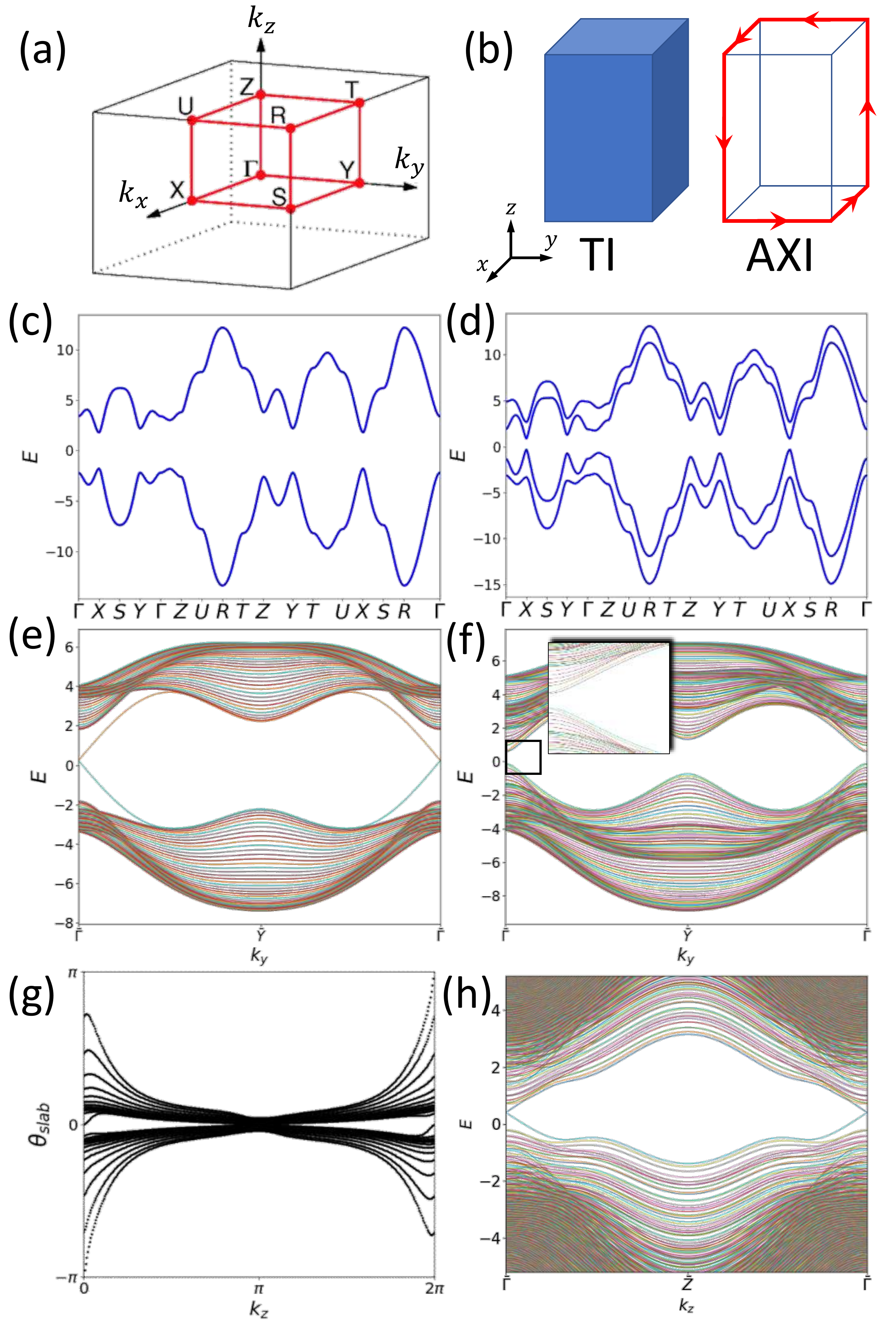}
\caption{(a) Bulk BZ of a primitive orthorhombic crystal~\cite{BCTBZ}.  (b)  The surfaces of a TI are gapless, but the surfaces of an AXI are gapped.  However, when an AXI is cleaved into an inversion- ($I$) -symmetric geometry, it exhibits a single chiral mode traversing half of its hinges~\cite{HOTIBernevig,AxionZhida,EslamInversion,FanHOTI,TMDHOTI,HarukiLayers,EzawaMagneticHOTI,VDBHOTI}.  (c) Bulk band structures of four-band tight-binding models of an $\mathcal{I}$-symmetric TI and (d) an AXI.  (e) The $(100)$-surface states of the TI in (c) are gapless, but (f) the $(100)$-surface states of the AXI in (d) are gapped.  (g) The Wilson loop over \emph{all} of the bands below $E=0$ of the gapped AXI \emph{slab} in (f) exhibits $C_{slab}=+1$ winding.  This confirms that even though individual surfaces of an AXI exhibit anomalous half-integer Hall conductivities~\cite{FuKaneMele,FuKaneInversion,AndreiInversion,QHZ,FanHOTI,VDBAxion,DiracInsulator,MulliganAnomaly,VDBHOTI}, an AXI slab acts as an isolated Chern insulator with a (non-anomalous) integer Hall conductivity.  (h) The bands of a z-directed rod of the AXI in (d) contain two oppositely chiral 1D modes localized on opposing hinges (Fig.~\ref{fig:2D}(b)).  All tight-binding calculations were performed using the~\textsc{PythTB} package~\cite{PythTB}.  All models and parameters are provided in Appendix~\ref{sec:tb}.}
\label{fig:TIAXI}
\end{figure}

The Wilson loop provides a natural machinery for answering this question: if the Wilson loop of a group of bands winds, then those bands are not Wannierizable~\cite{AndreiXiZ2,AlexeyVDBWannier,AlexeyVDBTI,Fidkowski2011,ArisInversion}.  However, and crucially, the converse statement is not always true.  In this letter, we reexamine a well-studied class of 3D inversion- ($\mathcal{I}$-) symmetric magnetic TIs~\cite{AFTI,AshvinMagnetic}, known as axion insulators (AXIs)~\cite{WilczekAxion,VDBAxion,QHZ,AndreiInversion,ChenGilbertChern,AshvinAxion1,AshvinAxion2,VDBHOTI}, and find that their Wilson loops do not generically wind.  The simplest AXIs have been proposed and experimentally verified by exposing 3D $\mathcal{I}$-symmetric TIs~\cite{VDBAxion,QHZ,AndreiInversion,ChenGilbertChern,AshvinAxion1,AshvinAxion2,WuAxionExp,EslamInversion,AxionZhida,OtherAxion1,OtherAxion2,OtherAxion3,OtherAxion4,AxionZahid1,AxionZahid2,HarukiLayers,AshvinMagnetic,VDBHOTI} or certain nodal-line semimetals~\cite{YoungkukLineNode,YoungkukMonopole,TMDHOTI,AlexeyMonopole} to an external magnetic field or doping them with an $\mathcal{I}$-symmetric configuration of magnetic atoms, such as Mn, Sm, and Cr.  This process breaks $\mathcal{T}$-symmetry while preserving $\mathcal{I}$, gapping the surface states of the TI and, as we show, for a large number of occupied bands removes the bulk Wilson loop winding.  A 3D TI carries a bulk response, the magnetoelectric polarizability $\theta$, which at low energies resembles the axion coupling in electrodynamics~\cite{WilczekAxion,VDBAxion,VDBHOTI}.  As $\mathcal{I}$ symmetry pins $\theta$ to the topological angle $\pi$,\cite{VDBAxion,QHZ,AndreiInversion,AshvinAxion1,AshvinAxion2,WuAxionExp,VDBHOTI}, then AXIs inheriting $\theta=\pi$ from a parent 3D TI are also not Wannierizable~\cite{VDBAxion,QHZ,AndreiInversion,AshvinAxion1,AshvinAxion2,VDBHOTI}, despite their absence of Wilson loop winding.  For some time, it has remained an open question as to whether a bulk topological calculation can explicitly demonstrate the absence of localized Wannier functions in AXIs.  

In this letter, we use analytic and tight-binding calculations to definitively answer this question.  We show that the recently discovered formulation of \emph{nested} Wilson loops and Berry phase~\cite{multipole,WladTheory,TMDHOTI,HingeSM,KoreanFragile} reveals higher-order Wannier flow in a generic AXI.  For an AXI with only two occupied bands, we show that the Wilson loop exhibits ``fragile''~\cite{AshvinFragile,JenFragile1,JenFragile2,HingeSM,AdrianFragile,BarryFragile,ZhidaBLG,AshvinBLG1,AshvinBLG2,AshvinFragile2,YoungkukMonopole,TMDHOTI,HarukiFragile,KoreanFragile,WiederDefect}  winding protected by bulk $\mathcal{I}$ symmetry.  However, under the addition of trivial bands below the Fermi energy, this winding is ``trivialized,'' allowing for the calculation of a nested Wilson loop, which we observe to wind.  Taken together, this implies that the AXI is equivalent to the cyclic pumping of a 2D fragile phase with anomalous corner charge~\cite{multipole,WladTheory,HOTIChen,WladPhotonCorner,ZeroBerry,FulgaAnon,EmilCorner,HingeSM,EzawaCorner,TMDHOTI,KoreanFragile,AshvinFragile2,HarukiFragile,WiederDefect}.  We observe the chiral flow of this charge along the hinges of the AXI, confirming the recent recognition that AXIs are, in fact, strong, symmetry-indicated magnetic HOTIs~\cite{HOTIBernevig,HOTIChen,WladTheory,AxionZhida,EslamInversion,FanHOTI,TMDHOTI,HarukiLayers,AshvinMagnetic,EzawaMagneticHOTI,VDBHOTI}.  We conclude by briefly generalizing our results to $\mathcal{T}$-symmetric HOTIs.  In Appendices~\ref{sec:tb} and~\ref{sec:detW2C2T}, we examine a closely-related crystalline AXI phase with gapless $\pm z$-normal surfaces and chiral hinge modes protected by the antiunitary combination of twofold $z$-axis rotation~\cite{FangFuNSandC2,ChenRotation,HarukiRotation,HOTIChen,HOTIBernevig} $C_{2z}$ and $\mathcal{T}$.  While the topology of this AXI cannot be diagnosed by symmetry eigenvalues, because it has no unitary crystal symmetries, it \emph{can} be identified using our methodology.  We show that this $C_{2z}\times\mathcal{T}$-symmetric AXI is also equivalent to the cyclic pumping of a 2D fragile phase~\cite{JenFragile1,BarryFragile,AdrianFragile,AshvinFragile,JenFragile2,HingeSM,ZhidaBLG,AshvinBLG1,AshvinBLG2,AshvinFragile2,HarukiFragile} with anomalous corner charges (Appendix~\ref{sec:C2TAXI}).  A similar $C_{2z}$- and $\mathcal{T}$-symmetric fragile phase has recently been identified in magic-angle twisted bilayer graphene (MABLG)~\cite{PabloMagic1,PabloMagic2,ZhidaBLG,AshvinBLG1,AshvinBLG2,KoreanFragile}.  As the nearly particle-hole-symmetric energy spectrum of MABLG consists of only fragile bands with odd winding and unobstructed atomic limits~\cite{ZhidaBLG,AshvinBLG1,AshvinBLG2}, then it should also exhibit anomalous, spin-degenerate~\cite{GrapheneReview}, midgap corner states.  The details of our extensive tight-binding calculations and proofs of the constraints imposed by crystal symmetry on the (nested) Wilson loop are provided in Appendices~\ref{sec:tb} and~\ref{sec:WilsonLoops}, respectively.

We begin by introducing an $\mathcal{I}$- and $\mathcal{T}$-symmetric 3D TI, modeled, for simplicity, with orthorhombic lattice vectors (Fig.~\ref{fig:TIAXI}(a)).  This TI is formed by placing $s$ and $p$ orbitals at the $1a$ Wyckoff position~\cite{BigBook,BCS1,BCS2} ($(x,y,z)=(0,0,0)$) of a unit cell with only $\mathcal{I}$ and $\mathcal{T}$ symmetries, and then inverting bands at the $\Gamma$ point.  The occupied bands of this TI (Fig.~\ref{fig:TIAXI}(c)) are doubly degenerate due to $\mathcal{I}\times\mathcal{T}$ symmetry~\cite{WiederLayers,Bandrep1,Bandrep2}, and carry the parity eigenvalues $(-,-)$ at $\Gamma$ and $(+,+)$ at all other time-reversal invariant momenta.  Through the Fu-Kane parity criterion~\cite{FuKaneInversion} and Wilson loop~\cite{AndreiXiZ2} (Appendix~\ref{sec:tbTI}), we confirm the $\mathbb{Z}_{2}$-nontrivial strong bulk topology of the two bands below $E=0$.  This topology can also be inferred using the framework of elementary band representations (EBRs)~\cite{ZakBandrep1,ZakBandrep2,QuantumChemistry,Bandrep1,Bandrep2,Bandrep3,JenFragile1,JenFragile2,BarryFragile} to determine that the two valence bands cannot be expressed as a sum (or difference) of Wannierizable bands, \emph{i.e.} topologically trivial bands from (obstructed) atomic limits.  The bulk topology necessitates the presence of unpaired (anomalous) linearly dispersing surface cones with a twofold degeneracy protected by $\mathcal{T}$ symmetry (Fig.~\ref{fig:TIAXI}(b,e)).  

\begin{figure}[t]
\includegraphics[width=\columnwidth]{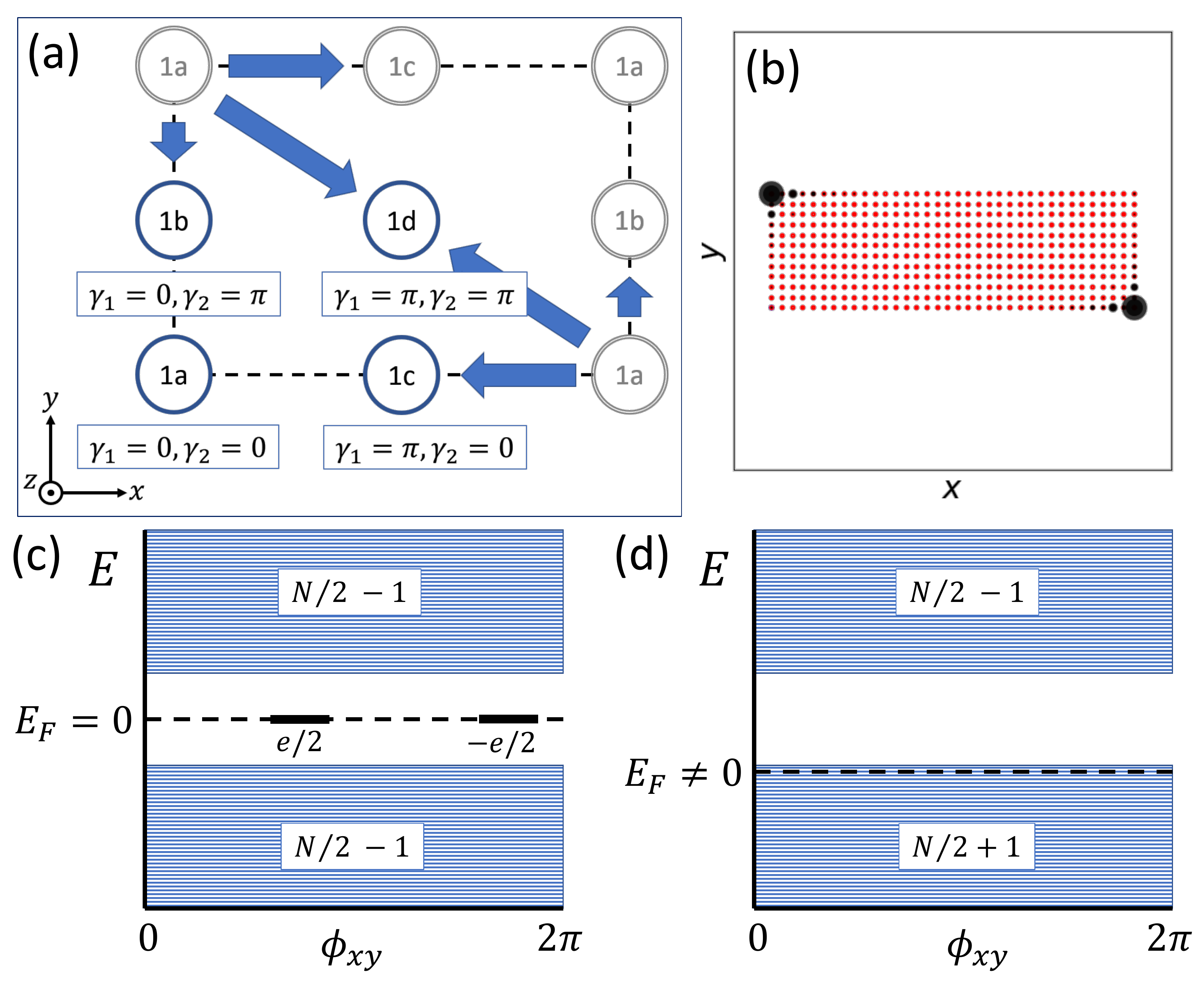}
\caption{(a) Maximal Wyckoff positions of a 2D magnetic layer (or wallpaper) group~\cite{BigBook,MagneticBook,subperiodicTables,WiederLayers,SteveMagnet,DiracInsulator,HingeSM,WiederDefect} with a symmetry that takes $(x,y)\rightarrow(-x,-y)$, \emph{e.g.} $\mathcal{I}$, $C_{2z}$, $C_{2z}\times\mathcal{T}$, or $\mathcal{I}\times\mathcal{T}$.  The number of Wannier orbitals at each Wyckoff position is indicated by the eigenvalues of the (nested) Wilson loop~\cite{multipole,WladTheory,HingeSM,WiederDefect,TMDHOTI} $W_{1,2}$  (Eqs.~(\ref{eq:WilsonMain}) and~(\ref{eq:nestedWilsonMain})).  The position of a single orbital is indicated by its (nested) Berry phase $\gamma_{1,2}$ (Eq.~(\ref{eq:mainBerryPhase})).  The blue arrows represent the obstructed atomic limit~\cite{QuantumChemistry} indicated by the Wilson loop over the fragile AXI bands at $k_{z}=0$ and a $p,\downarrow$ orbital at $1a$ (Fig.~\ref{fig:MainWilson}(c)).  (b) Corner modes of the $k_{z}=0$ plane in Fig.~\ref{fig:TIAXI}(h).  (c) These modes lie at the spectral center, and carry charges $\pm e/2$ as derived from $k\cdot p$ theory~\cite{TMDHOTI} and numerics~\cite{KoreanFragile,WladCorners}.  (d)  Breaking particle-hole symmetry while preserving $\mathcal{I}$ symmetry, the corner modes can float into the valence or conduction manifold, resulting in a measurable mismatch between the number of states above and below the gap~\cite{HingeSM}.}
\label{fig:2D}
\end{figure}

To induce the AXI phase, we introduce a bulk $z$-directed ferromagnetic potential (Appendix~\ref{sec:IAXI}) that splits the bulk bands (Fig.~\ref{fig:TIAXI}(d)) and gaps the 2D surface states (Fig.~\ref{fig:TIAXI}(b,f)).  Each gapped surface is expected to exhibit an anomalous half-integer Hall conductivity~\cite{FuKaneMele,FuKaneInversion,AndreiInversion,QHZ,FanHOTI,VDBAxion,DiracInsulator,MulliganAnomaly,VDBHOTI} $\sigma_{H}=e^{2}/(2h)$.  When a finite-sized AXI is cleaved in an $\mathcal{I}$-symmetric geometry, $\mathcal{I}$-related surfaces exhibit opposite signs of $\sigma_{H}$ (taken with respect to the surface normal), resulting in the presence of a single, chiral hinge mode traversing the AXI~\cite{AxionZhida,EslamInversion,FanHOTI,TMDHOTI,HarukiLayers,EzawaMagneticHOTI,VDBHOTI} (Fig.~\ref{fig:TIAXI}(b)).  Because it exhibits gapped surfaces and gapless hinges as a consequence of its bulk symmetry and topology, an AXI is therefore a magnetic HOTI~\cite{HOTIBernevig,HOTIChen,WladTheory,AxionZhida,EslamInversion,FanHOTI,TMDHOTI,HarukiLayers,AshvinMagnetic,EzawaMagneticHOTI,VDBHOTI}.  Furthermore, as the number of chiral hinge modes modulo two is indicated by bulk parity eigenvalues~\cite{VDBAxion,FanHOTI,AshvinMagnetic,VDBHOTI}, and as $\mathcal{I}$ symmetry is the only symmetry required to enforce the AXI phase, then an AXI represents the lowest-symmetry example of a symmetry-indicated TI, magnetic or otherwise; it requires neither the $\mathcal{T}$-symmetry of the $\mathcal{I}$-symmetric 3D TI, nor the combination of $\mathcal{T}$ and half-integer lattice translation of an antiferromagnetic TI~\cite{AFTI,AFTIExp}.  Calculating the band structure of a $z$-directed AXI rod (Fig.~\ref{fig:TIAXI}(h)), \emph{i.e.} a tight-binding model finite in the $x$ and $y$ directions and infinite in the $z$ direction~\cite{HOTIBernevig,HOTIChen,HingeSM,AshvinTCI}, we observe two oppositely chiral hinge modes localized on $\mathcal{I}$-related hinges (Fig.~\ref{fig:2D}(b)).  We also calculate the Wilson loop over \emph{all} of the occupied bands of the $x$-directed AXI slab in Fig.~\ref{fig:TIAXI}(f), and observe $C_{slab}=+1$ winding (Fig.~\ref{fig:TIAXI}(g)), demonstrating that an $\mathcal{I}$-symmetric AXI cleaved into a slab geometry is topologically equivalent to a $C=\pm 1$ Chern insulator~\cite{HOTIBernevig}.  

To develop a robust characterization of the bulk topology of an AXI, we begin by examining the $k_{z}=0$ plane of the model in Fig.~\ref{fig:TIAXI}(b,d,f,g,h).  This plane transforms under the same symmetries as a magnetic layer group~\cite{BigBook,MagneticBook,subperiodicTables,WiederLayers,SteveMagnet,DiracInsulator,HingeSM,WiederDefect} with only $\mathcal{I}$ and rectangular lattice translation symmetries (Fig.~\ref{fig:2D}(a)).  We observe two, 0D states localized on $\mathcal{I}$-related corners at $k_{z}=0$ (Fig.~\ref{fig:2D}(b)).  These two states appear at the center of the energy spectrum; there are exactly $(N/2)-1$ states in the manifolds above and below them in energy, where $N$ is the total number of states in the $k_{z}=0$ plane (Fig.~\ref{fig:2D}(c)).  The corner states are equivalent to (anti)solitons of the gapped edge~\cite{TMDHOTI,HingeSM,WiederDefect}, and are described by the same $k\cdot p$ theory (taken per spin) as the hinge states~\cite{TMDHOTI} of an $\mathcal{I}$- and $\mathcal{T}$-symmetric monopole nodal-line semimetal~\cite{FangWithWithout,YoungkukMonopole,TMDHOTI}.  Thus, when the system is half-filled and $\mathcal{I}$ symmetry is ``softly'' broken, the corner states exhibit charges~\cite{SSH,RiceMele,NiemiSemenoff,GoldstoneWilczek,HingeSM,WiederDefect} $\pm e/2$ (Fig.~\ref{fig:2D}(c)).  If particle-hole symmetry is relaxed while preserving $\mathcal{I}$ symmetry, then the corner modes can float together away from $E=0$.  They will therefore either appear as midgap states, or there will be an imbalance in the number of states in the manifolds above and below the gap~\cite{HingeSM} (Fig.~\ref{fig:2D}(d)); thus, the corner modes are anomalous.  This is analogous to the $\mathcal{I}$-symmetric, particle-hole-asymmetric Su-Schrieffer-Heeger (SSH) chain~\cite{SSH}, in which a nontrivial Berry phase indicates the presence of end charge, but not necessarily midgap end modes~\cite{NiemiSemenoff}.  This can also be understood from a field-theory perspective by considering the Goldstone-Wilczek formulation of a Jackiw-Rebbi domain wall~\cite{JackiwRebbi} with a complex mass~\cite{WilczekAxion,GoldstoneWilczek,NiemiSemenoff}, and is explored in more detail for related fragile phases in Refs.~\onlinecite{HingeSM,WiederDefect}.  In contrast to the $k_{z}=0$ plane, the $k_{z}=\pi$ plane of Fig.~\ref{fig:TIAXI}(h) exhibits no midgap corner modes, and its valence and conduction manifolds each contain $N/2$ states.  Therefore, the AXI is equivalent to the cyclic chiral pumping of a 2D (fragile) phase with anomalous corner charge.  

\begin{figure}[t]
\includegraphics[width=\columnwidth]{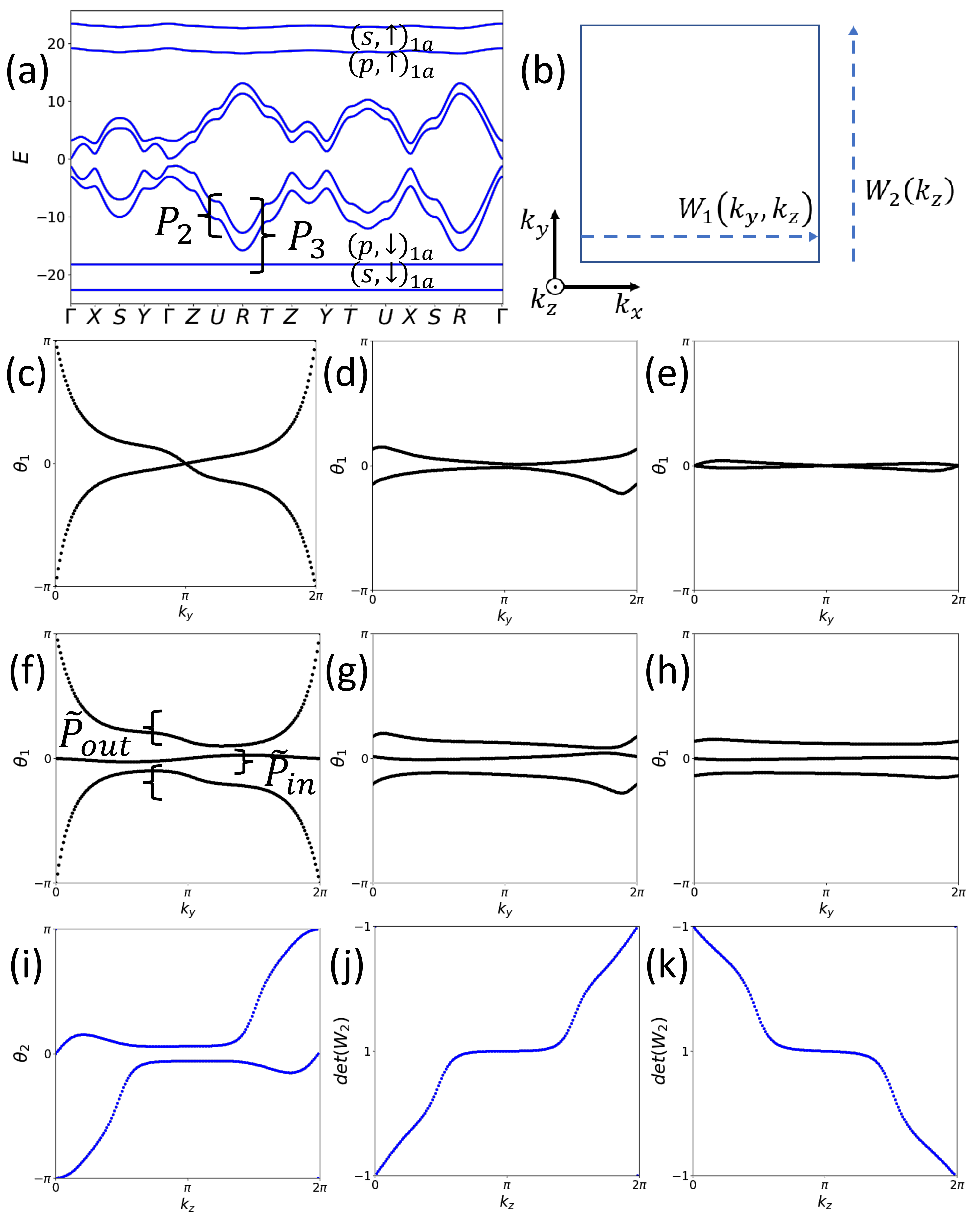}
\caption{(a) Bands of the AXI from Fig.~\ref{fig:TIAXI}(a) coupled to additional spin-split $s$ and $p$ orbitals at $1a$ ($(0,0,0)$).  (b) Directions of the (nested) Wilson loops (Eqs.~(\ref{eq:WilsonMain}) and~(\ref{eq:nestedWilsonMain})) in (c-k).  (c-e) $x$-directed Wilson loops over the $P_{2}$ bands in (a) at (c) $k_{z}=0$, (d) a generic value, and (e) $\pi$; the winding in (c) is representative of fragile topology (Eq.~(\ref{eq:2Dfragilebands})).  (f-h) $x$-directed Wilson loops over the $P_{3}$ bands in (a) at (f) $k_{z}=0$, (g) a generic value, and (h) $\pi$; this Wilson loop is gapped at all values of $k_{z}$.  (i) The nested Wilson loop over the outer Wilson bands in (f) and (j) its determinant exhibit $C_{\gamma_{2}}=+1$ winding.  (k) The determinant of the nested Wilson loop over the inner Wilson band in (f) exhibits $C_{\gamma_{2}}=-1$ winding.  In combination with bulk $\mathcal{I}$ symmetry, (i-k) indicate a $\mathbb{Z}_{2}$-nontrivial AXI topology (Eq.~(\ref{eq:MainThetaInv})).  Model details and supplemental calculations are provided in Appendix~\ref{sec:tb}.}
\label{fig:MainWilson}
\end{figure}

To relate this pumping to bulk topology, we develop a (nested) Wilson loop characterization of the AXI, beginning with the $k_{z}=0$ plane.  First, consider a 2D layer (or wallpaper) group~\cite{BigBook,MagneticBook,subperiodicTables,WiederLayers,SteveMagnet,DiracInsulator,HingeSM,WiederDefect} with four maximal Wyckoff positions, \emph{i.e.}, the centers of a symmetry that takes $(x,y)\rightarrow (-x,-y)$, such as $\mathcal{I}$, $C_{2z}$, $C_{2z}\times\mathcal{T}$, or $\mathcal{I}\times\mathcal{T}$ (Fig.~\ref{fig:2D}(a)).  We then place a Wannier orbital at one of these positions without hopping, resulting in a flat band in the energy spectrum.  Then, using only this band, we calculate the discretized~\cite{ArisInversion,Cohomological,DiracInsulator} $x$-directed Wilson loop~\cite{Fidkowski2011,AndreiXiZ2,ArisInversion,Cohomological,HourglassInsulator,DiracInsulator,BarryFragile}:
\begin{equation}
W_{1}(k_{y},k_{z}) = W_{1}(k_{\perp}) = V(2\pi\hat{x})\Pi(k_{x0},k_{y},k_{z}),
\label{eq:WilsonMain}
\end{equation}
where $V(2\pi\hat{x})$ is a sewing matrix that enforces the basepoint independence of Eq.~(\ref{eq:WilsonMain}), $\Pi(k_{x0},k_{y},k_{z})$ is the product of projectors onto the occupied bands at each $\textbf{k}$ point, and where Eq.~(\ref{eq:WilsonMain}) is provided in its 3D form for generality .  We then calculate the discretized $y$-directed nested Wilson loop~\cite{multipole,WladTheory} (Fig.~\ref{fig:MainWilson}(b)):
\begin{equation}
W_{2}(k_{z}) = W_{2}(k_{\perp}) = \tilde{V}(2\pi\hat{y})\tilde{\Pi}(k_{x0},k_{y0},k_{z}),
\label{eq:nestedWilsonMain}
\end{equation}
where $\tilde{V}(2\pi\hat{y})$ is again a sewing matrix that enforces basepoint indepence and $\tilde{\Pi}(k_{x0},k_{y0},k_{z})$ is the product of projectors onto a set of \emph{Wilson} bands (further details and more formal expressions for Eqs.~(\ref{eq:WilsonMain}) and~(\ref{eq:nestedWilsonMain}) are provided in Appendix~\ref{sec:WilsonLoops}).  As the eigenvalues of $W_{1,2}(k_{\perp})$ are gauge-invariant phases $\exp(i\theta_{1,2}(k_{\perp}))$, the (nested) Berry phase~\cite{ZakPhase,VDBpolarization,multipole,WladTheory} $\gamma_{1}$ ($\gamma_{2}$) is related to Eq.~(\ref{eq:WilsonMain}) (Eq.~(\ref{eq:nestedWilsonMain})) by:
\begin{equation}
\det(W_{1,2}(k_{\perp})) = \exp(i\gamma_{1,2}(k_{\perp})).
\label{eq:mainBerryPhase}
\end{equation}
When there is only a single band in the system and Wilson projector,  $\Pi(k_{x0},k_{y},k_{z})\rightarrow 1$ in Eq.~(\ref{eq:WilsonMain}), and Eq.~(\ref{eq:nestedWilsonMain}) simplifies into the (non-nested) $y$-directed Wilson loop: thus, the $\mathbb{Z}_{2}\times\mathbb{Z}_{2}$ index $(\gamma_{1},\gamma_{2})$ indicates which of the four maximal Wyckoff positions is occupied by a Wannier orbital (Fig.~\ref{fig:2D}(a)).  When there are more bands in a Wannierizable system, $W_{1,2}$ calculated over energy and particle-hole-symmetric Wilson bands (Appendix~\ref{sec:detW2Inv}) gives a \emph{full eigenvalue classification} of the number of Wannier orbitals from those bands at each Wyckoff position.  Thus, taken over all of the valence and Wilson bands, the eigenvalues of $W_{1,2}$ indicate the bulk multipole moment and the presence of anomalous corner states.  In layer groups where $\gamma_{1,2}$ are quantized and the first and second Stiefel-Whitney invariants~\cite{YoungkukMonopole,KoreanFragile} $w_{1,2}$ are well-defined, the classifications are equivalent~\cite{KoreanFragile}.  However, the Wilson loop formulation provides an advantage over the Stiefel-Whitney classification: whereas $w_{1,2}$ are only discretely valued when well-defined, $\gamma_{1,2}$ remain gauge-invariant bulk quantities \emph{even when they are not quantized} (Appendix~\ref{sec:WilsonLoops}).  This allows us to quantitatively track $\gamma_{1,2}$ as they are pumped between $k_{z}=0,\pi$ in an AXI.  

With this formalism established, we now diagnose the bulk topology of the AXI in Fig.~\ref{fig:TIAXI}(b,d,f,g,h).  We take the previous four-band AXI (Fig.~\ref{fig:TIAXI}(a)) and add spin-split bands from additional $s$ and $p$ orbitals at the $1a$ position (Fig.~\ref{fig:MainWilson}(a) and Appendix~\ref{sec:oneBandInversion}).  We first calculate the $x$-directed Wilson loop (Fig.~\ref{fig:MainWilson}(c-e)) over just the original two valence bands ($P_{2}$).  The Wilson loop winds in the $k_{z}=0$ plane (c), but does not wind at other values of $k_{z}$ (d,e).  Treating the $k_{z}=0$ plane as an isolated 2D insulator, this winding can be explained by bulk parity eigenvalues~\cite{ArisInversion} or by the notion of fragile topology~\cite{AshvinFragile,JenFragile1,JenFragile2,HingeSM,AdrianFragile,BarryFragile,ZhidaBLG,AshvinBLG1,AshvinBLG2,AshvinFragile2,YoungkukMonopole,TMDHOTI,HarukiFragile,KoreanFragile,WiederDefect}  .  As first shown in Ref.~\onlinecite{ArisInversion}, an $\mathcal{I}$-symmetric 2D insulator with two occupied bands $F$ with $(-,-)$ parity eigenvalues at $k_{x}=k_{y}=0$ and $(+,+)$ elsewhere exhibits Wilson loop winding, even in the absence of symmetry-protected surface states~\cite{DiracInsulator}. $F$ is also ``irreducible-representation-equivalent''~\cite{JenFragile1,JenFragile2,BarryFragile,HingeSM,AshvinFragile2} ($\stackrel{I}{\equiv}$) to a sum and difference of 2D EBRs:
\begin{equation}
F \stackrel{I}{\equiv} (p)_{1b} \oplus (p)_{1c} \oplus (p)_{1d} \ominus (p)_{1a},
\label{eq:2Dfragilebands}
\end{equation}
where $(\rho)_{i}$ is the EBR~\cite{ZakBandrep1,ZakBandrep2,QuantumChemistry,Bandrep1,Bandrep2,Bandrep3,JenFragile1,JenFragile2,BarryFragile} induced from atomic orbital $\rho$ at Wyckoff position $i$ (Fig.~\ref{fig:2D}(a)).  The $\ominus$ in Eq.~(\ref{eq:2Dfragilebands}) indicates that $F$ represents fragile bands~\cite{JenFragile1,JenFragile2,BarryFragile,HingeSM,AshvinFragile2}.  From both perspectives, we deduce that the Wilson loop winding in Fig.~\ref{fig:MainWilson}(c) can be removed by the addition of appropriately chosen trivial bands to the Wilson projector.  

We therefore take the Wilson loop over the highest \emph{three} valence bands ($P_{3}$) in Fig.~\ref{fig:MainWilson}(a).  Observing that all of the winding has been trivialized (Fig.~\ref{fig:MainWilson}(f-h)),  we calculate the nested Wilson loop onto particle-hole symmetric~\cite{ZhidaBLG,TMDHOTI} (Appendix~\ref{sec:detW2Inv}) groupings of Wilson bands ($\tilde{P}_{out,in}$).  In the $k_{z}=0$ plane (Fig.~\ref{fig:MainWilson}(f)), the $\tilde{P}_{out,in}$ Wilson band(s) both exhibit $(\gamma_{1},\gamma_{2}) = (0,\pi)$.  As both $\tilde{P}_{out}$ Wilson bands lie near $\theta_{1}=\pi$, the $P_{3}$ valence bands at $k_{z}=0$ describe a 2D obstructed atomic limit with Wannier orbitals at $1b$, $1c$, and $1d$ and atoms at $1a$ (Fig.~\ref{fig:2D}(a)): \emph{i.e.} a shift of three charges from one half of the crystal to the other yielding six corner charges~\cite{WladCorners,HingeSM}, which can be deformed preserving $\mathcal{I}$ symmetry~\cite{HingeSM,WladCorners,TMDHOTI,WiederDefect} to the two 0D states of the fragile phase in Fig.~\ref{fig:2D}(b,c,d)).

We observe that the outer (inner) Wilson bands exhibit $C_{\gamma_{2}}=+1$ ($-1$) spectral flow (Fig.~\ref{fig:MainWilson}(i-k)).  As detailed in Appendix~\ref{sec:detW2Inv}, because a Wilson gap can close without an energy gap closing~\cite{Fidkowski2011,BarryFragile}, $C_{\gamma_{2}}$ only defines the bulk topology of an AXI modulo $2$, and can become trivial without closing an energy gap if $\mathcal{I}$ symmetry is relaxed.  Specifically, Wilson gap closures occur pairwise with bulk $\mathcal{I}$ symmetry, but are free to occur individually without it.  Recognizing that this three-band AXI inherits $\theta=\pi$ from its parent TI phase, we relate $C_{\gamma_{2}}$ to the magnetoelectric polarizability~\cite{WilczekAxion,VDBAxion,QHZ,AndreiInversion,ChenGilbertChern,AshvinAxion1,AshvinAxion2,VDBHOTI}:
\begin{equation}
\frac{\theta}{\pi} = C_{\gamma_{2}}\text{ mod } 2.
\label{eq:MainThetaInv}
\end{equation}
We have demonstrated that $\theta=\pi$ in an AXI is a consequence of the \emph{bulk} flow of nested Berry phase.  The (nested) Wilson loop formulation also allows us to conclude that when $\mathcal{I}$ symmetry is further relaxed, an AXI becomes Wannierizable, because the winding of $W_{2}(k_{z})$ can be removed by a gap closure in $W_{1}(k_{y},k_{z})$ that occurs without a corresponding gapless point in the energy spectrum.  In Appendix~\ref{sec:twoBandsInversion}, we show that Eq.~(\ref{eq:MainThetaInv}) remains valid under the addition of two or more trivial bands, and thus that it is a $\mathbb{Z}_{2}$ strong topological invariant.  

We briefly generalize to $\mathcal{T}$-symmetric insulators.  As an $\mathcal{I}$- and $\mathcal{T}$-symmetric HOTI is equivalent to a time-reversed pair of AXIs~\cite{HOTIBernevig,HOTIChen,AshvinIndicators,AshvinTCI,ChenTCI,TMDHOTI,BarryPrep,BarryConvo}, we infer that it is equivalent to a \emph{helical} pump of fragile-phase corner charge.  Though both trivial insulators and HOTIs exhibit $\theta\text{ mod } 2\pi=0$, they can be distinguished by parity eigenvalues~\cite{HOTIChen,AshvinIndicators,AshvinTCI,ChenTCI,TMDHOTI}, or by a helical generalization of the nested Wilson invariant introduced in Eq.~(\ref{eq:MainThetaInv}).  Conversely, in $C_{2z}$- and $\mathcal{T}$-symmetric insulators, which exhibit only one set of symmetry eigenvalues~\cite{TMDHOTI,ChenTCI,AshvinTCI} (Appendix~\ref{sec:C2TAXI}), the (nested) Wilson loop provides the only bulk indicator of (higher-order) topology.

\begin{acknowledgements}
The authors are greatly indebted to Barry Bradlyn for discussions~\cite{BarryConvo} regarding the adaptation of the nested Wilson loop \textsc{PythTB}~\cite{PythTB} library from Ref.~\onlinecite{HingeSM} to the 3D AXI models in this letter.  We also thank Chen Fang and Zhida Song for providing crucial theoretical insights during the early stages of this project.  We further acknowledge helpful discussions with David Vanderbilt, Charles L. Kane, Jennifer Cano, Zhijun Wang, Bohm-Jung Yang, Junyeong Ahn, Sungjoon Park, Taylor L. Hughes, and Wladimir A. Benalcazar.  B. J. W. and B. A. B. were supported by the Department of Energy Grant No. DE-SC0016239, the National Science Foundation EAGER Grant No. NOA-AWD1004957, Simons Investigator Grant No. ONR-N00014-14-1-0330, NSF Grant No. NSF-MRSEC DMR- 1420541, the Packard Foundation, and the Schmidt Fund for Innovative Research.  During the final stages of preparing this letter, an analysis of the first Wilson loop and hinge states in candidate AXIs in the pyrochlore iridates was performed in Ref.~\onlinecite{VDBHOTI}, confirming the results of our investigations.  During the preparation of this letter, corner modes resulting from fragile topology in MABLG were also predicted and analyzed in detail in Ref.~\onlinecite{KoreanFragile}.   
\end{acknowledgements}

\clearpage
\onecolumngrid
\begin{appendix}

\section{3D Tight-Binding Models}
\label{sec:tb}

\begin{figure}[h]
\centering
\includegraphics[width=0.5\textwidth]{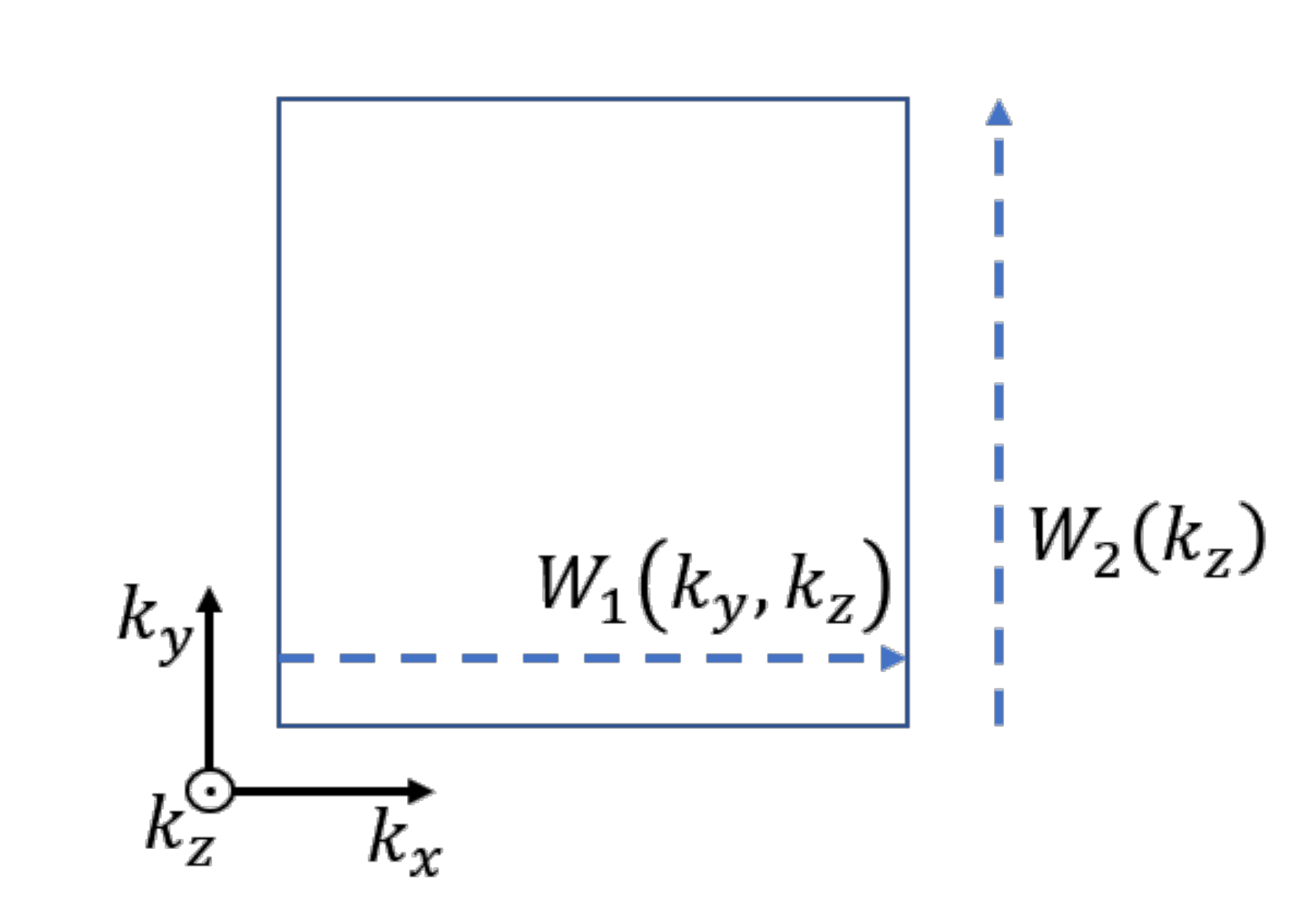}
\caption{Directions of the Wilson loops used in this appendix, unless otherwise specified.  For a 3D bulk crystal, we take the first Wilson loop $W_{1}(k_{y},k_{z})$ in the $x$ direction, and then take the nested Wilson loop $W_{2}(k_{z})$ in the $k_{y}$ direction.  Details of our Wilson loop calculations are provided in~\ref{sec:W2}.}
\label{fig:Wilson}
\end{figure}

In this section, we detail the parameters used to generate the numerical results shown in the main text, as well as present additional calculations demonstrating hinge states and nested Wilson loop flow in a $C_{2z}\times\mathcal{T}$-symmetric AXI that is only briefly discussed in the main text.  In~\ref{sec:tbTI}, we give the tight-binding models and parameters used to generate four band-models of a 3D topological insulator (TI), which we then use as the parent phase of two different axion insulators (AXIs).  In~\ref{sec:IAXI}, we model an AXI with only inversion ($\mathcal{I}$) symmetry, and in~\ref{sec:C2TAXI}, we model a topological crystalline AXI~\cite{FangFuNSandC2,ChenRotation,HarukiRotation,LiangTCI} with only the antiunitary symmetry $C_{2z}\times\mathcal{T}$, where $C_{2z}$ is twofold rotation about the $z$-axis.  As shown in the main text, both minimal models of AXIs, while exhibiting gapped (rod) surface states and chiral hinge states, nevertheless exhibit Wilson loop winding by the mechanisms detailed in Refs.~\onlinecite{ArisInversion,ArisNoSOC,YoungkukMonopole,TMDHOTI,KoreanFragile} for the $\mathcal{I}$-symmetric case and in Refs.~\onlinecite{JenFragile1,BarryFragile,AdrianFragile,AshvinFragile,JenFragile2,HingeSM,ZhidaBLG,AshvinBLG1,AshvinBLG2,AshvinFragile2,HarukiFragile} for the $C_{2z}\times\mathcal{T}$ case.  As this winding is only enforced by bulk symmetries and the limited number of bands in the Wilson projector (though in the $\mathcal{I}$-symmetric case, the winding can persist for an infinite number of occupied bands with appropriately chosen eigenvalues~\cite{ArisInversion}), it may be removed without closing a bulk gap or breaking any symmetries by adding extra bands (with appropriately chosen eigenvalues) to the Wilson projector.  This phenomenon was first characterized in $\mathcal{I}$-symmetric 2D models in Ref.~\onlinecite{ArisInversion}, and is referred to in recent works as ``fragile'' topology~\cite{AshvinFragile,JenFragile1,JenFragile2,HingeSM,AdrianFragile,BarryFragile,ZhidaBLG,AshvinBLG1,AshvinBLG2,AshvinFragile2,YoungkukMonopole,TMDHOTI,HarukiFragile,KoreanFragile,WiederDefect}.  We therefore add extra bands to these AXIs, and numerically demonstrate that for both the addition of one band to the Wilson projector (\ref{sec:oneBandInversion} and~\ref{sec:oneBandC2T}) and two bands (\ref{sec:twoBandsInversion} and~\ref{sec:twoBandsC2T}), the fragile Wilson loop winding of the AXIs is removed.  We also show, crucially, that this allows the calculation of a nested Wilson loop $W_{2}(k_{z})$ (Fig.~\ref{fig:Wilson}), a bulk quantity~\cite{multipole,WladTheory,HOTIBernevig,HingeSM,KoreanFragile} whose determinant is:
\begin{equation}
\det(W_{2}(k_{z})) = e^{i\gamma_{2}(k_{z})},
\label{eq:gamma2}
\end{equation}
where $\gamma_{2}(k_{z})$ is the nested Berry phase of a subset of Wilson bands within each $k_{z}$-indexed plane.  For both of the AXIs presented in this appendix, we demonstrate that $\gamma_{2}(k_{z})$ is chirally pumped.  Thus, the occupied bands of these AXIs do not admit a localized Wannier description~\cite{ThoulessWannier,ThoulessPump,TKNN,AndreiXiZ2,AlexeyVDBWannier,AlexeyVDBTI,HOTIChen,HOTIBernevig}, despite the absence of Wilson loop winding.  The lack of a localized Wannier representation is in agreement with the nonzero quantized magnetoelectric polarizability~\cite{WilczekAxion,VDBAxion,QHZ,AshvinAxion1,AshvinAxion2,ChenGilbertChern,VDBHOTI} $\theta=\pi$ that they inherit from the parent TI phase (\ref{sec:tbTI}).  In~\ref{sec:detW2Inv} and~\ref{sec:detW2C2T}, we explicitly relate $\theta$ to the winding of $W_{2}(k_{z})$ in the presence of bulk $\mathcal{I}$ and combined $C_{2z}\times\mathcal{T}$ symmetry, respectively.  

All calculations in this letter were performed employing the~\textsc{PythTB} package~\cite{PythTB}.  All nested Wilson loops were obtained by first calculating the $x$-directed Wilson loop $W_{1}(k_{y},k_{z})$, and then taking the $y$-directed nested Wilson loop $W_{2}(k_{z})$ (Fig.~\ref{fig:Wilson}).  The calculations of $\det(W_{2})$ were performed using a modified version of the code developed for Ref.~\onlinecite{HingeSM}; further details are provided in~\ref{sec:W2}.  We also show in~\ref{sec:detW2Inv} and~\ref{sec:detW2C2T} how crystal symmetries ($\mathcal{I}$ and $C_{2z}\times\mathcal{T}$, respectively) quantize $\det(W_{2})$ in 2D and make $\det(W_{2}(k_{z}))$ particle-hole symmetric in 3D.  

\subsection{Tight-Binding Model of a 3D Topological Insulator}
\label{sec:tbTI}

\begin{figure}[h]
\centering
\includegraphics[width=\textwidth]{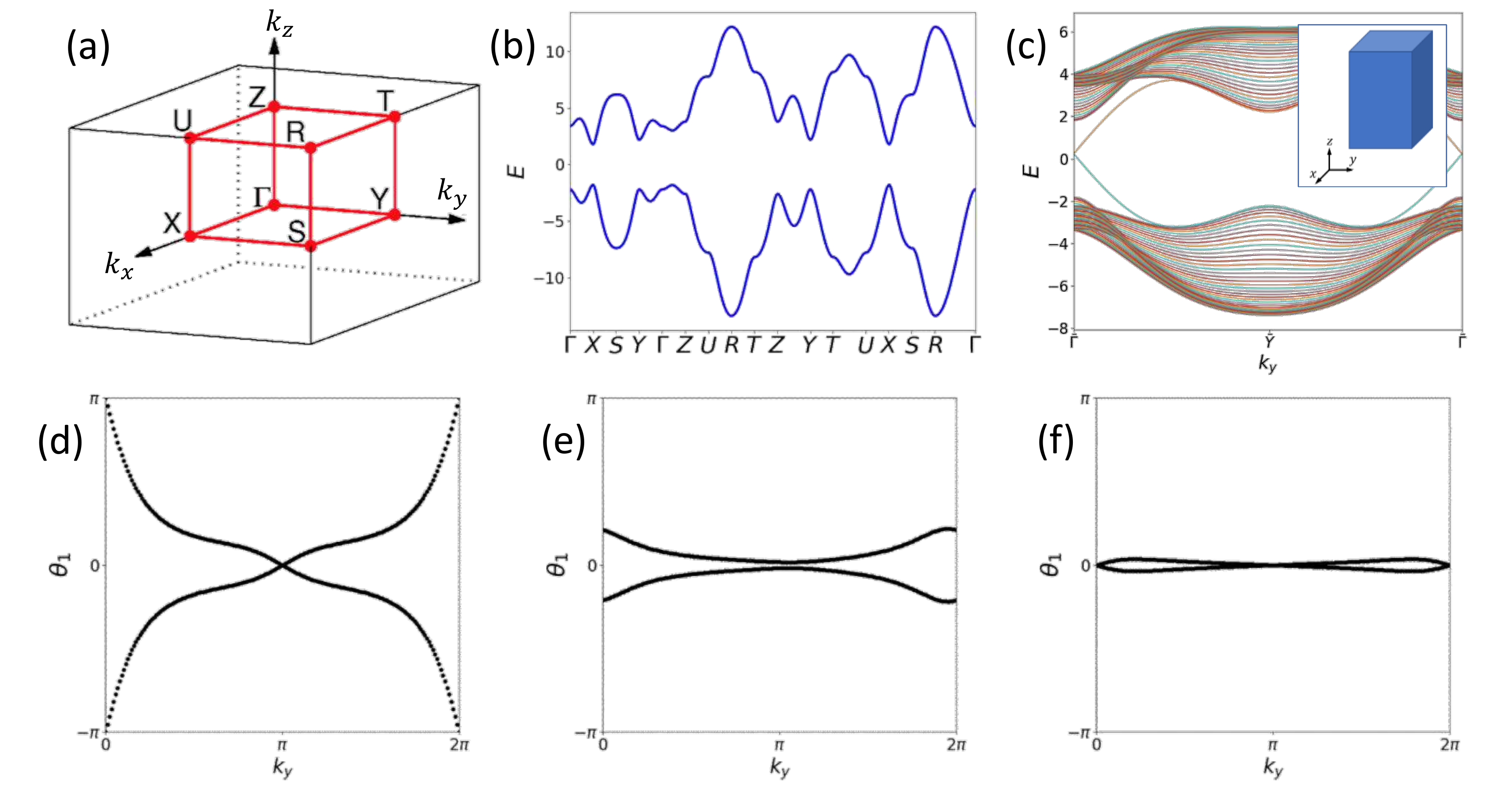}
\caption{(a) Bulk BZ and (b) bands of $\mathcal{H}_{TI}(\vec{k})$ (Eq.~(\ref{eq:TI})), a four-band model of a 3D TI, plotted using the parameters in Eq.~(\ref{eq:TIparams}).  (c)  The $x$-directed slab bands of this TI are gapless, and exhibit twofold linear degeneracies on the $\pm \hat{x}$-normal surfaces at $k_{y}=k_{z}=0$.  (d) The $x$-directed Wilson loop of the two bands below $E=0$ at $k_{z}=0$ exhibits helical winding.  (e) The $x$-directed Wilson loop at generic values of $k_{z}$ is gapped.  (f) The $x$-directed Wilson loop at $k_{z}=\pi$ does not wind, and exhibits Kramers degeneracies at $k_{y}=0,\pi$.}
\label{fig:TI}
\end{figure}

We begin by constructing a tight-binding model of a 3D TI~\cite{FuKaneMele,FuKaneInversion,AndreiInversion,CavaHasanTI,QHZ} in the (type-II Shubnikov~\cite{BigBook,MagneticBook,ITCA}) space group (SG) 2 $P\bar{1}1'$; \emph{i.e.}, one for which the only symmetries are lattice translation, time-reversal ($\mathcal{T}$), and inversion ($\mathcal{I}$).  To form our model, we place Kramers pairs of $s$ and $p$ orbitals at the $1a$ Wyckoff position~\cite{BCS1,BCS2} of SG 2 ($(x,y,z) = (0,0,0)$); we label the $s,p$-orbital (spin) degree of freedom with $\tau$ ($\sigma$).  In this representation, the momentum-space Hamiltonian for a 3D TI is:
\begin{eqnarray}
\mathcal{H}_{TI}(\vec{k}) &=& m\tau^{z} + \sum_{i=x,y,z}\left[t_{1,i}\tau^{z}\cos(k_{i}) + t_{PH,i}\mathds{1}_{\tau\sigma}\cos(k_{i}) + t_{2,i}\tau^{y}\sin(k_{i}) + v_{1,i}\tau^{x}\sigma^{i}\sin(k_{i})\right] \nonumber \\
&+& v_{2,xy}\tau^{x}\sigma^{z}(\sin(k_{x}) + \sin(k_{y})) + v_{2,z}\tau^{x}\sigma^{x}\sin(k_{z}), 
\label{eq:TI}
\end{eqnarray}
where $\mathds{1}_{\tau\sigma}$ is the $4\times 4$ identity in the space of $\tau$ and $\sigma$, only first-nearest-neighbor hopping is included, and where we have chosen orthorhombic lattice vectors for simplicity (Fig.~\ref{fig:TI}(a)).  We formulated Eq.~(\ref{eq:TI}) by adding additional terms to the Bernevig-Hughes-Zhang (BHZ) model of a 3D TI~\cite{AndreiTI,AndreiInversion} to break the extraneous rotation and mirror symmetries of that model.  Eq.~(\ref{eq:TI}) is invariant under the transformations~\cite{bernevigbook}: 
\begin{equation}
\mathcal{I}:\ \mathcal{H}_{TI}(\vec{k})\rightarrow \tau^{z}\mathcal{H}_{TI}(-\vec{k})\tau^{z},\ \mathcal{T}:\ \mathcal{H}_{TI}(\vec{k})\rightarrow \sigma^{y}\mathcal{H}_{TI}^{*}(-\vec{k})\sigma^{y},
\label{eq:TIsyms}
\end{equation}
and realizes a 3D TI when we choose the parameters:
\begin{eqnarray}
m &=& -5,\ t_{x}= 2.3,\ t_{y} = 2.5,\ t_{z} = 3,\ t_{PH,x}=t_{PH,y} = 0.3,\ t_{PH,z}=0,\ t_{2,x} = 0.9,\ t_{2,y} =  t_{2,z} = 0, \nonumber \\
 v_{1,x} &=& v_{1,y} = 3.2,\ v_{1,z} = 2.4,\ v_{2,xy} = 1.5,\ v_{2,z} = 0.4.
 \label{eq:TIparams}
\end{eqnarray}

In Fig.~\ref{fig:TI}(a-f), we show the bulk BZ, bulk bands, $x$-directed slab bands, and $x$-directed Wilson loops of $\mathcal{H}_{TI}(\vec{k})$ using the parameters in Eq.~(\ref{eq:TIparams}).  The $x$-directed slab bands are gapless, and exhibit Kramers pairs of linear surface degeneracies at $k_{z}=k_{y}=0$ (Fig.~\ref{fig:TI}(c)).  We have also verified that $y$- and $z$- directed slabs also exhibit the same unpaired surface cones.  Taking the $x$-directed Wilson loop $W_{1}(k_{y},k_{z})$ over the two bands below $E=0$ in Fig.~\ref{fig:TI}(a), we observe that it exhibits $\mathbb{Z}_{2}$-nontrivial winding at $k_{z}=0$, is trivial and gapped at values of $k_{z}$ away from $0$ and $\pi$, and is trivial and gapless at $k_{z}=\pi$ (Fig.~\ref{fig:TI}(d,e,f), respectively).  We have also verified that the Wilson loop also winds when taken in the $y$ and $z$ directions when taken in the $k_{x,z}=0$ and $k_{x,y}=0$ planes, respectively, as is expected for a 3D TI.

\subsection{Tight-Binding Model of an $\mathcal{I}$-Symmetric Axion Insulator}
\label{sec:IAXI}

To induce a transition from the 3D TI in~\ref{sec:tbTI} to an $\mathcal{I}$-symmetric AXI, we introduce the $\mathcal{T}$-breaking term:
\begin{equation}
V_{AXI}(\vec{k}) = m_{A}\sigma^{z} + m_{B}\tau^{z}\sigma^{z},
\end{equation}
to realize the Hamiltonian of an AXI:
\begin{equation}
\mathcal{H}_{AXI}(\vec{k}) = \mathcal{H}_{TI}(\vec{k}) + V_{AXI}(\vec{k}),
\label{eq:AXI}
\end{equation}
where $\mathcal{H}_{TI}(\vec{k})$ is the previous Hamiltonian for a 3D TI (Eq.~(\ref{eq:TI})), and where we realize the AXI phase by using the parameters:
\begin{equation}
m_{A} = 1.2,\ m_{B} = 0.3,
\label{eq:AXIparams}
\end{equation}
in addition to the previous parameters in Eq.~(\ref{eq:TIparams}).  As we have broken $\mathcal{T}$ symmetry while preserving translation and $\mathcal{I}$ symmetry, $\mathcal{H}_{AXI}(\vec{k})$ describes an AXI in the type-I magnetic SG $P\bar{1}$ (number 2.4 in the BNS setting~\cite{MagneticBook,BilbaoMagStructures}).  

\begin{figure}[h]
\centering
\includegraphics[width=\textwidth]{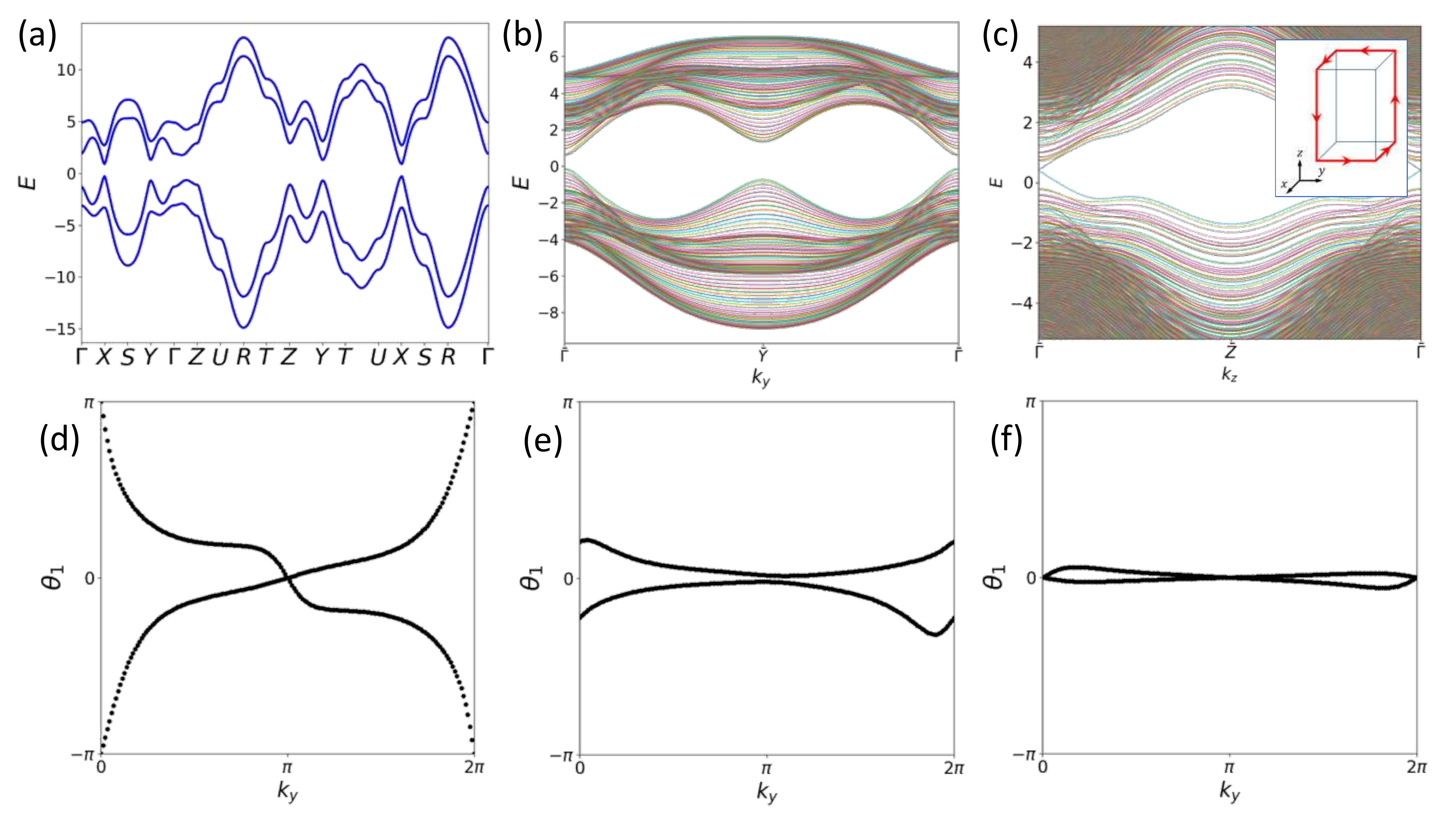}
\caption{(a) Bulk bands of $\mathcal{H}_{AXI}(\vec{k})$ (Eq.~(\ref{eq:AXI})), a four-band model of a 3D $\mathcal{I}$-symmetric AXI, plotted using the parameters in Eq.~(\ref{eq:AXIparams}).  (b) The $x$-directed slab bands of this AXI are fully gapped.  (c) The $z$-directed rod bands of Eq.~(\ref{eq:AXI}) exhibit two oppositely propagating chiral hinge modes, which are localized on opposing hinges (Fig.~\ref{fig:AXIcorners}(a)).  (d) The $x$-directed Wilson loop at $k_{z}=0$ exhibits helical winding.  (e) The $x$-directed Wilson loop at generic values of $k_{z}$ is gapped.  (f) The $x$-directed Wilson loop at $k_{z}=\pi$ does not wind, and exhibits Kramers degeneracies at $k_{y}=0,\pi$.  Unlike in the 3D TI in Fig.~\ref{fig:TI}, the Wilson degeneracies at $k_{y}=0,\pi$ in (d) and (f) are protected by bulk inversion symmetry and the presence of only two bands in the Wilson projector~\cite{ArisInversion,TMDHOTI,YoungkukMonopole,KoreanFragile,WiederDefect}.}
\label{fig:IAXI}
\end{figure}

\begin{figure}[h]
\centering
\includegraphics[width=\textwidth]{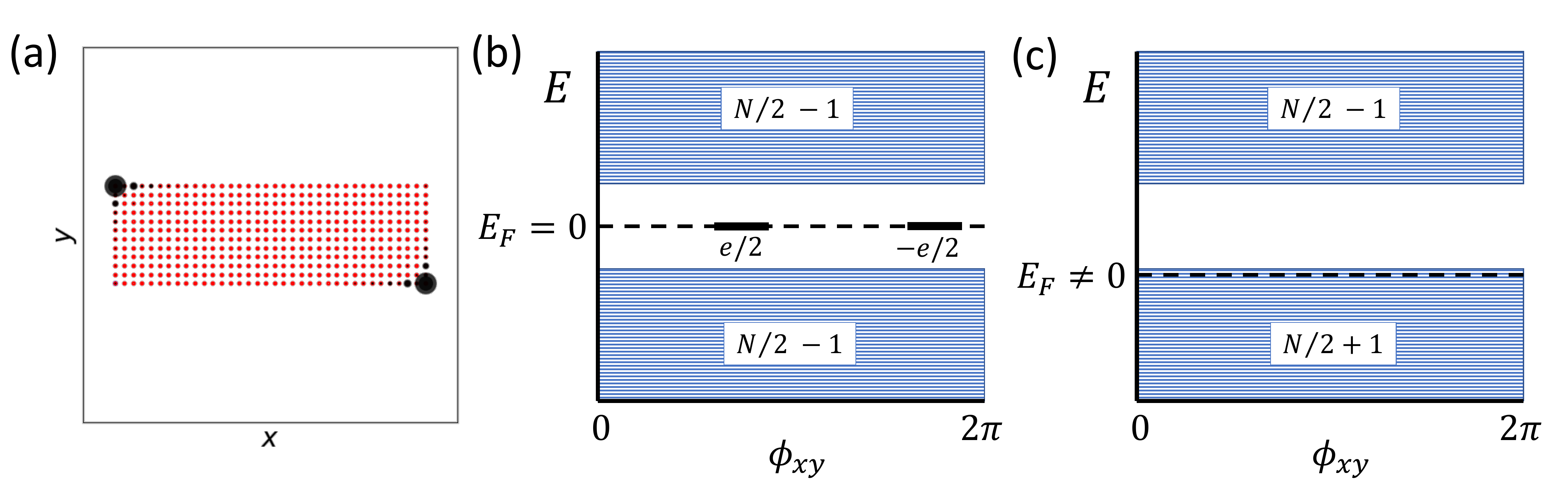}
\caption{(a) The localization in the $xy$-plane of the two hinge states at $k_{z}=0$ of the $z$-directed AXI rod in Fig.~\ref{fig:IAXI}(c).  The $k_{z}=0$ plane of Eq.~(\ref{eq:AXI}) with the parameters in Eq.~(\ref{eq:AXIparams}) is equivalent to a 2D insulator with two corner modes of opposite charge related by $\mathcal{I}$ symmetry.  The presence of these modes is indicated in a magnetic 2D insulator with $\mathcal{I}$ symmetry with two occupied bands by the Wilson loop winding in Fig.~\ref{fig:IAXI}(d), and in an insulator with more occupied bands by a nontrivial nested Berry phase $\gamma_{2}$ (\ref{sec:oneBandInversion},~\ref{sec:twoBandsInversion}, and Ref.~\onlinecite{TMDHOTI}).  (b) If the system is particle-hole-symmetric, then the two corner modes appear as anomalous, zero-energy, midgap states, and the valence and conduction manifolds are each missing one state.  In (b) and (c), we depict the spectrum of the Hamiltonian of the $k_{z}=0$ plane terminated in a disc geometry~\cite{HingeSM,TMDHOTI} plotted as a function of the polar angle $\phi_{xy}$ of the position-space $xy$-plane.  As shown in Ref.~\onlinecite{TMDHOTI}, the midgap 0D states at $\mathcal{I}$-related angles $\phi$ and $\phi+\pi$ are (anti)solitions of the gapped edge Hamiltonian, and thus, under softly broken $\mathcal{I}$ symmetry, can be assigned charges~\cite{SSH,RiceMele,multipole,NiemiSemenoff,GoldstoneWilczek,HingeSM} $\pm e/2$.  (c) If particle-hole symmetry is relaxed without closing a bulk or edge gap, then these modes can float in energy into the valence or conduction manifold.  However, this process donates two extra states to one of the manifolds, resulting in a measurable imbalance in the number of states above and below the (bulk and edge) gap~\cite{HingeSM}.  If exactly half of the states are still occupied, then the Fermi level will now lie just inside the valence or conduction manifold; here, it is depicted near the top of the valence manifold.}
\label{fig:AXIcorners}
\end{figure}

In Fig.~\ref{fig:IAXI}(a-f), we show the bulk bands, $x$-directed slab bands, $z$-directed rod bands, and $x$-directed Wilson loops of $\mathcal{H}_{AXI}(\vec{k})$ using the parameters in Eq.~(\ref{eq:AXIparams}).  The bulk bands are now generically singly degenerate, and the gap at half-filling is preserved by the introduction $V_{AXI}(\vec{k})$ with the parameters in Eq.~(\ref{eq:AXIparams}) (Fig.~\ref{fig:IAXI}(a)).  The $x$-directed slab bands are fully gapped (Fig.~\ref{fig:IAXI}(c)).  We have also verified that $y$- and $z$- directed slabs of $\mathcal{H}_{AXI}(\vec{k})$ are also fully gapped.  This allows us to calculate the bands of a $z$-directed rod of $\mathcal{H}_{AXI}(\vec{k})$, \emph{i.e.}, a tight-binding model that is finite in the $x$ and $y$ directions and infinite in the $z$ direction (Fig.~\ref{fig:IAXI})(c)).  We observe that the rod bands exhibit two 1D chiral modes propagating in the $z$ direction, with oppositely chiral modes localized on opposing hinges (Fig.~\ref{fig:IAXI}(c) and Fig.~\ref{fig:AXIcorners}(a)).  This confirms the recent recognition that 3D $\mathcal{I}$-symmetric AXIs are in fact symmetry-indicated magnetic higher-order topological insulators (HOTIs)~\cite{AshvinMagnetic,FanHOTI,HarukiLayers,HOTIBernevig,EzawaMagneticHOTI,TMDHOTI,WiederDefect,VDBHOTI}.  It is crucial to note that there is no symmetry or topology requirement for the corner modes of the $k_{z}=0$ plane (Fig.~\ref{fig:AXIcorners}(a)) to appear as midgap states~\cite{HOTIChen,KoreanFragile,WladCorners,HingeSM}.  Without closing a bulk or edge gap, the modes at $k_{z}=0$ in Fig.~\ref{fig:IAXI}(c) can float into the valence or conduction manifold (Fig.~\ref{fig:AXIcorners}(c)).  However, because the corner modes lie at the spectral center (\emph{i.e.}, there are exactly $N/2 -1$ states above and below them in energy (Fig.~\ref{fig:AXIcorners}(b))), then floating them into the valence or conduction manifold will result in a state imbalance above and below the bulk and edge gap (Fig.~\ref{fig:AXIcorners}(c)).  The presence of this state imbalance is indicated in a two-band model by the winding Wilson loop in Fig.~\ref{fig:IAXI}(d), and in a many-band model by $\det(W_{2})=-1$, \emph{i.e.}, that the nested Berry phase $\gamma_{2}$ (Eq.~(\ref{eq:gamma2})) is nontrivial (\ref{sec:oneBandInversion},~\ref{sec:twoBandsInversion}, and Ref.~\onlinecite{TMDHOTI}).  This is analogous to the $\mathcal{I}$-symmetric, particle-hole-asymmetric Su-Schrieffer-Heeger chain~\cite{SSH}, in which a nontrivial Berry phase $\gamma_{1}$ indicates the presence of end charge, but not necessarily midgap end modes~\cite{NiemiSemenoff}.  This can also be understood from a field-theory perspective by considering the Goldstone-Wilczek formulation of a Jackiw-Rebbi domain wall~\cite{JackiwRebbi} with a complex mass~\cite{WilczekAxion,GoldstoneWilczek,NiemiSemenoff}, and is explored in more detail for related fragile phases in Refs.~\onlinecite{HingeSM,WiederDefect}. 

\begin{figure}[h]
\centering
\includegraphics[width=0.45\textwidth]{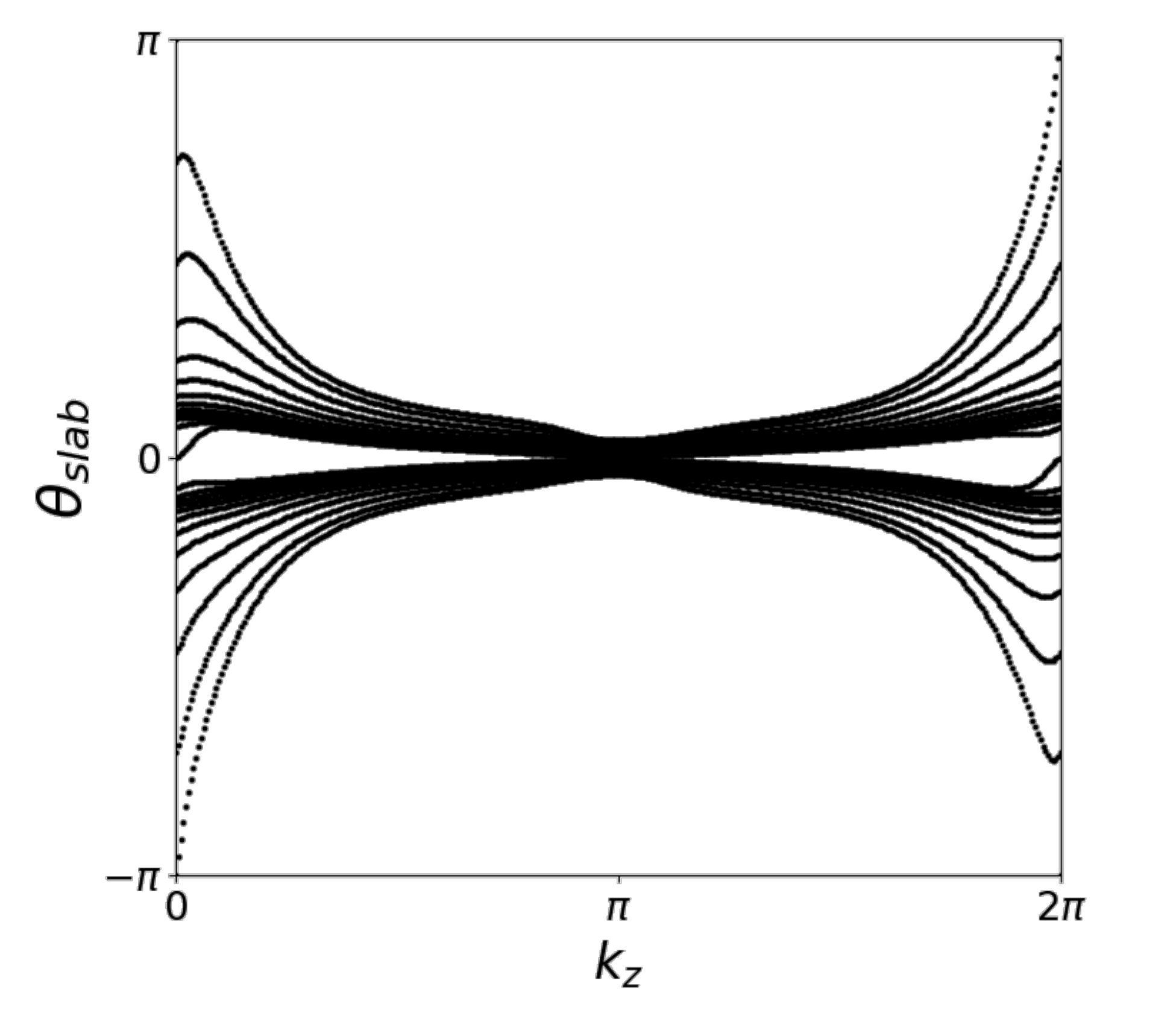}
\caption{The $y$-directed Wilson bands of \emph{all} of the bands below the gap in the $x$-directed AXI slab in Fig.~\ref{fig:IAXI}(b).  This Wilson loop exhibits $C_{slab}=+1$ winding, confirming that in a slab of an $\mathcal{I}$-symmetric AXI, the two opposing surfaces with the same half-integer anomalous Hall conductivity~\cite{FuKaneMele,FuKaneInversion,AndreiInversion,QHZ,FanHOTI,VDBAxion,DiracInsulator,MulliganAnomaly,VDBHOTI} combine to form an effective Chern insulator with an \emph{integer} Hall conductivity~\cite{HOTIBernevig}.}
\label{fig:slabInversion}
\end{figure}

The $x$-directed Wilson loop of $\mathcal{H}_{AXI}(\vec{k})$ displays the same winding as that of the 3D TI in Fig.~\ref{fig:TI}(d-f): it winds at $k_{z}=0$, is gapped and trivial at generic values of $k_{z}$, and is gapless and trivial at $k_{z}=\pi$ (Fig.~\ref{fig:IAXI}(d,e,f), respectively).  However, the gapless points at $k_{y}=0,\pi$ in the winding (trivial) Wilson spectrum at $k_{z} = 0$ ($k_{z}=\pi$) are not protected by $\mathcal{T}$ symmetry, as they previously were in Fig.~\ref{fig:TI}(d-f).  Instead, they arise because in the four-band model of an AXI described by Eqs.~(\ref{eq:AXI}) and~(\ref{eq:AXIparams}), the inversion eigenvalues of the two lower bands (Fig.~\ref{fig:IAXI}(a)) are $+1$ at all TRIM points except $\Gamma$, where they are both $-1$.  As detailed in Ref.~\onlinecite{ArisInversion}, this mandates that the Wilson spectrum in the $k_{z}=0,\pi$ planes exhibits gapless points, and in particular that the two-band Wilson spectrum at $k_{z}=0$ winds.  However, as we will see in~\ref{sec:oneBandInversion} and~\ref{sec:twoBandsInversion}, the introduction of trivial bands to the Wilson projector can lift these Wilson degeneracies~\cite{ArisInversion,TMDHOTI,YoungkukMonopole,KoreanFragile,WiederDefect}.  We have also verified that the $y$- and $z$-directed Wilson spectra exhibit the same behavior.   

We also confirm that opposing surfaces of the AXI described Eq.~(\ref{eq:AXI}) exhibit the same, anomalously quantized half-integer Hall conductivity $\sigma_{H}= e^{2}/(2h)$ (when measuring with respect to the positive $x,y,z$ directions for all surfaces (Fig.~\ref{fig:IAXI}(c)))~\cite{FuKaneMele,FuKaneInversion,AndreiInversion,QHZ,FanHOTI,VDBAxion,DiracInsulator,MulliganAnomaly,VDBHOTI}.  In Fig.~\ref{fig:slabInversion}, we plot the $y$-directed Wilson bands taken over \emph{all} of the bands below the gap in the $x$-directed slab shown in Fig.~\ref{fig:IAXI}(b); this Wilson loop is well-defined because the slab itself is just a gapped tight-binding model that is periodic in two directions ($x$ and $y$) with $4N_{layers}$ orbitals per $y$- and $z$-indexed unit cell.  The slab Wilson loop in Fig.~\ref{fig:slabInversion} exhibits an overall $C_{slab}=+1$ winding, confirming that when an $\mathcal{I}$-symmetric AXI is cleaved into a slab geometry, it is topologically equivalent to a $C=\pm 1$ Chern insulator~\cite{HOTIBernevig}, \emph{i.e.}, an isolated quasi-2D insulator with an \emph{integer} Hall conductivity.  

\subsubsection{Coupling One Additional Band to an $\mathcal{I}$-Symmetric AXI}
\label{sec:oneBandInversion}

We now demonstrate that the presence of appropriately chosen additional bands can remove all of the Wilson loop winding of an $\mathcal{I}$-symmetric AXI for a Wilson loop taken over all of the occupied bands, \emph{even if the additional bands lie far below the Fermi energy}.  We begin by placing two additional Kramers pairs of $s$ orbitals at the general Wyckoff position~\cite{BCS1,BCS2,BilbaoMagStructures} ($2i$) of the unit cell of $\mathcal{H}_{AXI}(\vec{k})$ (Eq.~(\ref{eq:AXI})).  The positions of these two pairs of orbitals are related by inversion about the $1a$ position ($(x,y,z)=(0,0,0)$):
\begin{equation}
\vec{r}_{3,\sigma} = (u_{3x},u_{3y},u_{3z}),\ \vec{r}_{4,\sigma} = -\vec{r}_{3,\sigma} = (-u_{3x},-u_{3y},-u_{3z}),
\label{eq:IorbitalPos}
\end{equation}
where we index the two additional Kramers pairs of orbitals as $3$ and $4$ to distinguish them from the spin-split pairs of $s$ and $p$ orbitals already present at the $1a$ position of $\mathcal{H}_{AXI}(\vec{k})$ (Eq.~(\ref{eq:AXI})), and where $\sigma$ denotes the spin degree of freedom.  As extensively detailed in Refs.~\onlinecite{ZakBandrep1,ZakBandrep2,QuantumChemistry,Bandrep1,Bandrep2,Bandrep3,JenFragile1,JenFragile2,BarryFragile}, this is equivalent to placing additional spin-degenerate pairs of $s$ and $p$ orbitals at the $1a$ position, where the effective orbitals at the $1a$ position are formed from the (anti)bonding combinations of orbitals $3$ and $4$.  We choose to place Kramers pairs $3$ and $4$ at the general position so that the Wilson spectrum taken over all of the bands from these orbitals can be gapped, even when they are uncoupled to other orbitals.  To summarize, we are choosing parameters to ensure that (within each $\sigma^{z}$ spin sector) bands from orbitals $3$ and $4$ carry the same parity eigenvalues as $s$ and $p$ orbitals at the $1a$ position, and that the two-band Wilson loop over bands from the $3,\downarrow$ and $4,\downarrow$ orbitals is gapped, even when those orbitals are decoupled from the original $1a$ $s$ and $p$ orbitals.  To understand the requirement of nonzero $\vec{r}_{3/4,\sigma}$ in Eq.~(\ref{eq:IorbitalPos}), we first note that taking the $x$-directed Wilson loop over an entire set of bands that are decoupled from the rest of a system simply yields flat Wilson bands at $\theta_{1}=u_{ix}\pi$, where $u_{ix}$ is the $x$ coordinate of the $i^{\text{th}}$ orbital (Eq.~(\ref{eq:IorbitalPos}) and taking the Wilson projector in Appendix~\ref{sec:W2} to be the identity).  For example, if spinless $s$ and $p$ orbitals are placed exactly at the $1a$ position of an $\mathcal{I}$-symmetric space group, and are the only orbitals in the entire Hilbert space, then the Wilson loop over both bands from these orbitals will not be gapped, even though the parity eigenvalue criterion from Ref.~\onlinecite{ArisInversion} states that the Wilson bands should generically be split.  We therefore must move orbitals $3$ and $4$ off of the origin for them to acquire nonzero Berry phases when they are uncoupled to other orbitals.  In this work, this will primarily apply to the four-band Wilson projectors in~\ref{sec:twoBandsInversion}, which, in the decoupled limit, contain the entire Hilbert space of bands from the $3,\downarrow$ and $4,\downarrow$ orbitals.

To split Kramers pairs $3$ and $4$ into effective $s$ and $p$ bands induced from the $1a$ position, we introduce the simple coupling between the two new orbitals within the same unit cell $i$:
\begin{eqnarray}
V_{sp} &=& t_{sp}\sum_{\sigma=\uparrow,\downarrow}\left(c^{\dag}_{3\sigma,i}c_{4,\sigma,i} + c^{\dag}_{4\sigma,i}c_{3,\sigma,i}\right) + t_{1}\left(c^{\dag}_{3\sigma,i}c_{3,\sigma,i} + c^{\dag}_{4\sigma,i}c_{4,\sigma,i}\right) \nonumber \\
V_{sp}(\vec{k}) &=& t_{sp}\left(\begin{array}{cccc}
0 & 0 & 0 & 0 \\
0 & 0 & 0 & 0 \\
0 & 0 & 0 & 1 \\
0 & 0 & 1 & 0\end{array}\right)\otimes\mathds{1}_{\sigma} + t_{1}\left(\begin{array}{cccc}
0 & 0 & 0 & 0 \\
0 & 0 & 0 & 0 \\
0 & 0 & 1 & 0 \\
0 & 0 & 0 & 1 \end{array}\right)\otimes\mathds{1}_{\sigma},
\label{eq:vsp}
\end{eqnarray}
where $\mathds{1}_{\sigma}$ denotes the $2\times 2$ identity in the spin subspace, and where the $4\times 4$ matrix is indexed by $(c_{s}\ c_{p}\ c_{3}\ c_{4})$ where $c_{s,p}$ denote the original pairs of $s$ and $p$ orbitals in $\mathcal{H}_{AXI}(\vec{k})$ (Eq.~(\ref{eq:AXI})), respectively.  We then introduce simple ferromagnetic splitting for the orbitals at the general position:
\begin{equation}
U_{s^{z}}(\vec{k}) = u_{s^{z}}\left(\begin{array}{cccc}
0 & 0 & 0 & 0 \\
0 & 0 & 0 & 0 \\
0 & 0 & 1 & 0 \\
0 & 0 & 0 & 1 \end{array}\right)\otimes\sigma^{z}.
\label{eq:Usz}
\end{equation}
Next, we form a new Hamiltonian in this expanded $8\times 8$ space:
\begin{equation}
\mathcal{H}_{U}(\vec{k}) = \mathcal{H}_{AXI}(\vec{k}) + V_{sp}(\vec{k}) + U_{s^{z}}(\vec{k}),
\label{eq:IUncoupled}
\end{equation}
where $\mathcal{H}_{AXI}(\vec{k})$ is the previous $4\times 4$ Hamiltonian of an $\mathcal{I}$-symmetric AXI with two occupied bands (Eq.~(\ref{eq:AXI})).  Eq.~(\ref{eq:IUncoupled}) remains invariant under $\mathcal{I}$ symmetry, which now takes the form:
\begin{equation}
\mathcal{I}:\ \mathcal{H}_{U}(\vec{k})\rightarrow \left[\left(\begin{array}{cccc}
1 & 0 & 0 & 0 \\
0 & -1 & 0 & 0 \\
0 & 0 & 0 & 1 \\
0 & 0 & 1 & 0\end{array}\right)\otimes\mathds{1}_{\sigma}\right]\mathcal{H}_{U}(-\vec{k}) \left[\left(\begin{array}{cccc}
1 & 0 & 0 & 0 \\
0 & -1 & 0 & 0 \\
0 & 0 & 0 & 1 \\
0 & 0 & 1 & 0\end{array}\right)\otimes\mathds{1}_{\sigma}\right].
\label{eq:8inv}
\end{equation}

\begin{figure}[h]
\centering
\includegraphics[width=\textwidth]{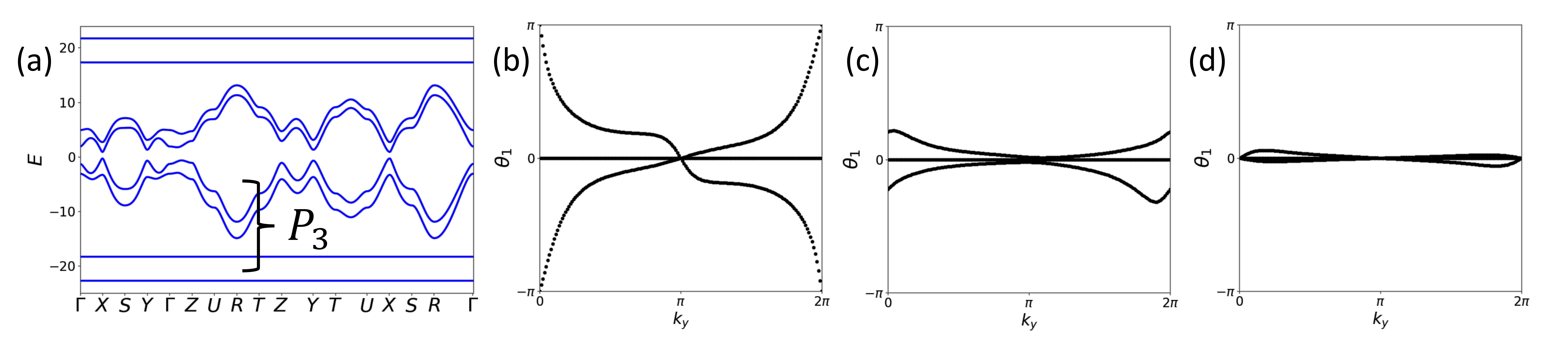}
\caption{(a) Bulk bands of the eight-band, uncoupled Hamiltonian $\mathcal{H}_{U}(\vec{k})$ (Eq.~(\ref{eq:IUncoupled})) plotted with the parameters in Eqs.~(\ref{eq:TIparams}),~(\ref{eq:AXIparams}), and~(\ref{eq:oneBandParamsI}).  (b-d) The $x$-directed Wilson loops at $k_{z} = 0$, a generic value, and $\pi$, taken over the three bands in the projector $P_{3}$ in (a); the lowest band in energy in $P_{3}$ has the same inversion eigenvalues as a band induced from a $p,\downarrow$ orbital at the $1a$ position~\cite{ZakBandrep1,ZakBandrep2,QuantumChemistry,Bandrep1,Bandrep2,Bandrep3,JenFragile1,JenFragile2,BarryFragile}.  Because the $p,\downarrow$ band is decoupled from the other two bands in the projector, the Wilson spectra in (b-d) decompose into the superpositions of the previous Wilson spectra in Fig.~\ref{fig:IAXI}(d-f) and an additional flat Wilson band at the Wilson energy $\theta_{1}(k_{y},k_{z})=0$.}
\label{fig:oneUncoupled}
\end{figure}

\begin{figure}[h]
\centering
\includegraphics[width=\textwidth]{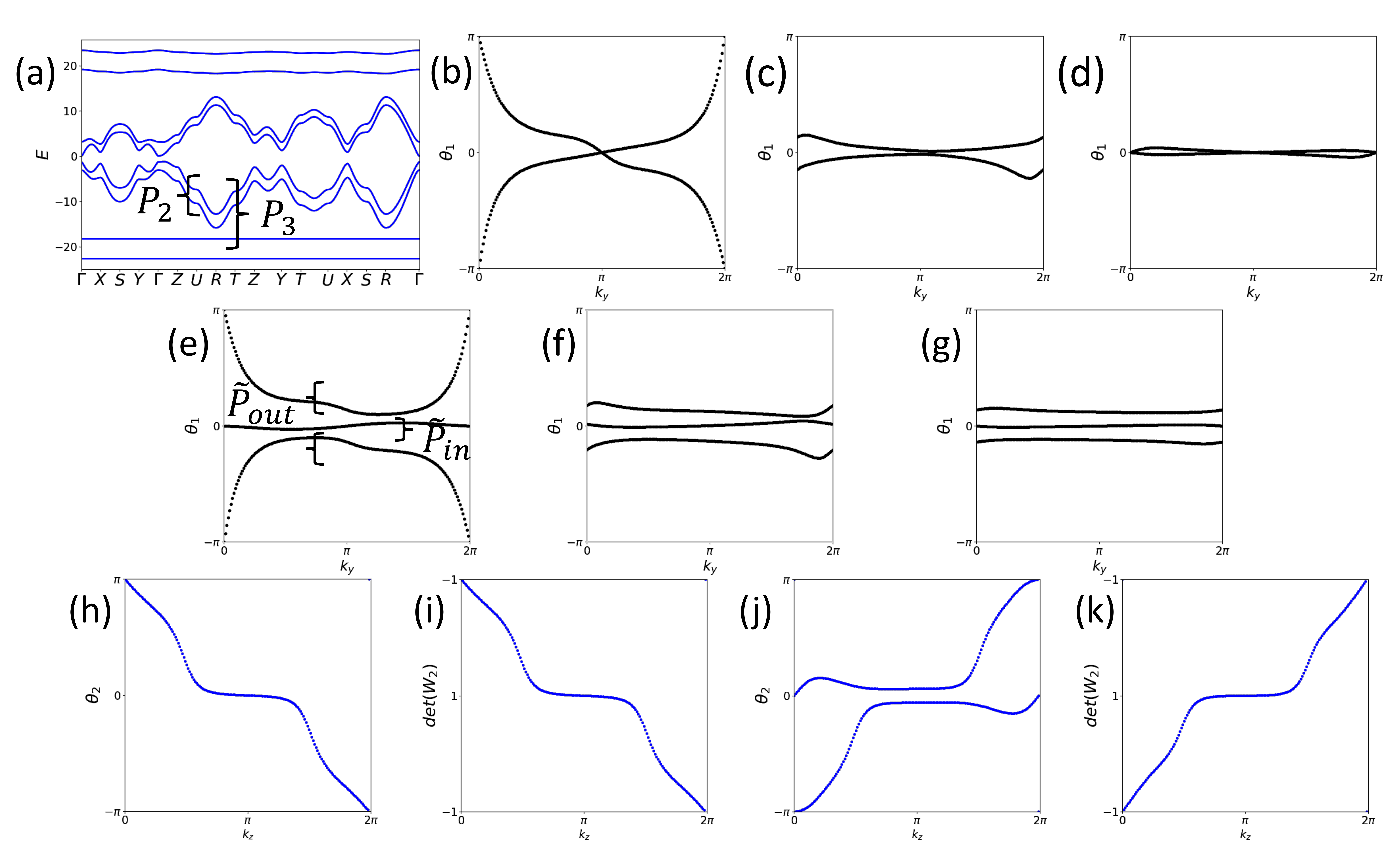}
\caption{(a) Bulk bands of the eight-band, coupled Hamiltonian $\mathcal{H}_{C}(\vec{k})$ (Eq.~(\ref{eq:ICoupled})) plotted with the parameters in Eqs.~(\ref{eq:TIparams}),~(\ref{eq:AXIparams}),~(\ref{eq:oneBandParamsI}), and~(\ref{eq:coupledParamsOneI}).  (b-d) The $x$-directed Wilson loops at $k_{z} = 0$, a generic value, and $\pi$, taken over the two bands in the projector $P_{2}$; these are the original fragile valence bands of the four-band AXI model in Eq.~(\ref{eq:AXI}).  Even though these bands are coupled to other bands below the Fermi energy, as they are isolated from them by an energy gap, they still have a well-defined nontrivial (fragile) topology indicated by the winding of the $x$- (and $y$- and $z$-) directed Wilson loop~\cite{BarryFragile}.  (e-g) The $x$-directed Wilson loops at $k_{z} = 0$, a generic value, and $\pi$, taken over the three bands in the projector $P_{3}$ in (a); the lowest band in energy in $P_{3}$ has the same inversion eigenvalues as a band induced from a $p,\downarrow$ orbital at the $1a$ position~\cite{ZakBandrep1,ZakBandrep2,QuantumChemistry,Bandrep1,Bandrep2,Bandrep3,JenFragile1,JenFragile2,BarryFragile}.  Here, unlike in Fig.~\ref{fig:oneUncoupled}, the Wilson spectrum is gapped and free of winding, allowing us to define particle-hole symmetric nested Wilson projectors onto the inner and outer Wilson bands~\cite{TMDHOTI,ZhidaBLG}.  (h) The $y$-directed nested Wilson loop of the $\tilde{P}_{in}$ Wilson band, plotted as a function of $k_{z}$; it exhibits $C_{\gamma_{2}}=-1$ spectral flow.  (i) $\det(W_{2}(k_{z}))$ of the $\tilde{P}_{in}$ Wilson band; it exhibits $C_{\gamma_{2}}=-1$ spectral flow and is quantized at $-1$ ($+1$) at $k_{z}=0$ $(k_{z}=\pi$) (\ref{sec:detW2Inv}).   (j) The $y$-directed nested Wilson loop of the $\tilde{P}_{out}$ Wilson bands, plotted as a function of $k_{z}$; it exhibits $C_{\gamma_{2}}=+1$ spectral flow.  (k) $\det(W_{2}(k_{z}))$ of the $\tilde{P}_{out}$ Wilson bands; it exhibits $C_{\gamma_{2}}=+1$ spectral flow and is quantized at $-1$ ($+1$) at $k_{z}=0$ $(k_{z}=\pi$) (\ref{sec:detW2Inv}).}
\label{fig:oneCoupled}
\end{figure}

In Fig.~\ref{fig:oneUncoupled}, we plot the bulk bands (a) and $x$-directed Wilson loops (b-d) of Eq.~(\ref{eq:IUncoupled}) using the parameters in Eqs.~(\ref{eq:TIparams}) and~(\ref{eq:AXIparams}), as well as:
\begin{equation}
u_{3x} = 0.35,\ u_{3y} = 0.15,\ u_{3z} = 0.12,\ t_{sp} = 20,\ t_{1} = -0.5,\ u_{s^{z}} = 2.2,
\label{eq:oneBandParamsI}
\end{equation}
in Eqs.~(\ref{eq:IorbitalPos}),~(\ref{eq:vsp}) and~(\ref{eq:Usz}).  In this limit, the original bands near $E=0$ are completely decoupled from the new bands induced from the orbitals at the general position.  For the Wilson loops in Fig.~\ref{fig:oneUncoupled}(b-d), we include three bands in the Wilson projector $P_{3}$: the original two valence bands of $\mathcal{H}_{AXI}(\vec{k})$ (Eq.~(\ref{eq:AXI})), and a band from the new orbitals which has the same symmetry eigenvalues as a band induced from a $p,\downarrow$ orbital at the $1a$ position~\cite{ZakBandrep1,ZakBandrep2,QuantumChemistry,Bandrep1,Bandrep2,Bandrep3,JenFragile1,JenFragile2,BarryFragile}.  In the uncoupled limit of Eq.~(\ref{eq:IUncoupled}), the Wilson loop just decomposes into the Wilson spectra of the original valence bands of $\mathcal{H}_{AXI}(\vec{k})$ (Fig.~\ref{fig:IAXI}(d-f)) and an additional flat Wilson band superposed at the Wilson energy $\theta_{1}(k_{y},k_{z})=0$.

We now couple the new orbitals to the original four $s$ and $p$ orbitals in $\mathcal{H}_{AXI}(\vec{k})$ (Eq.~(\ref{eq:AXI})) by introducing the term:
\begin{equation}
V_{C}(\vec{k}) = v_{C}\left(\begin{array}{cccc}
0 & 0 & 1 & 1 \\
0 & 0 & 0 & 0 \\
1 & 0 & 0 & 0 \\
1 & 0 & 0 & 0\end{array}\right)\otimes\mathds{1}_{\sigma}.
\label{eq:vc}
\end{equation}
We then form a coupled Hamiltonian by adding $V_{C}$ to Eq.~(\ref{eq:IUncoupled}):
\begin{equation}
\mathcal{H}_{C}(\vec{k}) = \mathcal{H}_{U}(\vec{k}) + V_{C}(\vec{k}), 
\label{eq:ICoupled}
\end{equation}
which we confirm is also invariant under $\mathcal{I}$ symmetry in the expanded $8\times 8$ basis (Eq.~(\ref{eq:8inv})).  In Fig.~\ref{fig:oneCoupled}, we plot the bulk bands (a), $x$-directed Wilson loops (b-g) using two choices of Wilson projectors, and the $y$-directed nested Wilson loops (h-k) (Fig.~\ref{fig:Wilson}) of Eq.~(\ref{eq:ICoupled}) using the parameters in Eqs.~(\ref{eq:TIparams}),~(\ref{eq:AXIparams}), and~(\ref{eq:oneBandParamsI}), as well as:
\begin{equation}
V_{C} = 4,
\label{eq:coupledParamsOneI}
\end{equation}
in Eq.~(\ref{eq:vc}).   In the band structure of $\mathcal{H}_{C}(\vec{k})$ (Fig.~\ref{fig:oneCoupled}(a)), the original two fragile valence bands of Eq.~(\ref{eq:AXI}) remain separated from the new bands by a gap below $E=0$, allowing us to define two Wilson projectors: a projector $P_{3}$ onto the three valence bands closest to $E=0$, and a projector $P_{2}$ onto the original fragile valence bands.  As shown in Ref.~\onlinecite{BarryFragile}, as long as $P_{2}$ is well defined, the Wilson loop of the fragile bands will still wind.  We confirm this in Fig.~\ref{fig:oneCoupled}(b-d), in which we observe the $x$-directed Wilson loop of the bands in $P_{2}$ to exhibit the same winding as the fragile bands of the original four-band AXI model (Fig.~\ref{fig:IAXI}(d-f)); we also confirm that the $y$- and $z$-directed Wilson loops of the $P_{2}$ bands also wind.  However, when we calculate the $x$-directed (and $y$- and $z$-directed) Wilson loops of the $P_{3}$ bands in Fig.~\ref{fig:oneCoupled}(a) (Fig.~\ref{fig:oneCoupled}(e-g)), \emph{all of the Wilson loop winding has been removed}.  Though we have only shown in Fig.~\ref{fig:IAXI}(f) the $x$-directed Wilson loop of the $P_{3}$ bands at one generic value of $k_{z}$, we have confirmed that there are no discernible Wilson crossings at all values of $k_{z}$.  This is in agreement with the absence of symmetries at generic values of $k_{y,z}$ in the $x$-directed Wilson loop.

One might naively deduce from this that an AXI with three valence bands can be described by maximally-localized, symmetric Wannier functions.  However, as the three bands in Wilson projector $P_{3}$ cannot be expressed as a linear combination of elementary band representations (EBRs), we know from the results of Refs.~\onlinecite{ZakBandrep1,ZakBandrep2,QuantumChemistry,Bandrep1,Bandrep2,Bandrep3,JenFragile1,JenFragile2,BarryFragile} that they are not Wannierizable.  We resolve this by calculating the \emph{nested} Wilson loop, and demonstrating that there is still $k_{z}$-dependent \emph{nested} Wannier flow.  Taking the split Wilson spectrum in Fig.~\ref{fig:oneCoupled}(e-g), we define projectors $\tilde{P}_{in,out}$ onto groupings of Wilson bands related by Wilson particle-hole symmetry, which we use to calculate the $y$-directed nested Wilson loop $W_{2}(k_{z})$ (Fig.~\ref{fig:Wilson} and~\ref{sec:W2}).  As $\mathcal{I}$ acts on the Wilson loop as a unitary particle-hole symmetry~\cite{ArisInversion,ArisNoSOC,Cohomological,HourglassInsulator,DiracInsulator} that takes $\theta_{1}(k_{y},k_{z})$ to $-\theta_{1}(-k_{y},-k_{z})$, then, as shown in Refs.~\onlinecite{TMDHOTI,ZhidaBLG} and in~\ref{sec:detW2Inv}, $\mathcal{I}$ remains a symmetry of $W_{2}(k_{z})$ under this choice of nested Wilson projectors.  In Fig.~\ref{fig:oneCoupled}(h,i) (Fig.~\ref{fig:oneCoupled}(j,k)), we plot the nested Wilson eigenvalues $\theta_{2}(k_{z})$ and $\det(W_{2}(k_{z}))$ for the inner (outer) Wilson band(s) in Fig.~\ref{fig:oneCoupled}(e-g), respectively.  Both exhibit spectral flow, with the inner (outer) Wilson band(s) exhibiting $C_{\gamma_{2}}=-1$ ($C_{\gamma_{2}}=+1$) winding.  Because the inner and outer Wilson bands exhibit opposite odd-integer Wilson Chern numbers $C_{\gamma_{2}}$, then, as $C_{\gamma_{2}}\text{ mod } 2$ is robust to Wilson gap closures in the presence of bulk $\mathcal{I}$ symmetry (\ref{sec:detW2Inv} and Fig.~\ref{fig:oddWindI}), it here indicates a well-defined nontrivial bulk topology.  Furthermore, in the $k_{z}=0,\pi$ planes, which are left invariant under the action of $\mathcal{I}$, the nested Berry phase $\gamma_{2}(k_{z})$ is quantized (Fig.~\ref{fig:oneCoupled}(i,k) and~\ref{sec:detW2Inv}), with the $k_{z} = 0$ ($k_{z}=\pi$) plane exhibiting a nontrivial (trivial) nested Berry phase:
\begin{equation}
\gamma_{2}(0) = \pi,\ \gamma_{2}(\pi) = 0.
\label{eq:oneBandsGamma2}
\end{equation}
The AXI with three valence bands is therefore \emph{not} a trivial, Wannierizable insulator, but is rather a magnetic HOTI characterized by the cyclic pumping of a 2D fragile phase with trivialized Wilson loop winding.  Specifically, the $\mathcal{I}$-symmetric AXI with three valence bands represents the chiral pumping of nested Berry phase from $0$ to $\pm\pi$ within each particle-hole symmetric grouping of Wilson bands.  The inner and outer Wilson bands exhibit opposite odd-integer Wilson Chern numbers $C_{\gamma_{2}}$ (\ref{sec:detW2Inv}); in a $z$-directed rod of an $\mathcal{I}$-symmetric AXI, this corresponds to the oppositely chiral flow of charge on hinges related by $\mathcal{I}$ symmetry (Fig.~\ref{fig:IAXI}(c) and Fig.~\ref{fig:AXIcorners}(a)).  

\subsubsection{Coupling Two Additional Bands to an $\mathcal{I}$-Symmetric AXI}
\label{sec:twoBandsInversion}

\begin{figure}[h]
\centering
\includegraphics[width=\textwidth]{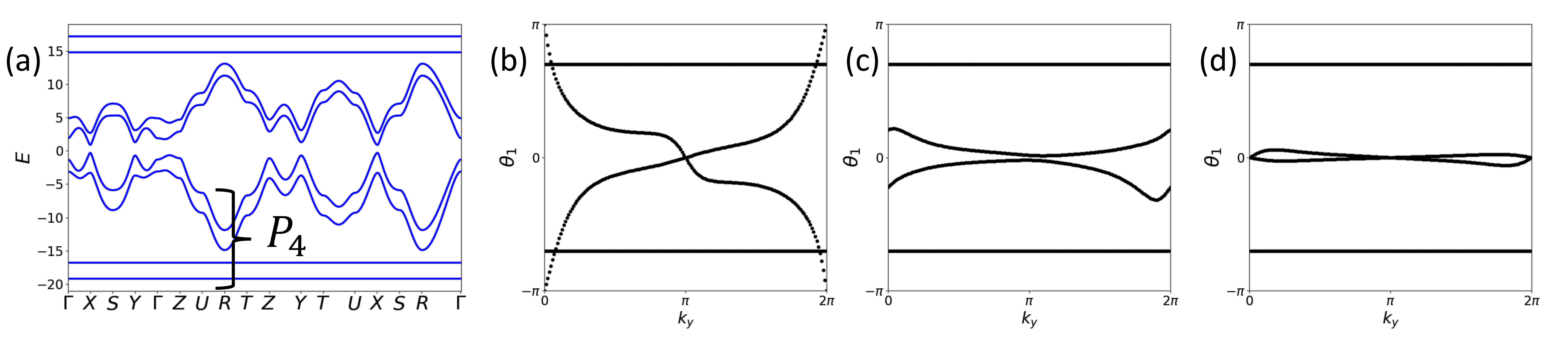}
\caption{(a) Bulk bands of the eight-band, uncoupled Hamiltonian $\mathcal{H}_{U}(\vec{k})$ (Eq.~(\ref{eq:IUncoupled})) plotted with the parameters in Eqs.~(\ref{eq:TIparams}),~(\ref{eq:AXIparams}), and~(\ref{eq:twoBandParamsI}).  (b-d) The $x$-directed Wilson loops at $k_{z} = 0$, a generic value, and $\pi$, taken over the four bands in the projector $P_{4}$ in (a); the lowest two bands in energy in $P_{4}$ have the same inversion eigenvalues as bands induced from $s,\downarrow$ and $p,\downarrow$ orbitals at the $1a$ position~\cite{ZakBandrep1,ZakBandrep2,QuantumChemistry,Bandrep1,Bandrep2,Bandrep3,JenFragile1,JenFragile2,BarryFragile}.  Because the $s,\downarrow$ and $p,\downarrow$ bands are decoupled from the other two bands in the projector, the Wilson spectra in (b-d) decompose into the superpositions of the previous Wilson spectra in Fig.~\ref{fig:IAXI}(d-f) and two additional flat Wilson bands at a particle-hole-symmetric pair of Wilson energies.}
\label{fig:twoUncoupled}
\end{figure}

\begin{figure}[h]
\centering
\includegraphics[width=\textwidth]{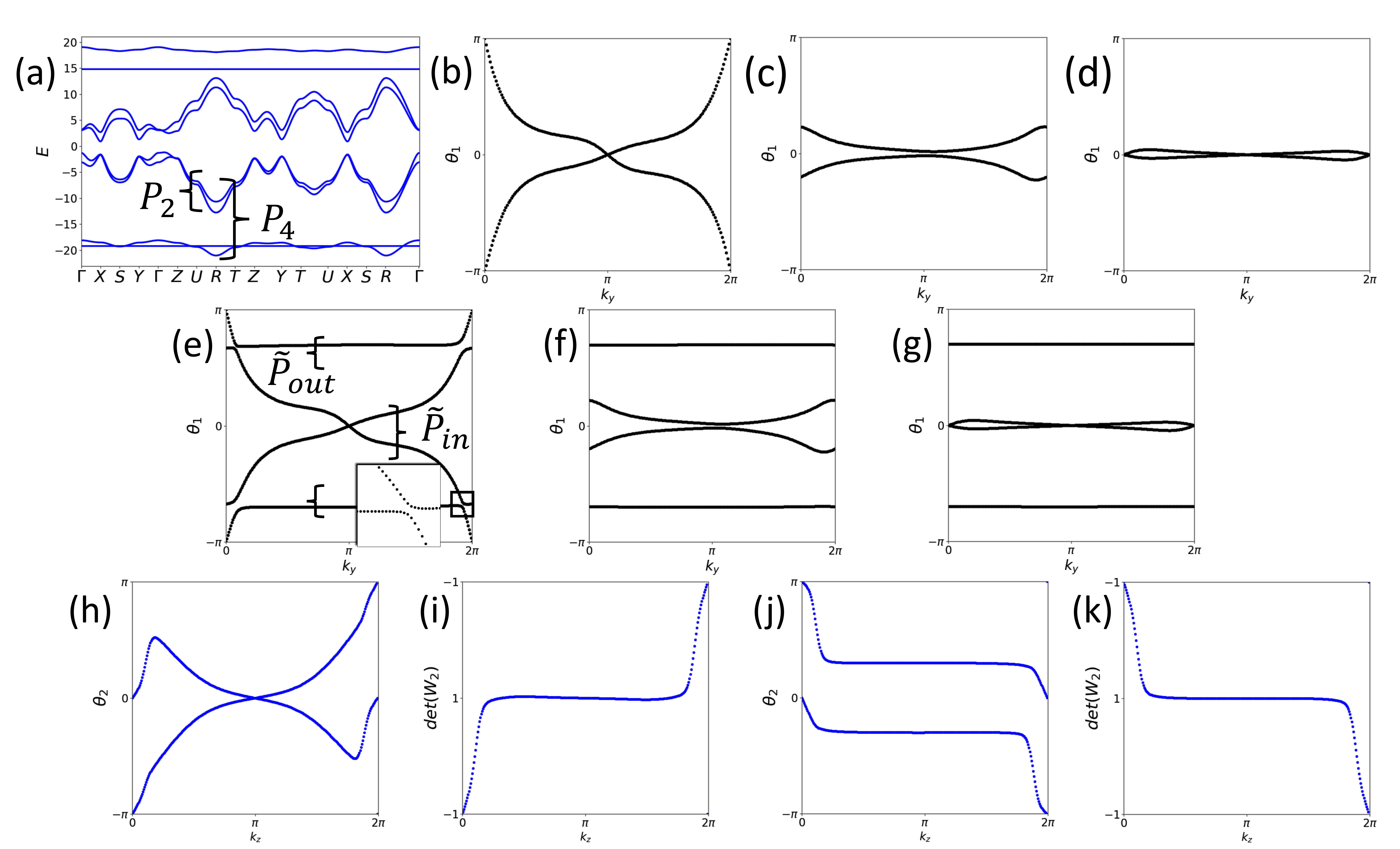}
\caption{(a) Bulk bands of the eight-band, coupled Hamiltonian $\mathcal{H}_{C}(\vec{k})$ (Eq.~(\ref{eq:ICoupled})) plotted with the parameters in Eqs.~(\ref{eq:TIparams}),~(\ref{eq:AXIparams}),~(\ref{eq:twoBandParamsI}), and~(\ref{eq:coupledParamsTwoI}).  (b-d) The $x$-directed Wilson loops at $k_{z} = 0$, a generic value, and $\pi$, taken over the two bands in the projector $P_{2}$; these are the original fragile valence bands of the four-band AXI model in Eq.~(\ref{eq:AXI}).  Even though these bands are coupled to other bands below the Fermi energy, as they are isolated from them by an energy gap, they still have a well-defined nontrivial (fragile) topology indicated by the winding of the $x$- (and $y$- and $z$-) directed Wilson loop~\cite{BarryFragile}.  (e-g) The $x$-directed Wilson loops at $k_{z} = 0$, a generic value, and $\pi$, taken over the four bands in the projector $P_{4}$ in (a); the lowest two bands in energy in $P_{4}$ have the same inversion eigenvalues as bands induced from $s,\downarrow$ and $p,\downarrow$ orbitals at the $1a$ position~\cite{ZakBandrep1,ZakBandrep2,QuantumChemistry,Bandrep1,Bandrep2,Bandrep3,JenFragile1,JenFragile2,BarryFragile}.  Here, unlike in Fig.~\ref{fig:twoUncoupled}, the Wilson spectrum is gapped and free of winding (the inset in (e) shows a narrowly avoided Wilson band crossing), allowing us to define particle-hole symmetric nested Wilson projectors onto the inner and outer Wilson bands~\cite{TMDHOTI,ZhidaBLG}.  (h) The $y$-directed nested Wilson loop of the $\tilde{P}_{in}$ Wilson bands, plotted as a function of $k_{z}$; it exhibits $C_{\gamma_{2}}=+1$ spectral flow.  (i) $\det(W_{2}(k_{z}))$ of the $\tilde{P}_{in}$ Wilson bands; it exhibits $C_{\gamma_{2}}=+1$ spectral flow and is quantized at $-1$ ($+1$) at $k_{z}=0$ $(k_{z}=\pi$) (\ref{sec:detW2Inv}).   (j) The $y$-directed nested Wilson loop of the $\tilde{P}_{out}$ Wilson bands, plotted as a function of $k_{z}$; it exhibits $C_{\gamma_{2}}=-1$ spectral flow.  (k) $\det(W_{2}(k_{z}))$ of the $\tilde{P}_{out}$ Wilson bands; it exhibits $C_{\gamma_{2}}=-1$ spectral flow and is quantized at $-1$ ($+1$) at $k_{z}=0$ $(k_{z}=\pi$) (\ref{sec:detW2Inv}).}
\label{fig:twoCoupled}
\end{figure}

We next demonstrate that the addition of two bands to the Wilson projector also removes the Wilson loop winding of an $\mathcal{I}$-symmetric AXI, and also reveals that this AXI is characterized by a cyclic pump of nested Berry phase.  This can be achieved by using the previous models from~\ref{sec:oneBandInversion}, and simply choosing different hopping parameters and Wilson projectors.    We again begin with the uncoupled, eight-band Hamiltonian $\mathcal{H}_{U}(\vec{k})$ (Eq.~(\ref{eq:IUncoupled})).  In Fig.~\ref{fig:twoUncoupled}, we plot the bulk (a) and $x$-directed Wilson (b-d) bands of $\mathcal{H}_{U}(\vec{k})$ using the parameters in Eqs.~(\ref{eq:TIparams}) and~(\ref{eq:AXIparams}) in addition to:
\begin{equation}
t_{sp} = 1.2,\ t_{1} = -1,\ u_{s^{z}} = 17,
\label{eq:twoBandParamsI}
\end{equation}
and where, in contrast to Fig.~\ref{fig:oneUncoupled}, the Wilson projector $P_{4}$ here includes all four of the bands below $E=0$ in Fig.~\ref{fig:twoUncoupled}(a).  These four bands are composed of the original two fragile valence bands of $\mathcal{H}_{AXI}(\vec{k})$ (Eq.~(\ref{eq:AXI})) as well as two bands with the same symmetry eigenvalues as $s,\downarrow$ and $p,\downarrow$ bands induced from the $1a$ position~\cite{ZakBandrep1,ZakBandrep2,QuantumChemistry,Bandrep1,Bandrep2,Bandrep3,JenFragile1,JenFragile2,BarryFragile}.   In the uncoupled limit of Eq.~(\ref{eq:IUncoupled}), the Wilson loop just decomposes into the Wilson spectra of the original valence bands of $\mathcal{H}_{AXI}(\vec{k})$ (Fig.~\ref{fig:IAXI}(d-f)) and additional flat Wilson bands superposed at a particle-hole-symmetric pair of Wilson energies.

We next, as in~\ref{sec:oneBandInversion}, again couple the bands far below $E=0$ to the fragile bands above them in energy.  In Fig.~\ref{fig:twoCoupled}, we plot the bulk bands (a), $x$-directed Wilson loops (b-g) using two choices of Wilson projectors, and the $y$-directed nested Wilson loops (h-k) (Fig.~\ref{fig:Wilson}) of Eq.~(\ref{eq:ICoupled}) using the parameters in Eqs.~(\ref{eq:TIparams}),~(\ref{eq:AXIparams}), and~(\ref{eq:twoBandParamsI}), as well as:
\begin{equation}
V_{C} = 3.6,
\label{eq:coupledParamsTwoI}
\end{equation}
in Eq.~(\ref{eq:vc}).   In the band structure of $\mathcal{H}_{C}(\vec{k})$ (Fig.~\ref{fig:twoCoupled}(a)), the original two fragile valence bands of Eq.~(\ref{eq:AXI}) remain separated from the new bands by a gap below $E=0$, allowing us to define two Wilson projectors: a projector $P_{4}$ onto all four of the bands below $E=0$, and a projector $P_{2}$ onto the original fragile valence bands.  As shown in Ref.~\onlinecite{BarryFragile}, as long as $P_{2}$ is well defined, the Wilson loop of the fragile bands will still wind.  We confirm this in Fig.~\ref{fig:twoCoupled}(b-d), in which we observe the $x$-directed Wilson loop of the bands in $P_{2}$ to exhibit the same winding as the fragile bands of the original four-band AXI model (Fig.~\ref{fig:IAXI}(d-f)); we also confirm that the $y$- and $z$-directed Wilson loops of the $P_{2}$ bands also wind.  However, when we calculate the $x$-directed (and $y$- and $z$-directed) Wilson loops of the $P_{4}$ bands in Fig.~\ref{fig:twoCoupled}(a) (Fig.~\ref{fig:twoCoupled}(e-g)), all of the Wilson loop winding has again been removed, as it was in~\ref{sec:oneBandInversion}.  

By calculating the nested Wilson loop, we can again demonstrate that there is still $k_{z}$-dependent Wannier flow, despite the absence of $x$-, $y$-, and $z$-directed Wilson loop winding.  Taking the split Wilson spectrum in Fig.~\ref{fig:twoCoupled}(e-g), we again define projectors $\tilde{P}_{in,out}$ onto groupings of Wilson bands related by Wilson particle-hole symmetry~\cite{TMDTI,ZhidaBLG}, which we use to calculate the $y$-directed nested Wilson loop $W_{2}(k_{z})$ (Fig.~\ref{fig:Wilson} and~\ref{sec:W2}).  In Fig.~\ref{fig:twoCoupled}(h,i) (Fig.~\ref{fig:twoCoupled}(j,k)), we plot the nested Wilson eigenvalues $\theta_{2}(k_{z})$ and $\det(W_{2}(k_{z}))$ for the inner (outer) Wilson bands in Fig.~\ref{fig:twoCoupled}(e-g), respectively.  Both exhibit spectral flow, with the inner (outer) Wilson bands exhibiting $C_{\gamma_{2}}=+1$ ($C_{\gamma_{2}}=-1$) winding.  Because the inner and outer Wilson bands exhibit opposite odd-integer Wilson Chern numbers $C_{\gamma_{2}}$, then, as $C_{\gamma_{2}}\text{ mod } 2$ is robust to Wilson gap closures in the presence of bulk $\mathcal{I}$ symmetry (\ref{sec:detW2Inv} and Fig.~\ref{fig:oddWindI}), it here indicates a well-defined nontrivial bulk topology.  Furthermore, in the $k_{z}=0,\pi$ planes, the nested Berry phase $\gamma_{2}(k_{z})$ is quantized (Fig.~\ref{fig:twoCoupled}(i,k) and~\ref{sec:detW2Inv}), with the $k_{z} = 0$ ($k_{z}=\pi$) plane exhibiting a nontrivial (trivial) nested Berry phase:
\begin{equation}
\gamma_{2}(0) = \pi,\ \gamma_{2}(\pi) = 0.
\label{eq:twoBandsGamma2}
\end{equation}
The AXI with four valence bands is therefore also not a trivial, Wannierizable insulator.  Like the AXI with three valence bands in~\ref{sec:oneBandInversion}, it is instead a magnetic HOTI characterized by the cyclic pumping of a 2D fragile phase with trivialized Wilson loop winding.  As we have shown that adding one or two bands to the valence manifold of an $\mathcal{I}$-symmetric HOTI does not trivialize the nested Wannier flow, we can conclude that there is no linear combination of trivial bands that can be added to an $\mathcal{I}$-symmetric AXI to localize its Wannier functions.  Therefore, the $\mathcal{I}$-symmetric AXI is a strong magnetic TI~\cite{QuantumChemistry}.  

\subsection{Tight-Binding Model of a $C_{2z}\times\mathcal{T}$-Symmetric Crystalline Axion Insulator}
\label{sec:C2TAXI}

\begin{figure}[h]
\centering
\includegraphics[width=\textwidth]{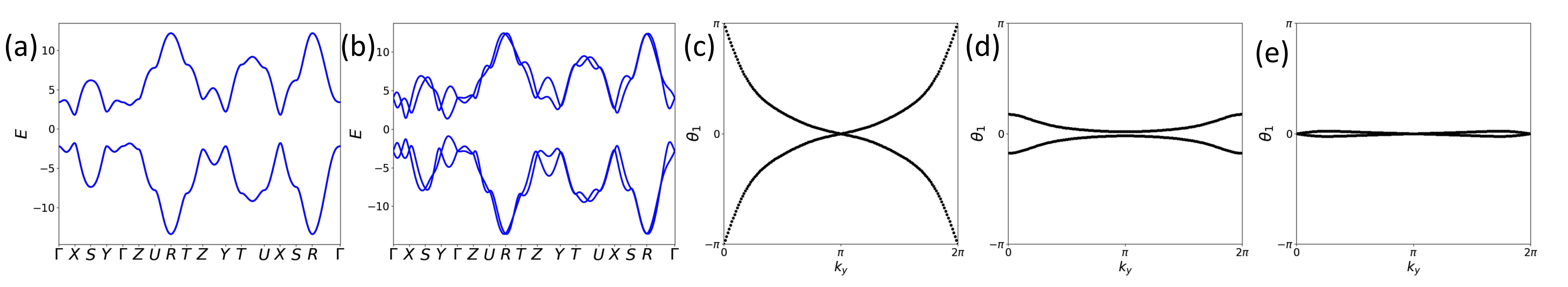}
\caption{(a) Bulk bands of $\mathcal{H}_{TI}$ (Eq.~(\ref{eq:TI})) using the parameters in Eqs.~(\ref{eq:TIparams}) and~(\ref{eq:C2Tparams1}).  This TI is invariant under lattice translation, $\mathcal{I}$, $C_{2z}$, and $\mathcal{T}$, and is thus characterized by~\cite{BigBook,MagneticBook,ITCA} SG 10 $P2/m1'$.  (b) Bulk bands of the same TI under the addition of the $\mathcal{I}$- and $M_{z}$-breaking potential $V_{M_{z}}(\vec{k})$ in Eq.~(\ref{eq:breakMz}); this TI is described by $\mathcal{H}_{2}(\vec{k})$ (Eq.~(\ref{eq:C2andT})), plotted with the additional parameters in Eq.~(\ref{eq:C2Tparams2}).  $\mathcal{H}_{2}(\vec{k})$ is only invariant under lattice translation, $C_{2z}$, and $\mathcal{T}$, and is thus characterized by~\cite{BigBook,MagneticBook,ITCA} SG 3 $P21'$.  (c-e) The $x$-directed Wilson loops of the two bands below $E=0$ of (b), taken at (c) $k_{z}=0$, (d) a generic value of $k_{z}$, and (e) at $k_{z}=\pi$. The Wilson loop still winds, because the bulk remains a 3D TI.  However, because bulk $\mathcal{I}$ is broken, this TI phase is no longer indicated by bulk inversion eigenvalues~\cite{FuKaneMele,FuKaneInversion,AndreiInversion}, and the Wilson particle-hole symmetry in (c-e) is enforced by $C_{2z}\times\mathcal{T}$ symmetry (\ref{sec:detW2C2T}), rather than by the previous $\mathcal{I}$ symmetry (\ref{sec:detW2Inv}) of Eq.~(\ref{eq:TI}).}
\label{fig:C2andT}
\end{figure}

We next demonstrate that a related 3D AXI phase, invariant under the magnetic antiunitary crystal symmetry $C_{2z}\times\mathcal{T}$, \emph{i.e.} the combination of a twofold rotation about the $z$-axis ($C_{2z})$ and $\mathcal{T}$, is also equivalent to the cyclic pumping of a 2D fragile phase.  To construct a model of a 3D insulator in a $C_{2z}\times\mathcal{T}$-symmetric AXI phase, we begin by returning to the model of an $\mathcal{I}$- and $\mathcal{T}$--symmetric TI in Eq.~(\ref{eq:TI}).  Starting with the parameters in Eq.~(\ref{eq:TIparams}), we tune:
\begin{equation}
t_{2,x}\rightarrow 0,\ v_{2,xy}\rightarrow 0,\ v_{2,z}\rightarrow 0.
\label{eq:C2Tparams1}
\end{equation}
In the limit of Eq.~(\ref{eq:C2Tparams1}), the bulk TI gap remains open (Fig.~\ref{fig:C2andT}(a)), and the Wilson loops remain identical to those in Fig.~\ref{fig:TI}(d-f).  In the limit of Eq.~(\ref{eq:C2Tparams1}), the bulk Hamiltonian $\mathcal{H}_{TI}(\vec{k})$ is now invariant under the action of:
\begin{align}
C_{2z}:\ \mathcal{H}_{TI}(k_{x},k_{y},k_{z})\rightarrow \sigma^{z}\mathcal{H}_{TI}(-k_{x},-k_{y},k_{z})\sigma^{z}, 
\label{eq:C2Tsym} 
\end{align}
in addition to $\mathcal{I}$ and $\mathcal{T}$ (Eq.~(\ref{eq:TIsyms})).  Eq.~(\ref{eq:C2Tsym}) also implies $M_{z}=\mathcal{I}C_{2z}$ symmetry, and therefore in the limit of Eq.~(\ref{eq:C2Tparams1}), $\mathcal{H}_{TI}(\vec{k})$ is characterized by~\cite{BigBook,MagneticBook,ITCA} SG 10 $P2/m1'$.  As the $k_{z}=0,\pi$ planes are now invariant under $M_{z}$, these two planes can additionally be characterized (if topologically nontrivial) as 2D topological crystalline insulators (TCIs)~\cite{TeoFuKaneTCI,NagaosaDirac,LiangTCI}, with the $k_{z}=0$ ($k_{z}=\pi$) planes exhibiting mirror Chern numbers $C_{M_{z}}=1$ ($C_{M_{z}}=0$).  

We next introduce the $\mathcal{T}$-symmetric, $\mathcal{I}$- and $M_{z}$-breaking term:
\begin{equation}
V_{M_{z}}(\vec{k}) = \tau^{y}\sigma^{z}(v_{x,M_{z}}\cos(k_{x}) + v_{y,M_{z}}\cos(k_{y})),
\label{eq:breakMz}
\end{equation}
and form the Hamiltonian:
\begin{equation}
\mathcal{H}_{2}(\vec{k}) = \mathcal{H}_{TI}(\vec{k}) + V_{M_{z}}(\vec{k}),
\label{eq:C2andT}
\end{equation}
which, when $v_{x/y,M_{z}}\neq 0$, is only invariant under lattice translation, $\mathcal{T}$, and $C_{2z}$, such that it is characterized by~\cite{BigBook,MagneticBook,ITCA} SG 3 $P21'$.  In Fig.~\ref{fig:C2andT}(b) we plot the bulk bands of Eq.~(\ref{eq:C2andT}) using the parameters in Eqs.~(\ref{eq:TIparams}) and~(\ref{eq:C2Tparams1}), as well as:
\begin{equation}
v_{x,M_{z}}= 2, v_{y,M_{z}} = 0.
\label{eq:C2Tparams2} 
\end{equation}
As we have not closed a bulk gap from the previous $\mathcal{I}$-symmetric TI phase, and have preserved $\mathcal{T}$-symmetry, this system is still a 3D TI, though one whose topology is no longer indicated by $\mathcal{I}$ symmetry~\cite{FuKaneMele,FuKaneInversion,AndreiInversion}.  To confirm the topology of $\mathcal{H}_{2}(\vec{k})$, we plot in Fig.~\ref{fig:C2andT}(c-e) the $x$-directed Wilson loops.  They exhibit the same winding and connectivities as those of the $\mathcal{I}$-symmetric TI in Fig.~\ref{fig:TI}(d-f); they also exhibit a Wilson particle-hole symmetry, which is here enforced by $C_{2z}\times\mathcal{T}$ (\ref{sec:detW2C2T}), rather than by the previous $\mathcal{I}$ symmetry (\ref{sec:detW2Inv}) of Eq.~(\ref{eq:TI}). 

\begin{figure}[h]
\centering
\includegraphics[width=\textwidth]{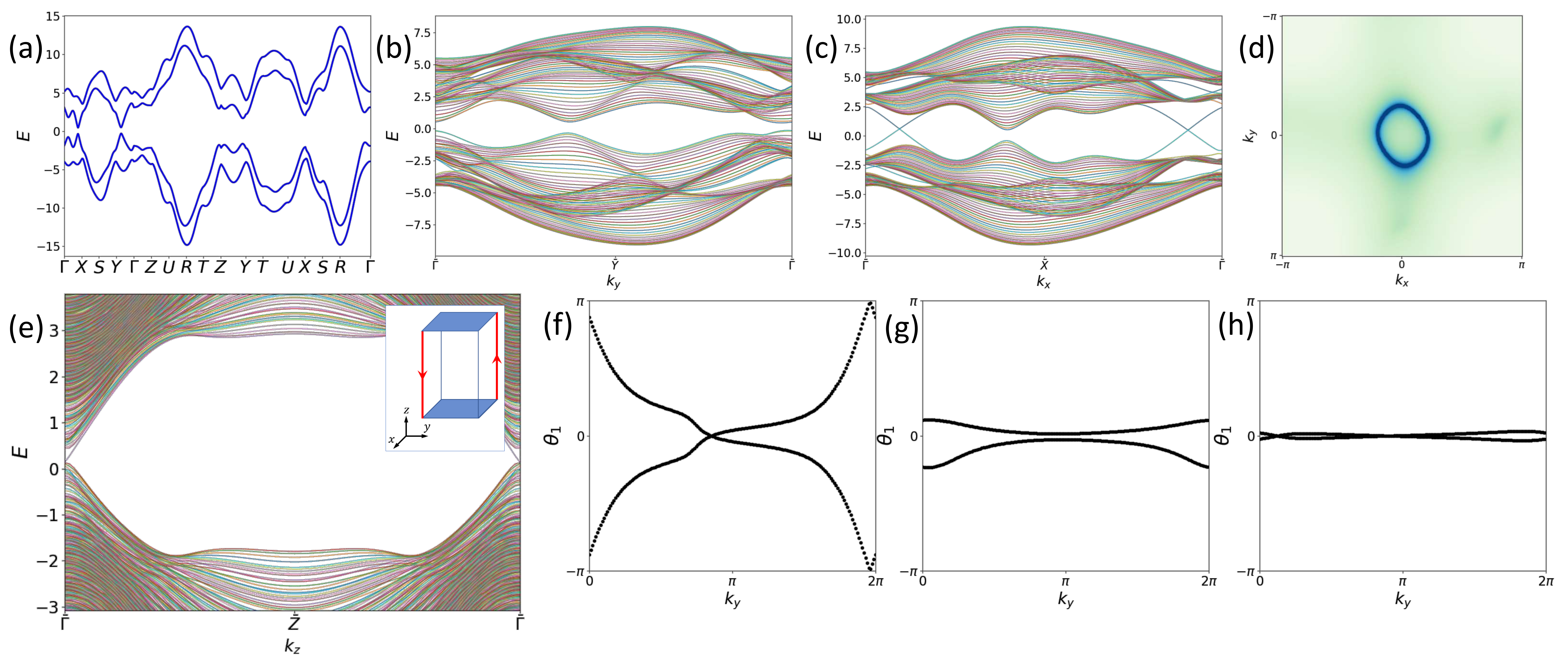}
\caption{(a) Bulk bands of $\mathcal{H}_{C2T}(\vec{k})$ (Eq.~(\ref{eq:C2T})), a four-band model of a 3D $C_{2z}\times\mathcal{T}$-symmetric AXI, plotted using the parameters in Eq.~(\ref{eq:C2Tparams3}).  (b) The $x$-directed slab bands of this AXI are fully gapped.  (c)  However, the $z$-directed slab bands are gapless, featuring two, twofold linear degeneracies protected by surface $C_{2z}\times\mathcal{T}$ symmetry~\cite{YoungkukLineNode,FangFuNSandC2,ChenRotation,HarukiRotation,LiangTCI}.  (d) The Green's function of the $(001)$-surface indicates that just one twofold cone sits on this surface; the other can be inferred to be localized on the $(00\bar{1})$-surface.  (e) The $z$-directed rod bands of Eq.~(\ref{eq:C2T}) exhibit two oppositely propagating chiral hinge modes, which are localized on opposing hinges (Fig.~\ref{fig:C2Tcorners}).  (f) The $x$-directed Wilson loop at $k_{z}=0$ exhibits helical winding.  (g) The $x$-directed Wilson loop at generic values of $k_{z}$ is gapped.  (h) The $x$-directed Wilson loop at $k_{z}=\pi$ does not wind, and exhibits a pair of twofold degeneracies.  Unlike in the 3D TI in Fig.~\ref{fig:TI}, the Wilson degeneracies in (d) and (f) are protected by bulk $C_{2z}\times\mathcal{T}$ symmetry and the presence of only two bands in the Wilson projector~\cite{JenFragile1,JenFragile2,AdrianFragile,KoreanFragile,BarryFragile,ZhidaBLG,AshvinBLG1,AshvinBLG2,AshvinFragile2,HarukiFragile}.}
\label{fig:C2T}
\end{figure}

\begin{figure}[h]
\centering
\includegraphics[width=0.55\textwidth]{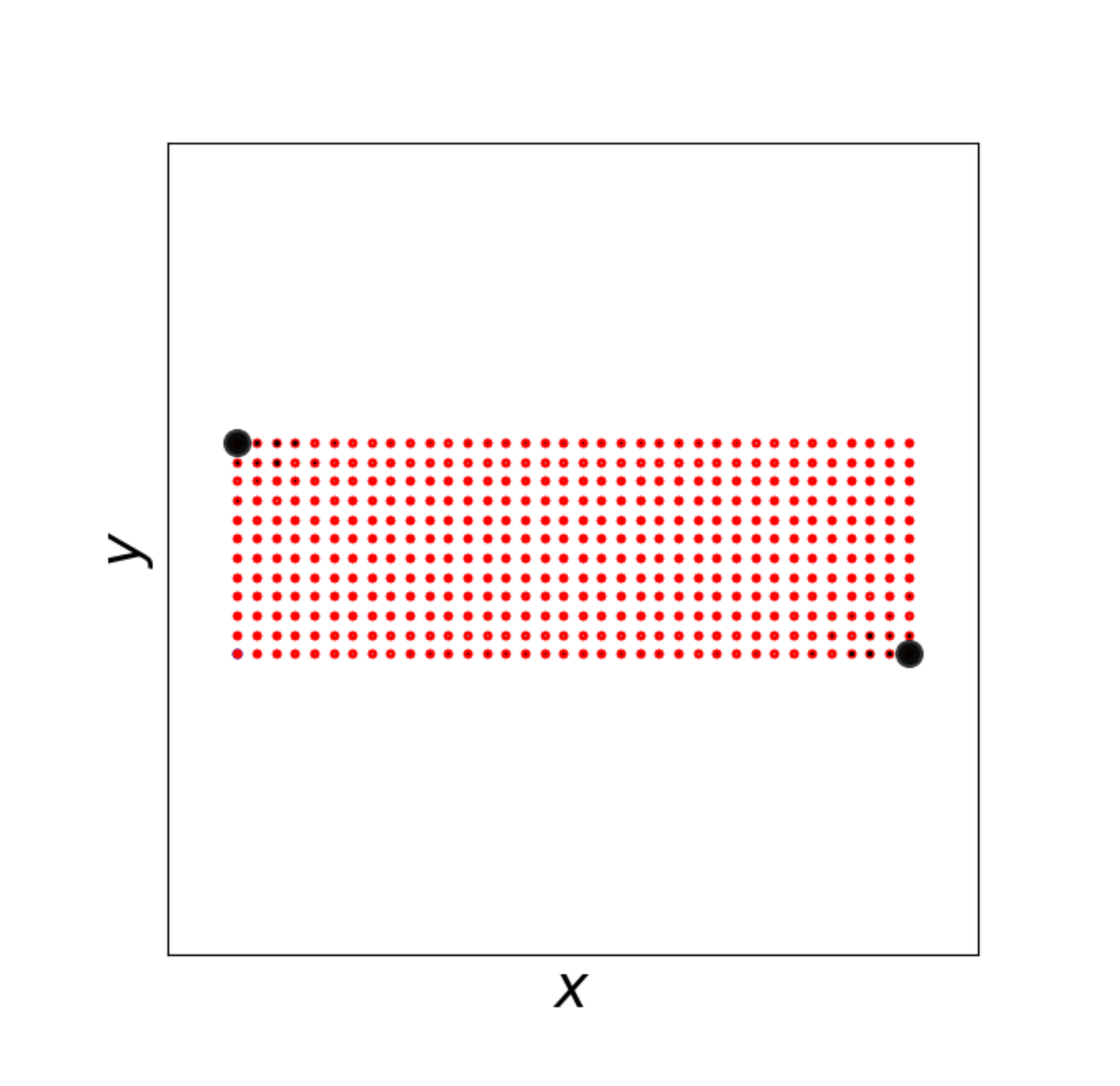}
\caption{The localization in the $xy$-plane of the two hinge states at $k_{z}=0$ of the $z$-directed $C_{2z}\times\mathcal{T}$-symmetric AXI rod in Fig.~\ref{fig:C2T}(e).  The $k_{z}=0$ plane of Eq.~(\ref{eq:C2T}) with the parameters in Eq.~(\ref{eq:C2Tparams3}) is equivalent to a 2D insulator with two corner modes of opposite charge related by $C_{2z}\times\mathcal{T}$ symmetry.  The presence of these modes is indicated in a magnetic 2D insulator with $\mathcal{C}_{2z}\times\mathcal{T}$ symmetry with two occupied bands by the Wilson loop winding in Fig.~\ref{fig:C2T}(f), and in an insulator with more occupied bands by a nontrivial nested Berry phase $\gamma_{2}$ (\ref{sec:oneBandC2T},~\ref{sec:twoBandsC2T}, and Refs.~\onlinecite{TMDHOTI,KoreanFragile}).}
\label{fig:C2Tcorners}
\end{figure}

\begin{figure}[h]
\centering
\includegraphics[width=\textwidth]{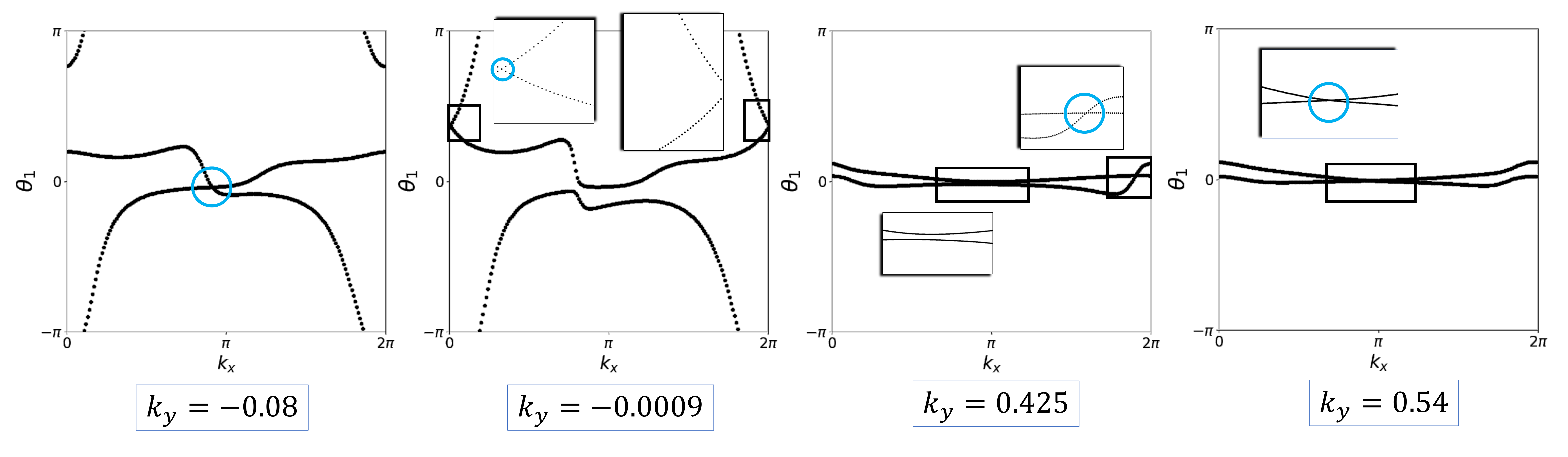}
\caption{The $z$-directed Wilson loop of the lower two bands of $\mathcal{H}_{C2T}(\vec{k})$, using the parameters in Eq.~(\ref{eq:C2Tparams3}), also winds, but across the entire 2D BZ, instead of along high-symmetry lines, as it did for the 3D TI in Fig.~\ref{fig:TI}(d) and Fig.~\ref{fig:C2andT}(c).  This winding is characterized by four Wilson degeneracies, which we label in blue circles.  As shown in detail in Refs.~\onlinecite{JenFragile1,BarryFragile} and in~\ref{sec:zW1C2T}, they are protected by bulk $C_{2z}\times\mathcal{T}$.}
\label{fig:C2TwilsonZ2}
\end{figure}
To induce a transition from the $\mathcal{I}$-broken TI in Eq.~(\ref{eq:C2andT}) to a $C_{2z}\times\mathcal{T}$-symmetric AXI, we introduce the term:
\begin{equation}
V_{C2T}(\vec{k}) = \tilde{m}_{A}\sigma^{x} + \tilde{m}_{B}\sigma^{y},
\end{equation}
to realize the Hamiltonian:
\begin{equation}
\mathcal{H}_{C2T}(\vec{k}) = \mathcal{H}_{2}(\vec{k}) + V_{C2T}(\vec{k}),
\label{eq:C2T}
\end{equation}
where $\mathcal{H}_{2}(\vec{k})$ is the previous Hamiltonian for a $C_{2z}$-symmetic 3D TI (Eq.~(\ref{eq:C2andT})), and where we realize the crystalline AXI phase by using the parameters:
\begin{equation}
\tilde{m}_{A} = 1,\ \tilde{m}_{B} = 0.8,
\label{eq:C2Tparams3}
\end{equation}
in addition to the previous parameters in Eqs.~(\ref{eq:TIparams}),~(\ref{eq:C2Tparams1}), and~(\ref{eq:C2Tparams2}).  As we have broken $C_{2z}$ and $\mathcal{T}$ symmetry while preserving their product $C_{2z}\times\mathcal{T}$, $\mathcal{H}_{C2T}(\vec{k})$ describes a crystalline AXI in the type-III magnetic SG $P2'$ (number 3.3 in the BNS setting~\cite{MagneticBook,BilbaoMagStructures}).  

In Fig.~\ref{fig:C2T}(a-h), we show the bulk bands, $x$-directed slab bands, $z$-directed slab bands, $(001)$-surface states, $z$-directed rod bands, and $x$-directed Wilson loops of $\mathcal{H}_{C2T}(\vec{k})$ using the parameters in Eq.~(\ref{eq:C2Tparams3}).  The bulk bands are now generically singly degenerate, and the gap at half-filling is preserved under the introduction $V_{C2T}(\vec{k})$ with the parameters in Eq.~(\ref{eq:C2Tparams3}) (Fig.~\ref{fig:C2T}(a)).  The $x$-directed slab bands are fully gapped (Fig.~\ref{fig:C2T}(b)).  We have also verified that $y$-directed slabs of $\mathcal{H}_{C2T}(\vec{k})$ are also fully gapped.  Unlike in~\ref{sec:IAXI}, the $z$-directed slab bands are \emph{not} gapped (Fig.~\ref{fig:C2T}(c)).  Instead, the $(001)$ and $(00\bar{1})$ surfaces exhibit twofold linear degneracies at generic $z$-surface crystal momenta (Fig.~\ref{fig:C2T}(c,d)).  These degeneracies are inherited from the parent TI phase (Eq.~(\ref{eq:C2andT})). However, they are now allowed to move off the TRIM point, and are protected by the magnetic antiunitary crystal symmetry~\cite{YoungkukLineNode,FangFuNSandC2,ChenRotation,HarukiRotation,LiangTCI} $C_{2z}\times\mathcal{T}$.  In this sense, $\mathcal{H}_{C2T}(\vec{k})$ is not just an AXI, but is also the simplest example of a ``rotation anomaly'' TCI~\cite{FangFuNSandC2,ChenRotation,HarukiRotation,HOTIChen,HOTIBernevig}.  Though a $z$-terminated geometry of $\mathcal{H}_{C2T}(\vec{k})$ thus exhibits gapless surfaces, a $z$-directed rod, which is infinite in only the $z$ direction, will be gapped.  We therefore plot in Fig.~\ref{fig:C2T}(e) the bands of a $z$-directed rod of $\mathcal{H}_{C2T}(\vec{k})$.   We observe that, like the $\mathcal{I}$-symmetric AXI in Fig.~\ref{fig:IAXI}(c), the rod bands of $\mathcal{H}_{C2T}(\vec{k})$ exhibit two 1D chiral modes propagating in the $z$ direction, with oppositely chiral modes localized on opposing hinges (Fig.~\ref{fig:C2Tcorners}).  This confirms that rotation-anomaly TCIs exhibit not just gapless surfaces, but also gapless hinges along rods cleaved parallel to their rotation axes~\cite{ChenRotation,HarukiRotation,HOTIChen,HOTIBernevig}.  It is again crucial to note that there is no symmetry or topology requirement for the corner modes of the $k_{z}=0$ plane (Fig.~\ref{fig:C2Tcorners}) to appear as midgap states~\cite{HOTIChen,KoreanFragile}.  Without closing a bulk or edge gap, the modes at $k_{z}=0$ in Fig.~\ref{fig:C2T}(c) can float into the valence or conduction manifold (indeed, they are already nearly absorbed into the valence manifold) .  However, because the corner modes lie at the spectral center (\emph{i.e.}, there are exactly $N/2 -1$ states above and below them in energy), then floating them into the valence or conduction manifold will result in a state imbalance above and below the bulk gap.  This process is depicted for the $\mathcal{I}$-symmetric case in Fig.~\ref{fig:AXIcorners}(b,c); the state counting is identical for the $C_{2z}\times\mathcal{T}$-symmetric case.  As discussed in detail in~\ref{sec:IAXI}, the presence of this state imbalance is indicated in a two-band model by the winding Wilson loop in Fig.~\ref{fig:C2T}(f), and in a many-band model by $\det(W_{2})=-1$, \emph{i.e.}, that the nested Berry phase $\gamma_{2}$ (Eq.~(\ref{eq:gamma2})) is nontrivial (\ref{sec:oneBandC2T},~\ref{sec:twoBandsC2T}, and Refs.~\onlinecite{ZhidaBLG,KoreanFragile,AshvinBLG1,AshvinBLG2,AshvinFragile2,HarukiFragile}).   

The $x$-directed Wilson loop of $\mathcal{H}_{C2T}(\vec{k})$ displays the same winding as that of the 3D TI in Fig.~\ref{fig:TI}(d-f): it winds at $k_{z}=0$, is gapped and trivial at generic values of $k_{z}$, and is gapless and trivial at $k_{z}=\pi$ (Fig.~\ref{fig:C2T}(f,g,h), respectively).  However, at $k_{z}=0,\pi$, the gapless points in the Wilson loop have become unpinned from $k_{y}=0,\pi$; these crossings are instead protected by $C_{2z}\times\mathcal{T}$ symmetry and the presence of only two bands in the Wilson projector~\cite{AshvinFragile,JenFragile1,JenFragile2,AdrianFragile,KoreanFragile,BarryFragile,ZhidaBLG,AshvinBLG1,AshvinBLG2,AshvinFragile2,HarukiFragile}.  As we will see in~\ref{sec:oneBandC2T} and~\ref{sec:twoBandsC2T}, the introduction of trivial bands to the Wilson projector can lift these Wilson degeneracies~\cite{AshvinFragile,JenFragile1,JenFragile2,AdrianFragile,KoreanFragile,BarryFragile,ZhidaBLG,AshvinBLG1,AshvinBLG2,AshvinFragile2,HarukiFragile}.  We have also verified that the $y$-directed Wilson loop exhibits the same behavior.   

Crucially, the $z$-directed Wilson loop also winds, but in a more subtle manner.  As shown in detail in~\ref{sec:zW1C2T}, $C_{2z}\times\mathcal{T}$ symmetry protects twofold Wilson degeneracies at generic momenta $(k_{x},k_{y})$.  Specifically, $C_{2z}\times\mathcal{T}$ acts as an antiunitary symmetry that preserves $k_{x,y}$ and squares to $+1$.  Both 2D Wilson (\ref{sec:zW1C2T}) and energy Hamiltonians~\cite{YoungkukLineNode,FangFuNSandC2,ChenRotation,HarukiRotation,LiangTCI} with this symmetry can feature robust topological twofold linear degeneracies at generic values of $k_{x,y}$.  As the $z$-directed Wilson loop of a 3D TI with band inversion at $\Gamma$ (Fig.~\ref{fig:TI}(a)) features four twofold linear crossings (two at Wilson energy $\theta_{1}(k_{x},k_{y})=0$ at $k_{x}=0,\pi$, $k_{y}=\pi$, one at $\theta_{1}(k_{x},k_{y})=0$ at $k_{x}=\pi$, $k_{y}=0$, and one at $\theta_{1}(k_{x},k_{y})=\pi$ at $k_{x}=0$, $k_{y}=0$), and because we have only weakly broken $\mathcal{T}$ symmetry with the parameters in Eq.~(\ref{eq:C2Tparams3}), then the $z$-directed Wilson loop of $\mathcal{H}_{C2T}(\vec{k})$ should still exhibit four Wilson degeneracies, but with no restrictions on their crystal momenta and Wilson energies.  In Fig.~\ref{fig:C2TwilsonZ2}, we plot the $k_{y}$-indexed, $z$-directed Wilson loops featuring each of these four degeneracies, and label them with blue circles.  Because $C_{2z}\times\mathcal{T}$ symmetry is a symmetry of the $(100)$-surface wallpaper group~\cite{WiederLayers,DiracInsulator}, the Wilson loop winding in Fig.~\ref{fig:C2TwilsonZ2} is \emph{strong}, and not fragile.  We will see in~\ref{sec:oneBandC2T} and~\ref{sec:twoBandsC2T} the that the winding of the $z$-directed Wilson loop is indeed preserved under the addition of trivial bands to the Wilson projector. 

\begin{figure}[h]
\centering
\includegraphics[width=0.45\textwidth]{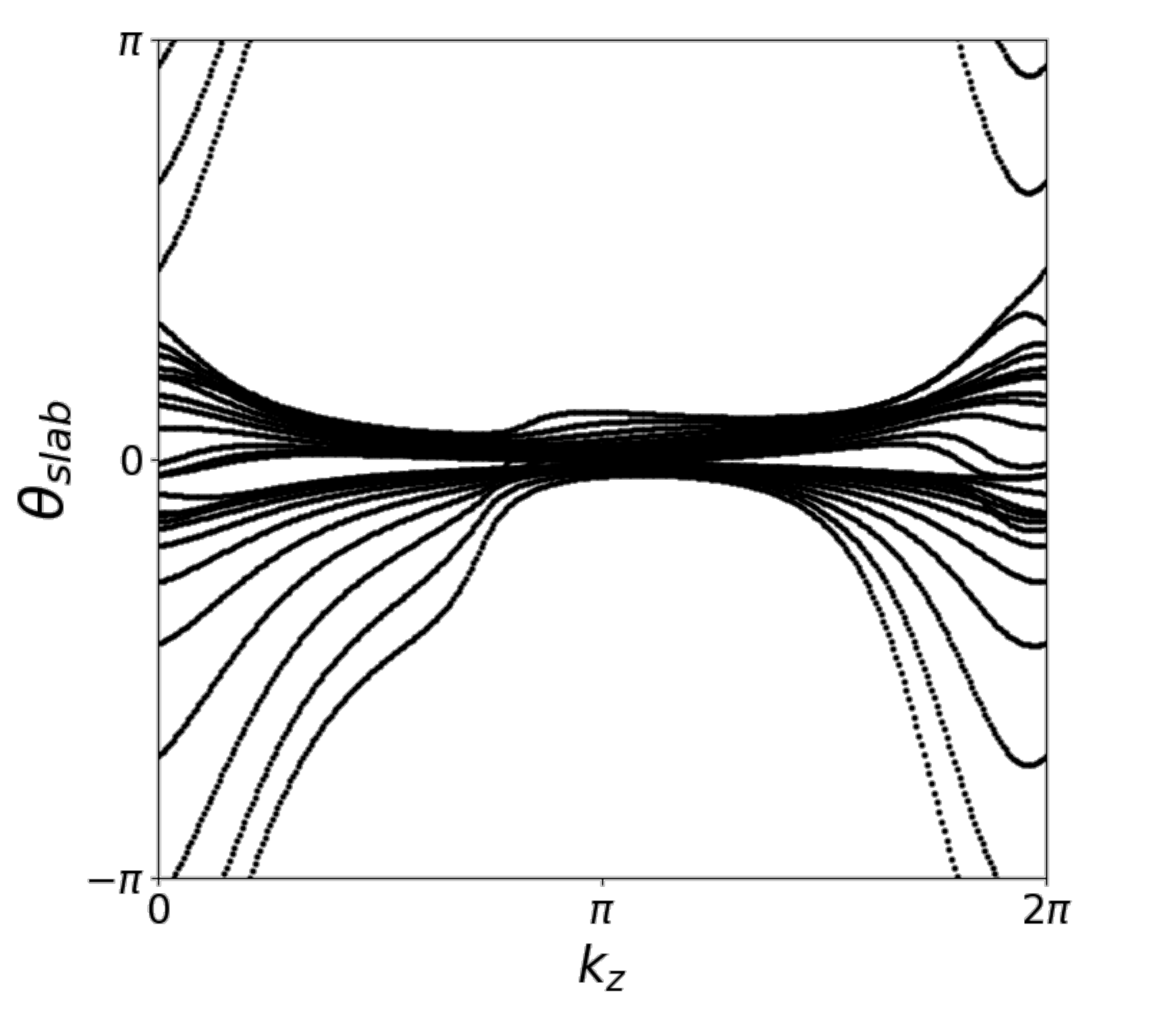}
\caption{The $y$-directed Wilson bands of \emph{all} of the bands below the gap in the $x$-directed $C_{2z}\times\mathcal{T}$-symmetric AXI slab in Fig.~\ref{fig:C2T}(b).  This Wilson loop exhibits {$C_{slab}=+1$} winding, confirming that in a slab directed along an axis perpendicular to the $z$-axis of a $C_{2z}\times\mathcal{T}$-symmetric AXI, the two opposing surfaces with the same half-integer anomalous Hall conductivity~\cite{QHZ,ChenRotation,HarukiRotation,HOTIChen,HOTIBernevig,VDBAxion,DiracInsulator,MulliganAnomaly,VDBHOTI} combine to form an effective Chern insulator with an \emph{integer} Hall conductivity~\cite{HOTIBernevig,ChenRotation,HOTIChen,HarukiRotation}.}
\label{fig:slabC2T}
\end{figure}

We also confirm that opposing gapped surfaces of the $C_{2z}\times\mathcal{T}$-symmetric AXI described Eq.~(\ref{eq:C2T}) also exhibit the same, anomalously quantized half-integer Hall conductivity $\sigma_{H}= e^{2}/(2h)$ (when measuring with respect to the positive $x$ and $y$ directions for all in-plane surfaces; the $z$-normal surfaces are gapless (Fig.~\ref{fig:C2T}(c,d)))~\cite{QHZ,ChenRotation,HarukiRotation,HOTIChen,HOTIBernevig,VDBAxion,DiracInsulator,MulliganAnomaly,VDBHOTI}.  In Fig.~\ref{fig:slabC2T}, we plot the $y$-directed Wilson bands taken over \emph{all} of the bands below the gap in the $x$-directed slab shown in Fig.~\ref{fig:C2T}(b).  The slab Wilson loop in Fig.~\ref{fig:slabC2T} exhibits an overall $C_{slab}=+1$ winding, confirming that when a $C_{2z}\times\mathcal{T}$-symmetric AXI is cleaved into a slab geometry along an axis normal to the $z$-axis (\emph{i.e.} the $x$- or $y$-axis), it is topologically equivalent to a $C=\pm 1$ Chern insulator~\cite{HOTIBernevig,ChenRotation,HOTIChen,HarukiRotation}, \emph{i.e.}, an isolated quasi-2D insulator with an \emph{integer} Hall conductivity. 

\subsubsection{Coupling One Additional Band to a $C_{2z}\times\mathcal{T}$-Symmetric AXI}
\label{sec:oneBandC2T}

We now demonstrate that the presence of additional bands in the Wilson projector can remove the winding of Wilson loops directed in the $xy$-plane over the occupied bands of a $C_{2z}\times\mathcal{T}$-symmetric AXI, \emph{even if those bands lie far below the Fermi energy}.  We begin by placing two additional Kramers pairs of $s$ orbitals at the general Wyckoff position~\cite{BCS1,BCS2,BilbaoMagStructures} ($2e$) of the unit cell of $\mathcal{H}_{C2T}(\vec{k})$ (Eq.~(\ref{eq:C2T})).  The positions of these two pairs of orbitals are related by $C_{2z}\times\mathcal{T}$ about the $1a$ position of magnetic SG~\cite{BCS1,BCS2,BilbaoMagStructures} $P2'$ ($(x,y,z)=(0,y,0)$):
\begin{equation}
\vec{r}_{3,\sigma} = (\tilde{u}_{3x},\tilde{u}_{3y},\tilde{u}_{3z}),\ \vec{r}_{4,\sigma} =  (-\tilde{u}_{3x},-\tilde{u}_{3y},\tilde{u}_{3z}),
\label{eq:C2TorbitalPos}
\end{equation}
where we index the two additional Kramers pairs of orbitals as $3$ and $4$ to distinguish them from the spin-split pairs of $s$ and $p$ orbitals already present at the $1a$ position of $\mathcal{H}_{C2T}(\vec{k})$ (Eq.~(\ref{eq:C2T})), and where $\sigma$ denotes the spin degree of freedom.  Because magnetic SG $P2'$ has no unitary crystal symmetries, aside from lattice translations, then there are no symmetry eigenvalues with which to label bands, and all of the splitting of Wilson loops directed in the $xy$-plane can be attributed to the Berry phases of the new, trivial bands~\cite{ZakBandrep1,ZakBandrep2,QuantumChemistry,Bandrep1,Bandrep2,Bandrep3,JenFragile1,JenFragile2,BarryFragile}.  As previously detailed in~\ref{sec:oneBandInversion}, we choose to place Kramers pairs $3$ and $4$ at the general position so that the Wilson spectrum taken over all of the bands from these orbitals is gapped when they are uncoupled to other orbitals.  Summarizing the discussion in~\ref{sec:oneBandInversion}, this is because when the Wilson loop is taken over all of the bands in the Hilbert space (here, the decoupled subspace of orbitals $3$ and $4$), it decomposes into flat Wilson bands at the positions of the atoms.

To split Kramers pairs $3$ and $4$, we introduce the simple coupling between the two new orbitals within the same unit cell $i$:
\begin{eqnarray}
\tilde{V}_{sp} &=& \tilde{t}_{sp}\sum_{\sigma=\uparrow,\downarrow}\left(c^{\dag}_{3\sigma,i}c_{4,\sigma,i} + c^{\dag}_{4\sigma,i}c_{3,\sigma,i}\right) + \tilde{t}_{1}\left(c^{\dag}_{3\sigma,i}c_{3,\sigma,i} + c^{\dag}_{4\sigma,i}c_{4,\sigma,i}\right) \nonumber \\
\tilde{V}_{sp}(\vec{k}) &=& \tilde{t}_{sp}\left(\begin{array}{cccc}
0 & 0 & 0 & 0 \\
0 & 0 & 0 & 0 \\
0 & 0 & 0 & 1 \\
0 & 0 & 1 & 0\end{array}\right)\otimes\mathds{1}_{\sigma} + \tilde{t}_{1}\left(\begin{array}{cccc}
0 & 0 & 0 & 0 \\
0 & 0 & 0 & 0 \\
0 & 0 & 1 & 0 \\
0 & 0 & 0 & 1 \end{array}\right)\otimes\mathds{1}_{\sigma},
\label{eq:vspC2T}
\end{eqnarray}
where $\mathds{1}_{\sigma}$ denotes the $2\times 2$ identity in the spin subspace, and where the $4\times 4$ matrix is indexed by $(c_{s}\ c_{p}\ c_{3}\ c_{4})$ where $c_{s,p}$ denote the original pairs of $s$ and $p$ orbitals in $\mathcal{H}_{2}(\vec{k})$ (Eq.~(\ref{eq:C2andT})), respectively.  We then introduce simple ferromagnetic splitting for the orbitals at the general position:
\begin{equation}
\tilde{U}_{s^{x}}(\vec{k}) = \tilde{u}_{s^{x}}\left(\begin{array}{cccc}
0 & 0 & 0 & 0 \\
0 & 0 & 0 & 0 \\
0 & 0 & 1 & 0 \\
0 & 0 & 0 & 1 \end{array}\right)\otimes\sigma^{x}.
\label{eq:UsxC2T}
\end{equation}
Next, we form a new Hamiltonian in this expanded $8\times 8$ space:
\begin{equation}
\tilde{\mathcal{H}}_{U}(\vec{k}) = \mathcal{H}_{C2T}(\vec{k}) + \tilde{V}_{sp}(\vec{k}) + \tilde{U}_{s^{x}}(\vec{k}),
\label{eq:C2TUncoupled}
\end{equation}
where $\mathcal{H}_{C2T}(\vec{k})$ is the previous $4\times 4$ Hamiltonian of a $C_{2z}\times\mathcal{T}$-symmetric AXI with two occupied bands (Eq.~(\ref{eq:C2T})).  Eq.~(\ref{eq:C2TUncoupled}) remains invariant under $C_{2z}\times\mathcal{T}$ symmetry, which now takes the form:
\begin{equation}
C_{2z}\times\mathcal{T}:\ \tilde{\mathcal{H}}_{U}(k_{x}, k_{y}, k_{z})\rightarrow \left[\left(\begin{array}{cccc}
1 & 0 & 0 & 0 \\
0 & 1 & 0 & 0 \\
0 & 0 & 0 & 1 \\
0 & 0 & 1 & 0\end{array}\right)\otimes\sigma^{x}\right]\tilde{\mathcal{H}}^{*}_{U}(-k_{x}, -k_{y}, k_{z}) \left[\left(\begin{array}{cccc}
1 & 0 & 0 & 0 \\
0 & 1 & 0 & 0 \\
0 & 0 & 0 & 1 \\
0 & 0 & 1 & 0\end{array}\right)\otimes\sigma^{x}\right].
\label{eq:8C2T}
\end{equation}

\begin{figure}[h]
\centering
\includegraphics[width=\textwidth]{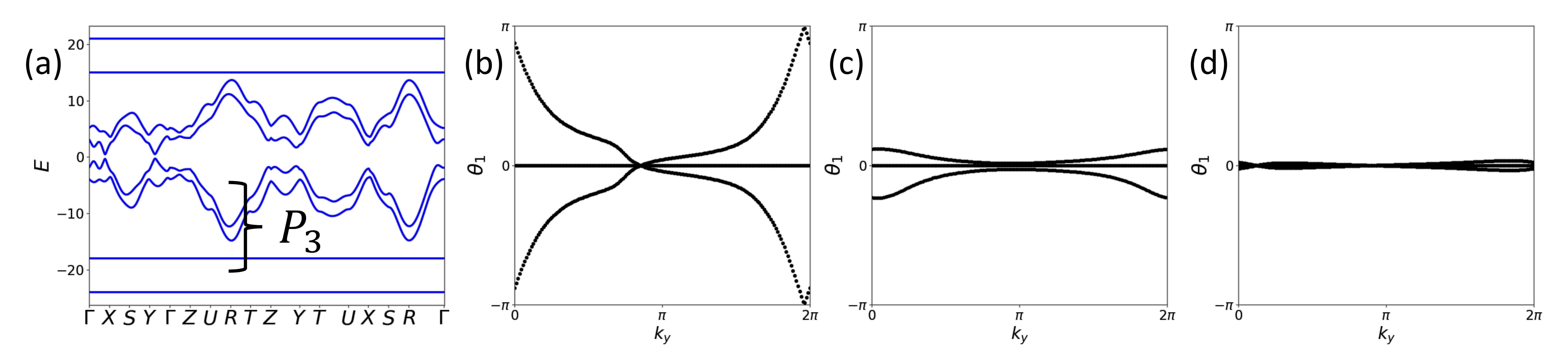}
\caption{(a) Bulk bands of the eight-band, uncoupled Hamiltonian $\tilde{\mathcal{H}}_{U}(\vec{k})$ (Eq.~(\ref{eq:C2TUncoupled})) plotted with the parameters in Eqs.~(\ref{eq:TIparams}),~(\ref{eq:C2Tparams1}),~(\ref{eq:C2Tparams2}),~(\ref{eq:C2Tparams3}), and~(\ref{eq:oneBandParamsC2T}).  (b-d) The $x$-directed Wilson loops at $k_{z} = 0$, a generic value, and $\pi$, taken over the three bands in the projector $P_{3}$ in (a).  Because the lowest band in energy in the projector is decoupled from the other two bands, the Wilson spectra in (b-d) decompose into the superpositions of the previous Wilson spectra in Fig.~\ref{fig:IAXI}(d-f) and an additional flat Wilson band at the Wilson energy $\theta_{1}(k_{y},k_{z})=0$.}
\label{fig:oneUncoupledC2T}
\end{figure}

\begin{figure}[h]
\centering
\includegraphics[width=\textwidth]{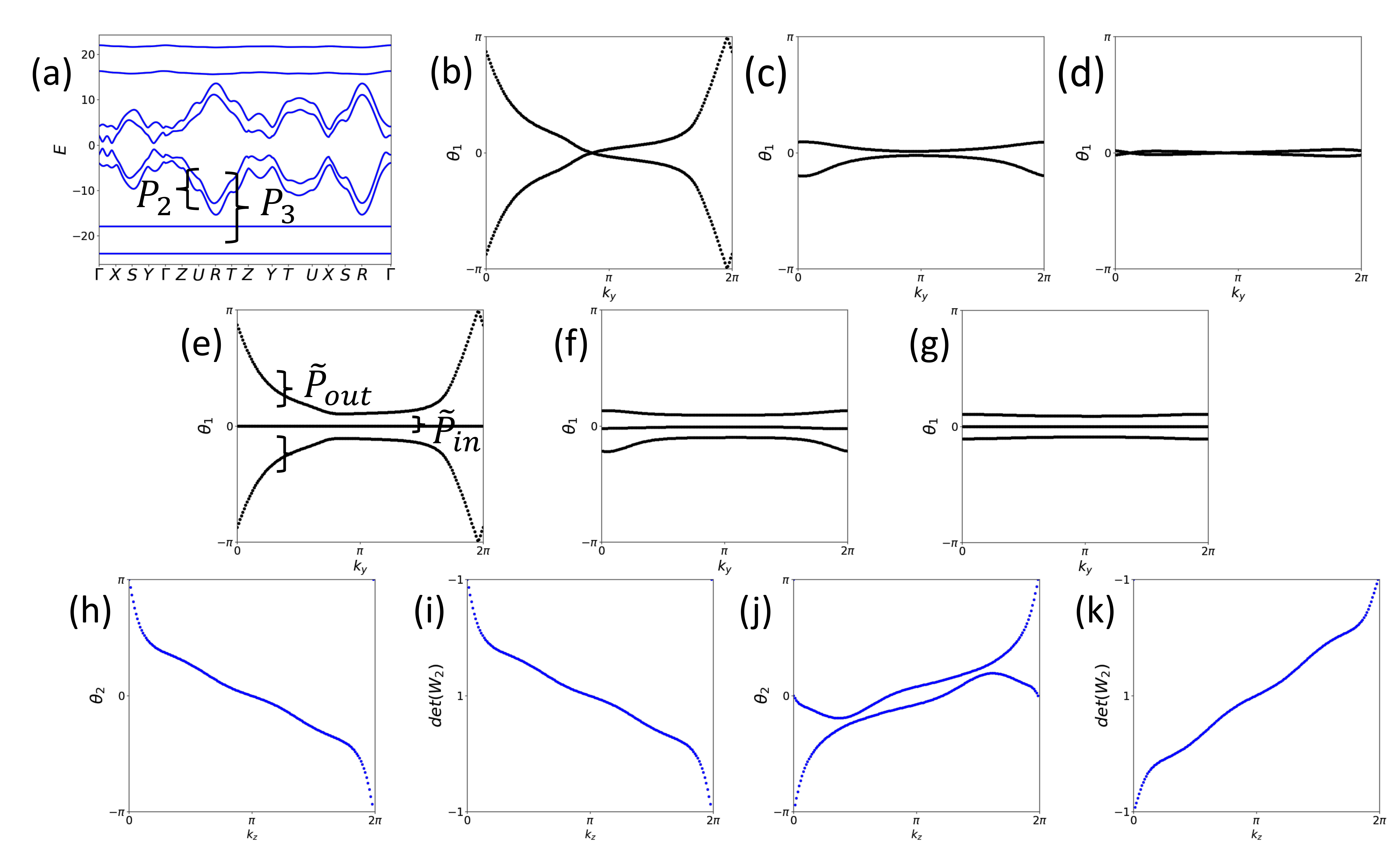}
\caption{(a) Bulk bands of the eight-band, coupled Hamiltonian $\tilde{\mathcal{H}}_{C}(\vec{k})$ (Eq.~(\ref{eq:C2TCoupled})) plotted with the parameters in  Eqs.~(\ref{eq:TIparams}),~(\ref{eq:C2Tparams1}),~(\ref{eq:C2Tparams2}),~(\ref{eq:C2Tparams3}),~(\ref{eq:oneBandParamsC2T}), and~(\ref{eq:coupledParamsOneC2T}).  (b-d) The $x$-directed Wilson loops at $k_{z} = 0$, a generic value, and $\pi$, taken over the two bands in the projector $P_{2}$; these are the original fragile valence bands of the four-band $C_{2z}\times\mathcal{T}$-symmetric AXI model in Eq.~(\ref{eq:C2T}).  Even though these bands are coupled to other bands below the Fermi energy, as they are isolated from them by an energy gap, they still have a well-defined nontrivial (fragile) topology indicated by the winding of the $x$- (and $y$-) directed Wilson loop~\cite{BarryFragile}.  (e-g) The $x$-directed Wilson loops at $k_{z} = 0$, a generic value, and $\pi$, taken over the three bands in the projector $P_{3}$ in (a).  Here, unlike in Fig.~\ref{fig:oneUncoupled}, the Wilson spectrum is gapped and free of winding, allowing us to define particle-hole symmetric nested Wilson projectors onto the inner and outer Wilson bands~\cite{TMDHOTI,ZhidaBLG}.  (h) The $y$-directed nested Wilson loop of the $\tilde{P}_{in}$ Wilson band, plotted as a function of $k_{z}$; it exhibits $C_{\gamma_{2}}=-1$ spectral flow.  (i) $\det(W_{2}(k_{z}))$ of the $\tilde{P}_{in}$ Wilson band; it exhibits $C_{\gamma_{2}}=-1$ spectral flow and is quantized at $-1$ ($+1$) at $k_{z}=0$ $(k_{z}=\pi$) (\ref{sec:detW2C2T}).   (j) The $y$-directed nested Wilson loop of the $\tilde{P}_{out}$ Wilson bands, plotted as a function of $k_{z}$; it exhibits $C_{\gamma_{2}}=+1$ spectral flow.  (k) $\det(W_{2}(k_{z}))$ of the $\tilde{P}_{out}$ Wilson bands; it exhibits $C_{\gamma_{2}}=+1$ spectral flow and is quantized at $-1$ ($+1$) at $k_{z}=0$ $(k_{z}=\pi$) (\ref{sec:detW2C2T}).}
\label{fig:oneCoupledC2T}
\end{figure}

\begin{figure}[h]
\centering
\includegraphics[width=\textwidth]{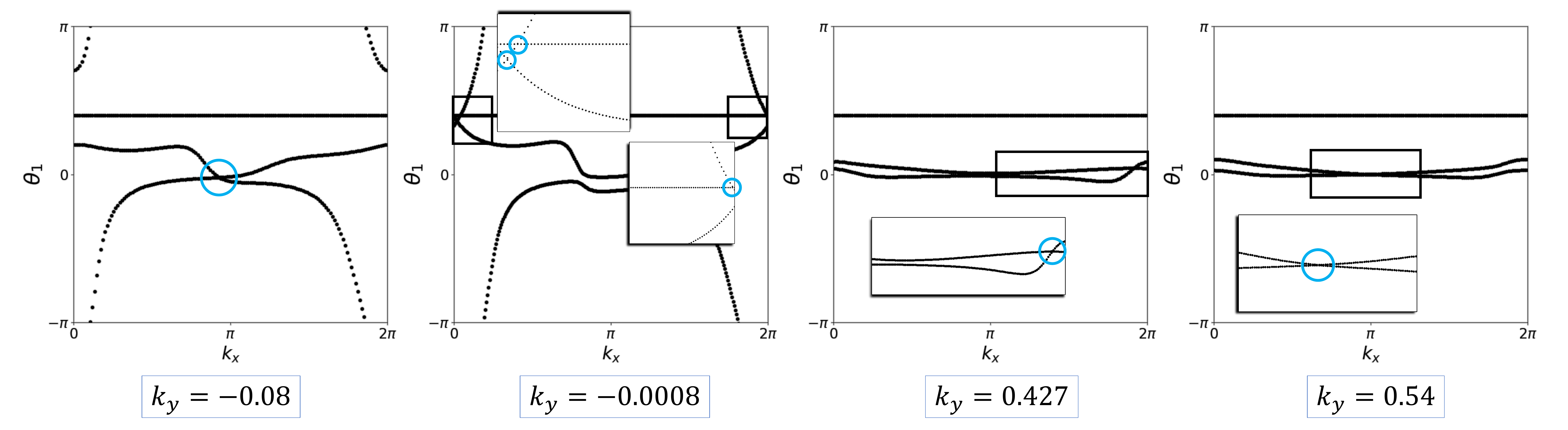}
\caption{The $z$-directed Wilson loop of the three bands in the Wilson projector $P_{3}$ in Fig.~\ref{fig:oneCoupledC2T}(a); the higher two bands in energy are the fragile valence bands of $\mathcal{H}_{C2T}(\vec{k})$ (Eq.~(\ref{eq:C2T})), and the third band comes from an atomic orbital at the general position (Eq.~(\ref{eq:oneBandParamsC2T})).  As shown previously in Fig.~\ref{fig:C2TwilsonZ2}, this Wilson loop also winds, but across the entire 2D BZ.  We highlight with blue circles the six Wilson band degeneracies that comprise this winding; four are inherited from the previous $z$-directed Wilson loop in Fig.~\ref{fig:C2TwilsonZ2}, and two additional ones originate from the intersection of the previous Wilson bands with the Wilson band of the new band in the Wilson projector (Fig.~\ref{fig:oneCoupledC2T}(a)).  As the Wilson crossings at generic crystal momenta and energies are protected by $C_{2z}\times\mathcal{T}$, \emph{without a restriction on the number of bands in the Wilson projector} (Refs.~\onlinecite{JenFragile1,BarryFragile} and in~\ref{sec:zW1C2T}), they persist under the addition of trivial bands to the Wilson projector.  The $z$-directed Wilson loop winding is therefore strong, and not fragile.}
\label{fig:C2TwilsonZOneBand}
\end{figure}

In Fig.~\ref{fig:oneUncoupledC2T}, we plot the bulk bands (a) and $x$-directed Wilson loops (b-d) of Eq.~(\ref{eq:C2TUncoupled}) using the parameters in Eqs.~(\ref{eq:TIparams}),~(\ref{eq:C2Tparams1}),~(\ref{eq:C2Tparams2}), and~(\ref{eq:C2Tparams3}), as well as:
\begin{equation}
\tilde{u}_{3x} = 0.34,\ \tilde{u}_{3y} = 0.25,\ \tilde{u}_{3z} = 0.2,\ \tilde{t}_{sp} = 19.5,\ \tilde{t}_{1} = -1.5,\ \tilde{u}_{s^{x}} = 3,
\label{eq:oneBandParamsC2T}
\end{equation}
in Eqs.~(\ref{eq:C2TorbitalPos}),~(\ref{eq:vspC2T}) and~(\ref{eq:UsxC2T}).  In this limit, the original bands near $E=0$ are completely decoupled from the new bands induced from the orbitals at the general position.  For the Wilson loops in Fig.~\ref{fig:oneUncoupledC2T}(b-d), we include three bands in the Wilson projector $P_{3}$: the original two valence bands of $\mathcal{H}_{C2T}(\vec{k})$ (Eq.~(\ref{eq:C2T})), and one of the four bands from the new orbitals.  In the uncoupled limit of Eq.~(\ref{eq:C2TUncoupled}), the Wilson loop just decomposes into the Wilson spectra of the original valence bands of $\mathcal{H}_{C2T}(\vec{k})$ (Fig.~\ref{fig:C2T}(d-f)) and an additional flat Wilson band at the Wilson energy $\theta_{1}(k_{y},k_{z})=0$.  We now couple the new orbitals to the original four $s$ and $p$ orbitals in $\mathcal{H}_{C2T}(\vec{k})$ (Eq.~(\ref{eq:C2T})) by introducing the term:
\begin{equation}
\tilde{V}_{C}(\vec{k}) = \tilde{v}_{C}\left(\begin{array}{cccc}
0 & 0 & 1 & 1 \\
0 & 0 & 0 & 0 \\
1 & 0 & 0 & 0 \\
1 & 0 & 0 & 0\end{array}\right)\otimes\mathds{1}_{\sigma}.
\label{eq:vcC2T}
\end{equation}
We then form a coupled Hamiltonian by adding $\tilde{V}_{C}$ to Eq.~(\ref{eq:C2TUncoupled}):
\begin{equation}
\tilde{\mathcal{H}}_{C}(\vec{k}) = \tilde{\mathcal{H}}_{U}(\vec{k}) + \tilde{V}_{C}(\vec{k}), 
\label{eq:C2TCoupled}
\end{equation}
which we confirm is also invariant under $C_{2z}\times\mathcal{T}$ symmetry in the expanded $8\times 8$ basis (Eq.~(\ref{eq:8C2T})).  In Fig.~\ref{fig:oneCoupledC2T}, we plot the bulk bands (a), $x$-directed Wilson loops (b-g) using two choices of Wilson projectors, and the $y$-directed nested Wilson loops (h-k) (Fig.~\ref{fig:Wilson}) of Eq.~(\ref{eq:C2TCoupled}) using the parameters in Eqs.~(\ref{eq:TIparams}),~(\ref{eq:C2Tparams1}),~(\ref{eq:C2Tparams2}),~(\ref{eq:C2Tparams3}), and~(\ref{eq:oneBandParamsC2T}), as well as:
\begin{equation}
\tilde{V}_{C} = 3,
\label{eq:coupledParamsOneC2T}
\end{equation}
in Eq.~(\ref{eq:vcC2T}).   In the band structure of $\tilde{\mathcal{H}}_{C}(\vec{k})$ (Fig.~\ref{fig:oneCoupledC2T}(a)), the original two fragile valence bands of Eq.~(\ref{eq:C2T}) remain separated from the new bands by a gap below $E=0$, allowing us to define two Wilson projectors: a projector $P_{3}$ onto the three valence bands closest to $E=0$, and a projector $P_{2}$ onto the original fragile valence bands.  As shown in Ref.~\onlinecite{BarryFragile}, as long as $P_{2}$ is well defined, the Wilson loop of the fragile bands will still wind.  We confirm this in Fig.~\ref{fig:oneCoupledC2T}(b-d), in which we observe the $x$-directed Wilson loop of the bands in $P_{2}$ to exhibit the same winding as the fragile bands of the original four-band $C_{2z}\times\mathcal{T}$-symmetric AXI model (Fig.~\ref{fig:C2T}(d-f)); we also confirm that the $y$-directed Wilson loop of the $P_{2}$ bands also winds.  However, when we calculate the $x$-directed (and $y$-directed) Wilson loop of the $P_{3}$ bands in Fig.~\ref{fig:oneCoupledC2T}(a) (Fig.~\ref{fig:oneCoupledC2T}(e-g)), the Wilson loop winding disappears. 

We are therefore able to calculate a $y$-directed nested Wilson loop of the gapped $x$-directed Wilson bands.  Taking the split Wilson spectrum in Fig.~\ref{fig:oneCoupledC2T}(e-g), we define projectors $\tilde{P}_{in,out}$ onto groupings of Wilson bands related by Wilson particle-hole symmetry, which we use to calculate the $y$-directed nested Wilson loop $W_{2}(k_{z})$ (Fig.~\ref{fig:Wilson} and~\ref{sec:W2}).  As $C_{2z}\times\mathcal{T}$ acts on the Wilson loop as an antiunitary particle-hole symmetry~\cite{JenFragile1,JenFragile2,BarryFragile,ZhidaBLG} that takes $\theta_{1}(k_{y},k_{z})$ to $-\theta(k_{y},-k_{z})$, then, as shown in Refs.~\onlinecite{TMDHOTI,ZhidaBLG} and in~\ref{sec:detW2C2T}, $C_{2z}\times\mathcal{T}$ remains a symmetry of $W_{2}(k_{z})$ under this choice of nested Wilson projectors.  In Fig.~\ref{fig:oneCoupledC2T}(h,i) (Fig.~\ref{fig:oneCoupledC2T}(j,k)), we plot the nested Wilson eigenvalues $\theta_{2}(k_{z})$ and $\det(W_{2}(k_{z}))$ for the inner (outer) Wilson band(s) in Fig.~\ref{fig:oneCoupledC2T}(e-g), respectively.  Both exhibit spectral flow, with the inner (outer) Wilson band(s) exhibiting $C_{\gamma_{2}}=-1$ ($C_{\gamma_{2}}=+1$) winding.  Because the inner and outer Wilson bands exhibit opposite odd-integer Wilson Chern numbers $C_{\gamma_{2}}$, then, as $C_{\gamma_{2}}\text{ mod } 2$ is robust to Wilson gap closures in the presence of bulk $C_{2z}\times\mathcal{T}$ symmetry (\ref{sec:detW2C2T} and Fig.~\ref{fig:oddWindC2T}), it here indicates a well-defined nontrivial bulk topology.  Furthermore, in the $k_{z}=0,\pi$ planes, which are left invariant under the action of $C_{2z}\times\mathcal{T}$, the nested Berry phase $\gamma_{2}(k_{z})$ is quantized (Fig.~\ref{fig:oneCoupledC2T}(i,k) and~\ref{sec:detW2C2T}), with the $k_{z} = 0$ ($k_{z}=\pi$) plane exhibiting a nontrivial (trivial) nested Berry phase:
\begin{equation}
\gamma_{2}(0) = \pi,\ \gamma_{2}(\pi) = 0.
\label{eq:oneBandsGamma2C2T}
\end{equation}

We also note that unlike the previous $\mathcal{I}$-symmetric AXI in~\ref{sec:oneBandInversion}, the $C_{2z}\times\mathcal{T}$-symmetric AXI \emph{still exhibits strong, $z$-directed Wilson loop winding}.  Previously, we demonstrated in Fig.~\ref{fig:C2TwilsonZ2} that for the two valence bands of $\mathcal{H}_{C2T}(\vec{k})$ (Eq.~(\ref{eq:C2T})), the $z$-directed Wilson loop exhibits four crossings across the $k_{x,y}$-plane, which are protected by $C_{2z}\times\mathcal{T}$ symmetry (\ref{sec:zW1C2T}).  Under the addition of one trivial band to the Wilson projector (Fig.~\ref{fig:oneCoupledC2T}(a)), two additional crossings are added to the $z$-directed Wilson loop, and the original four crossings from Fig.~\ref{fig:C2TwilsonZ2} become slightly shifted in Wilson energy and crystal momenta.  We label these six Wilson crossings in Fig.~\ref{fig:C2TwilsonZOneBand} with blue circles.  

Because the $C_{2z}\times\mathcal{T}$-symmetric AXI exhibits strong ($z$-directed) Wilson loop winding and surface states protected by its $(100)$-surface wallpaper group~\cite{WiederLayers,DiracInsulator,FangFuNSandC2,ChenRotation,HarukiRotation,LiangTCI} (Fig.~\ref{fig:C2T}(c,d)), it may be considered a first-order, non-symmetry-indicated, magnetic TCI, rather than a second-order magnetic TI~\cite{HOTIBernevig,HOTIChen}.  However, like the previous $\mathcal{I}$-symmetric AXI in~\ref{sec:oneBandInversion}, it is also equivalent to the cyclic pumping of a 2D fragile topological phase.  

\subsubsection{Coupling Two Additional Bands to a $C_{2z}\times\mathcal{T}$-Symmetric AXI}
\label{sec:twoBandsC2T}

\begin{figure}[h]
\centering
\includegraphics[width=\textwidth]{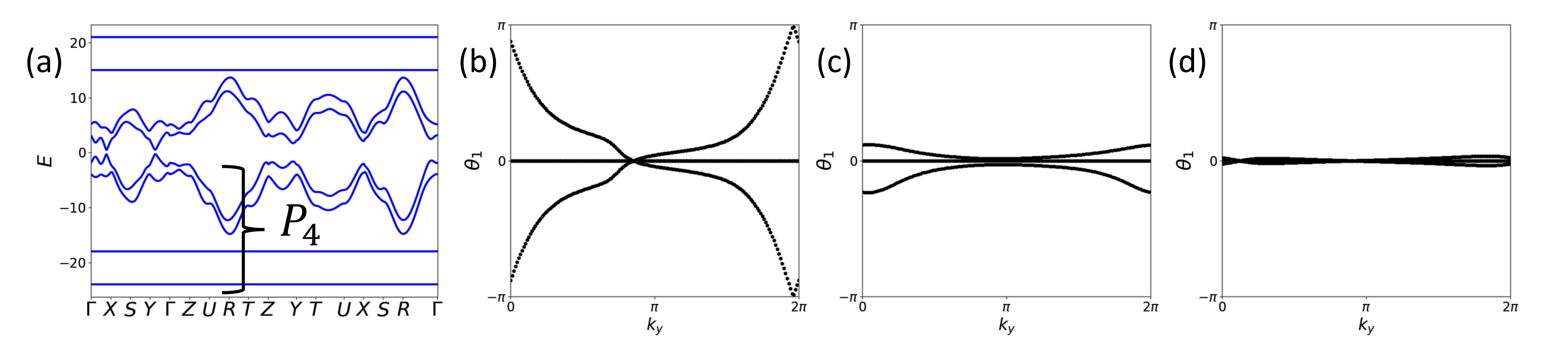}
\caption{(a) Bulk bands of the eight-band, uncoupled Hamiltonian $\tilde{\mathcal{H}}_{U}(\vec{k})$ (Eq.~(\ref{eq:C2TUncoupled})) plotted with the same parameters as Fig.~\ref{fig:oneUncoupledC2T}.  (b-d) The $x$-directed Wilson loops at $k_{z} = 0$, a generic value, and $\pi$, taken over the four bands in the projector $P_{4}$ in (a).  Because the lowest two bands in energy are decoupled from the other two bands in the projector, the Wilson spectra in (b-d) decompose into the superpositions of the previous Wilson spectra in Fig.~\ref{fig:C2T}(f-h) and two additional flat Wilson bands at a particle-hole-symmetric pair of Wilson energies.}
\label{fig:twoUncoupledC2T}
\end{figure}

\begin{figure}[h]
\centering
\includegraphics[width=\textwidth]{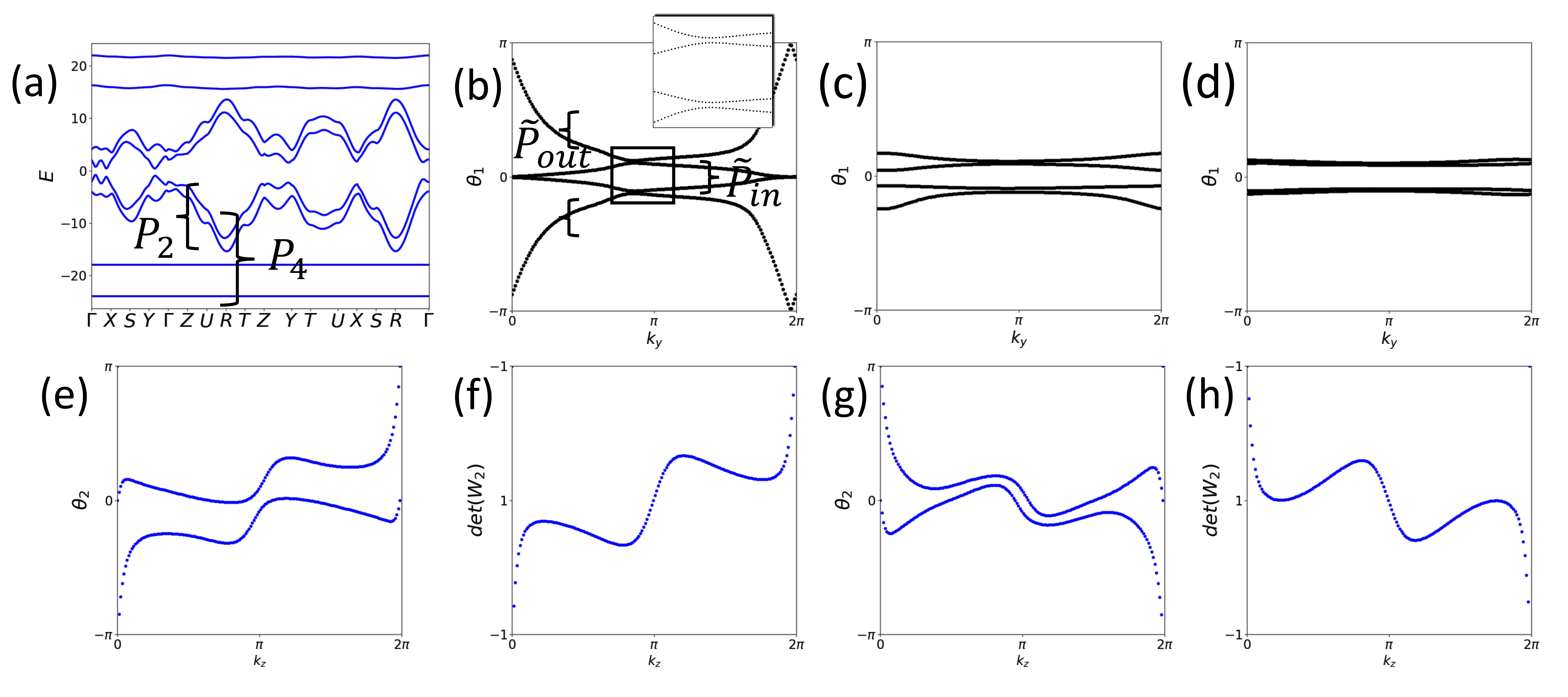}
\caption{(a) Bulk bands of the eight-band, coupled Hamiltonian $\tilde{\mathcal{H}}_{C}(\vec{k})$ (Eq.~(\ref{eq:C2TCoupled})) plotted with the same parameters as Fig.~\ref{fig:oneCoupledC2T}. The $x$-directed Wilson loops over the bands in projector $P_{2}$ in (a) are thus identical to those in Fig.~\ref{fig:oneCoupledC2T}(b-d).  (b-d) The $x$-directed Wilson loops at $k_{z} = 0$, a generic value, and $\pi$, taken over the four bands in the projector $P_{4}$ in (a).  Here, unlike in Fig.~\ref{fig:twoUncoupledC2T}, the $x$-directed Wilson spectrum is gapped and free of winding (the inset in (b) shows a pair of narrowly avoided Wilson band crossings), allowing us to define particle-hole-symmetric nested Wilson projectors onto the inner and outer Wilson bands~\cite{TMDHOTI,ZhidaBLG}.  (e) The $y$-directed nested Wilson loop of the $\tilde{P}_{in}$ Wilson bands, plotted as a function of $k_{z}$; it exhibits $C_{\gamma_{2}}=+1$ spectral flow.  (f) $\det(W_{2}(k_{z}))$ of the $\tilde{P}_{in}$ Wilson bands; it exhibits $C_{\gamma_{2}}=+1$ spectral flow and is quantized at $-1$ ($+1$) at $k_{z}=0$ $(k_{z}=\pi$) (\ref{sec:detW2C2T}).   (g) The $y$-directed nested Wilson loop of the $\tilde{P}_{out}$ Wilson bands, plotted as a function of $k_{z}$; it exhibits $C_{\gamma_{2}}=-1$ spectral flow.  (g) $\det(W_{2}(k_{z}))$ of the $\tilde{P}_{out}$ Wilson bands; it exhibits $C_{\gamma_{2}}=-1$ spectral flow and is quantized at $-1$ ($+1$) at $k_{z}=0$ $(k_{z}=\pi$) (\ref{sec:detW2C2T}).}
\label{fig:twoCoupledC2T}
\end{figure}

\begin{figure}[h]
\centering
\includegraphics[width=\textwidth]{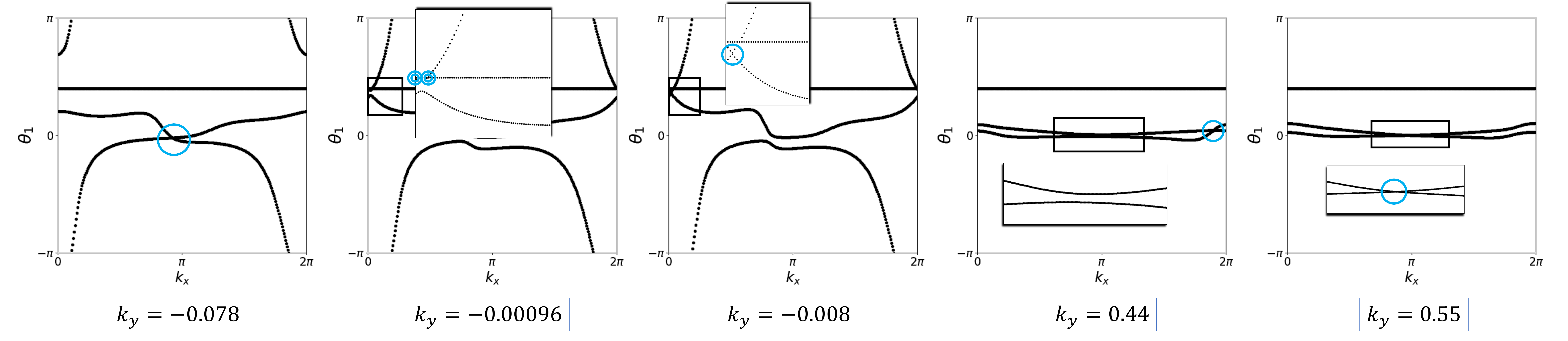}
\caption{The $z$-directed Wilson loop of the four bands in the Wilson projector $P_{4}$ in Fig.~\ref{fig:twoCoupledC2T}(a); the higher two bands in energy are the fragile valence bands of $\mathcal{H}_{C2T}(\vec{k})$ (Eq.~(\ref{eq:C2T})), and the lower two bands come from atomic orbitals at the general position (Eq.~(\ref{eq:oneBandParamsC2T})).  As shown previously in Fig.~\ref{fig:C2TwilsonZ2}, this Wilson loop also winds, but across the entire 2D BZ.  We highlight with blue circles the eight Wilson band degeneracies that comprise this winding; four are inherited from the previous $z$-directed Wilson loop in Fig.~\ref{fig:C2TwilsonZ2}, and four additional ones originate from the intersection of the previous Wilson bands with the Wilson bands of the new bands in the Wilson projector (Fig.~\ref{fig:twoCoupledC2T}(a)).  As both of the two trivial bands originate from atomic orbitals with the same $z$ position in the unit cell ($\tilde{u}_{3/4,z}$ in Eqs.~(\ref{eq:C2TorbitalPos}) and~\ref{eq:oneBandParamsC2T})), and as $\tilde{\mathcal{H}}_{C}(\vec{k})$ (Eq.~(\ref{eq:C2TCoupled})) does not contain all possible coupling terms, there are flat Wilson bands at $\theta_{1}(k_{x},k_{y})=\tilde{u}_{3}\pi$ for all values of $k_{y}$, representing an (artificial) plane degeneracy in the Wilson spectrum.  We therefore only label the points at which this twofold plane Wilson degeneracy intersects the other Wilson bands, which occur at $k_{y} = -0.00096$, and are labeled with double blue circles to represent the higher Wilson degeneracy.  As discussed in Fig.~\ref{fig:C2TwilsonZOneBand}, the point-like Wilson crossings at generic crystal momenta and energies are protected by $C_{2z}\times\mathcal{T}$, without a restriction on the number of bands in the Wilson projector (Refs.~\onlinecite{JenFragile1,BarryFragile} and in~\ref{sec:zW1C2T}), and thus, the $z$-directed Wilson loop winding is strong, and not fragile.}
\label{fig:C2TwilsonZTwoBands}
\end{figure}

We next demonstrate that the addition of two bands to the Wilson projector also removes the Wilson loop winding directed along an axis perpendicular to the $z$-axis of a $C_{2z}\times\mathcal{T}$-symmetric AXI, and also reveals that this AXI can be characterized by a cyclic pump of nested Berry phase.  As the magnetic SG~\cite{MagneticBook,BigBook,BilbaoMagStructures} of $\mathcal{H}_{C2T}(\vec{k})$ (Eq.~(\ref{eq:C2T})), $P2'$, has no unitary crystal symmetries (aside from lattice translation), all bands carry the same symmetry labels, and there are no eigenvalue restrictions on which trivial bands can trivialize the fragile Wilson loop winding~\cite{ZakBandrep1,ZakBandrep2,QuantumChemistry,Bandrep1,Bandrep2,Bandrep3,JenFragile1,JenFragile2,BarryFragile,AshvinIndicators,AshvinMagnetic,ArisInversion}.  We can therefore simply use all of the previous models and parameters from~\ref{sec:oneBandC2T}, and just choose Wilson projectors onto more bands.  We again begin with the uncoupled, eight-band Hamiltonian $\tilde{\mathcal{H}}_{U}(\vec{k})$ (Eq.~(\ref{eq:C2TUncoupled})).  In Fig.~\ref{fig:twoUncoupledC2T}, we plot the bulk (a) and $x$-directed Wilson (b-d) bands of $\tilde{\mathcal{H}}_{U}(\vec{k})$ using the same parameters as previously used in Fig.~\ref{fig:oneUncoupledC2T}, but instead here choose the Wilson projector $P_{4}$ to include all four bands below $E=0$.  These four bands are composed of the original two fragile valence bands of $\mathcal{H}_{C2T}(\vec{k})$ (Eq.~(\ref{eq:C2T})) as well as two bands from atomic orbitals at the general position (Eq.~(\ref{eq:oneBandParamsC2T})).   In the uncoupled limit of Eq.~(\ref{eq:C2TUncoupled}), the Wilson loop just decomposes into the Wilson spectra of the original valence bands of $\mathcal{H}_{C2T}(\vec{k})$ (Fig.~\ref{fig:C2T}(d-f)) and two additional flat Wilson bands superposed at a particle-hole-symmetric pair of Wilson energies.

We next, as in~\ref{sec:oneBandC2T}, again couple the bands far below $E=0$ to the fragile bands above them in energy.  In Fig.~\ref{fig:twoCoupledC2T}, we plot the bulk bands (a), $x$-directed Wilson loops (b-d) using the Wilson projector over all four bands below $E=0$ in (a), and the $y$-directed nested Wilson loops (e-h) (Fig.~\ref{fig:Wilson}) of Eq.~(\ref{eq:C2TCoupled}) using the same parameters used in Fig.~\ref{fig:oneCoupledC2T}.  The bulk band structure (Fig.~\ref{fig:twoCoupledC2T}(a)) is thus identical to the previous band structure in Fig.~\ref{fig:oneCoupledC2T}(a).  Here, however, we are choosing a Wilson projector onto all four of the bands below $E=0$, which include the the original two fragile valence bands of Eq.~(\ref{eq:C2T}), as well as two bands from atomic orbitals at the general position.  As in Fig.~\ref{fig:oneCoupledC2T}, the projector $P_{2}$ onto the original fragile valence bands of $\mathcal{H}_{C2T}(\vec{k})$ is still well-defined, and thus, the $x$-directed Wilson loop of the $P_{2}$ bands still winds~\cite{BarryFragile}, and remains unchanged from Fig.~\ref{fig:oneCoupledC2T}(b-d).  However, the $x$-directed (and $y$-directed) Wilson loop over the four $P_{4}$ bands in Fig.~\ref{fig:twoCoupledC2T}(a) (Fig.~\ref{fig:twoCoupledC2T}(b-d)), no longer winds, as previously occurred in~\ref{sec:oneBandC2T} with three bands in the Wilson projector.  

We are therefore able to again calculate the nested Wilson loop of the gapped, $x$-directed Wilson bands.  Taking the split Wilson spectrum in Fig.~\ref{fig:twoCoupledC2T}(b-d), we again define projectors $\tilde{P}_{in,out}$ onto groupings of Wilson bands related by Wilson particle-hole symmetry~\cite{TMDTI,ZhidaBLG}, which we use to calculate the $y$-directed nested Wilson loop $W_{2}(k_{z})$ (Fig.~\ref{fig:Wilson} and~\ref{sec:W2}).  In Fig.~\ref{fig:twoCoupledC2T}(e,f) (Fig.~\ref{fig:twoCoupledC2T}(g,h)), we plot the nested Wilson eigenvalues $\theta_{2}(k_{z})$ and $\det(W_{2}(k_{z}))$ for the inner (outer) Wilson bands in Fig.~\ref{fig:twoCoupledC2T}(b-d), respectively.  Both exhibit spectral flow, with the inner (outer) Wilson bands exhibiting $C_{\gamma_{2}}=+1$ ($C_{\gamma_{2}}=-1$) winding.  Because the inner and outer Wilson bands exhibit opposite odd-integer Wilson Chern numbers $C_{\gamma_{2}}$, then, as $C_{\gamma_{2}}\text{ mod } 2$ is robust to Wilson gap closures in the presence of bulk $C_{2z}\times\mathcal{T}$ symmetry (\ref{sec:detW2C2T} and Fig.~\ref{fig:oddWindC2T}), it here indicates a well-defined nontrivial bulk topology.  Furthermore, in the $k_{z}=0,\pi$ planes, the nested Berry phase $\gamma_{2}(k_{z})$ is quantized (Fig.~\ref{fig:twoCoupledC2T}(f,h) and~\ref{sec:detW2C2T}), with the $k_{z} = 0$ ($k_{z}=\pi$) plane exhibiting a nontrivial (trivial) nested Berry phase:
\begin{equation}
\gamma_{2}(0) = \pi,\ \gamma_{2}(\pi) = 0.
\label{eq:twoBandsGamma2C2T}
\end{equation}
Thus, a $C_{2z}\times\mathcal{T}$-symmetric AXI with four or more occupied bands is not Wannierizable, and is instead equivalent to the cyclic $z$-directed pumping of nested Berry phase.

Finally, as noted previously in~\ref{sec:oneBandC2T}, the $C_{2z}\times\mathcal{T}$-symmetric AXI still exhibits strong, $z$-directed Wilson loop winding.  Previously, we demonstrated in Fig.~\ref{fig:C2TwilsonZ2} that for the two occupied bands of $\mathcal{H}_{C2T}(\vec{k})$ (Eq.~(\ref{eq:C2T})), the $z$-directed Wilson loop exhibits four crossings across the $k_{x,y}$-plane, which are protected by $C_{2z}\times\mathcal{T}$ symmetry (\ref{sec:zW1C2T}).  Under the addition of two trivial bands to the Wilson projector (Fig.~\ref{fig:twoCoupledC2T}(a)), four additional crossings are added to the $z$-directed Wilson loop, and the original four crossings from Fig.~\ref{fig:C2TwilsonZ2} become slightly shifted in Wilson energy and crystal momenta.  We label these eight Wilson crossings in Fig.~\ref{fig:C2TwilsonZTwoBands} with blue circles.  This further reinforces that the $C_{2z}\times\mathcal{T}$-symmetric AXI, with any number of occupied bands, can either be described as a first-order, non-symmetry-indicated, magnetic TCI, or as a magnetic HOTI equivalent to the cyclic pumping of a 2D fragile topological phase.  

\section{Wilson Loops}
\label{sec:WilsonLoops}

\subsection{Definition of the Wilson Loop and Nested Wilson Loop}
\label{sec:W2}

Here, we define the discretized Wilson loop and nested Wilson loop.  For notational simplicity, we assume that the (nested) Wilson loop is calculated in a 3D crystal with mutually orthogonal crystal axes, \emph{i.e.} a primitive orthorhombic, tetragonal, or cubic crystal.  The Hamiltonian of this crystal $\mathcal{H}(k_{x},k_{y},k_{z})$ characterizes a set of energy bands; here we focus on a grouping of energy bands that is separated from the other bands by a gap at all crystal momenta.  We can then calculate the $x$-directed Wilson loop as it is defined in Refs.~\onlinecite{ArisInversion,Cohomological,DiracInsulator}:
\begin{align}
\left[ W_{1(k_{x0},k_\perp)}  \right]_{nm} & \equiv \left[ \mathcal{P} e^{i \int_{k_{x0}}^{k_{x0}+2\pi} dk_x A_x(k_{x0},k_\perp)}\right]_{nm}  \nonumber\\
&\approx  \left[ \mathcal{P} e^{i \frac{2\pi}{N}\sum_{j=1}^N A_x(k_{x0}+ \frac{2\pi j}{N},k_\perp) }\right]_{nm} \nonumber\\
\approx \langle u^n(k_{x0}+2\pi,k_\perp) |  \bigg[ \mathcal{P} \prod_{j=1}^N  P(k_{x0}+\frac{2\pi j}{N},k_\perp ) & \left( 1  - \frac{2\pi }{N} \partial_{k_x}|_{(k_{x0}+ \frac{2\pi j}{N},k_\perp) }\right) P(k_{x0}+\frac{2\pi j}{N},k_\perp ) \bigg] |u^m(k_{x0},k_\perp)\rangle \nonumber\\
&\approx \langle u^n(k_{x0},k_\perp)|V(2\pi\hat{x})\Pi(k_{x0},k_\perp ) |u^m(k_{x0},k_\perp) \rangle, 
\label{eq:wilsondisc}
\end{align}
where $k_\perp \equiv (k_y,k_z)$, $P(\mathbf{k})$ is the projector onto the occupied states (here the separated grouping of energy bands):
\begin{equation}
P(\mathbf{k}) = \sum_{n=1}^{n_{\rm occ}}|u^n(\mathbf{k})\rangle \langle u^n(\mathbf{k})|,
\label{eq:defproj}
\end{equation}
and where in the last line of Eq.~(\ref{eq:wilsondisc}), we have defined the ordered product of projectors,
\begin{equation} 
\Pi(k_{x0},k_\perp) \equiv P(k_{x0}  + 2\pi,k_\perp)P(k_{x0} +\frac{2\pi (N-1)}{N},k_\perp)\cdots P(k_{x0}+\frac{2\pi }{N},k_\perp).
\end{equation} 
Crucially, the discretized loop defined by the product of projectors in Eq.~(\ref{eq:wilsondisc}) is closed by a sewing matrix:
\begin{equation}
\left[V(2\pi\hat{x})\right]_{nm} =  |u^n(k_{x0},k_\perp)\rangle\langle u^m(k_{x0}+2\pi,k_\perp)|,
\label{eq:defsew}
\end{equation}
that enforces the gauge and basepoint ($k_{x0}$) invariance of the eigenvalues of Eq.~(\ref{eq:wilsondisc})~\cite{ArisInversion,Cohomological,DiracInsulator}.  Specifically, Eq.~(\ref{eq:defsew}) closes the Wilson loop with the overlap between states at the basepoint and states at the basepoint plus a reciprocal lattice vector; this allows Eq.~(\ref{eq:wilsondisc}) to be independent of the choice of basepoint or BZ, even if $|u^{n}(k_{x},k_\perp)\rangle$ is not $2\pi$-periodic.  If the eigenstates of $\mathcal{H}(k_{x},k_{y},k_{z})$ are chosen to be mutually orthogonal, then $V(2\pi\hat{x})$ simplifies into a diagonal matrix of phases~\cite{DiracInsulator}; however, whenever there is a band degeneracy at a point $\mathbf{k}$ within the bands in the Wilson projector $P(\mathbf{k})$ (Eq.~(\ref{eq:defproj})), some numerical diagonalization algorithms will not automatically generate an orthonormal basis of eigenstates at that point, resulting in the appearance of off-diagonal elements in $V(2\pi\hat{x})$.  The eigenvalues of $W_{1(k_{x0},k_\perp)}$ are gauge-independent~\cite{Fidkowski2011,ArisInversion,Cohomological,DiracInsulator} and take the form of phases $\exp(i\theta_{1}(k_{y},k_{z}))$.  We can thus define a Hermitian ``Wilson Hamiltonian,''
\begin{equation}
\left[H_{W_{1(k_{x0})}}(k_{y},k_{z})\right]_{nm} = \left[-i\ln(W_{1(k_{x0},k_{y},k_{z})})\right]_{nm} \equiv H_{W_{1}}(k_{y},k_{z})= -i\ln(W_{1}(k_{y},k_{z})),
\label{eq:WilsonHam}
\end{equation}
where in the equivalence, we define the less formal expressions for the $x$-directed Wilson Hamiltonian and loop with suppressed band indices used throughout this work.  The eigenvalues of $\mathcal{H}_{W_{1(k_{x0})}}(k_{y},k_{z})$ take the form of real angles $\theta_{1}(k_{y},k_{z})$ and form smooth and continuous ``Wilson bands''~\cite{ArisInversion,Cohomological,DiracInsulator}.  We refer to the values of $\theta_{1}(k_{y},k_{z})$ as ``Wilson energies.''

It was recently discovered that a \emph{nested} Wilson loop can be performed on a set of Wilson bands that is isolated in Wilson energy from all other Wilson bands~\cite{multipole,WladTheory}.  To calculate the $y$-directed nested Wilson loop (Fig.~\ref{fig:Wilson}), we perform the same steps employed in Eqs.~(\ref{eq:wilsondisc}) to~(\ref{eq:WilsonHam}), only this time projecting onto the eigenstates $|\tilde{u}^n(k_{x0},k_{y},k_{z})\rangle$ of $H_{W_{1(k_{x0})}}(k_{y},k_{z})$.  From a numerical perspective, it is crucial that the Wilson eigenstates $|\tilde{u}^n(k_{x0},k_{y},k_{z})\rangle$ are expressed in terms of the eigenstates $|u^n(k_{x},k_{y},k_{z})\rangle$ of the original Hamiltonian $\mathcal{H}(k_{x},k_{y},k_{z})$, as opposed to being numerically regenerated using only $H_{W_{1(k_{x0})}}(k_{y},k_{z})$, so that information regarding the positions of the atoms in the unit cell is not lost.  Specifically, to use $|\tilde{u}^n(k_{x0},k_{y},k_{z})\rangle$ to calculate a $y$-directed nested Wilson loop, one state $|\tilde{u}^n(k_{x0},k_{y0},k_{z})\rangle$ along the $k_{y}$ direction must contain the contribution to the (nested) Berry phase acquired from the relative $y$-axis coordinates of the atoms in the unit cell.  A rote numerical diagonalization of Eq.~(\ref{eq:WilsonHam}) does not automatically correctly assign these phases; they must be numerically re-embedded.  We implement this in~\text{PythTB} by forming a diagonal embedding matrix of phases $\exp(i\pi u_{iy})$ where $u_{iy}$ is the $y$ coordinate of the $i^{\text{th}}$ atomic orbital, and then right-multiply it with $|\tilde{u}^n(k_{x0},k_{y0},k_{z})\rangle$.

We define the $y$-directed nested Wilson loop:
\begin{align}
\left[ W_{2(k_{x0},k_{y0},k_{z})}  \right]_{nm} & \equiv \left[ \mathcal{P} e^{i \int_{k_{y0}}^{k_{y0}+2\pi} dk_y \tilde{A}_y(k_{x0},k_{y0},k_{z})}\right]_{nm}  \nonumber\\
&\approx  \left[ \mathcal{P} e^{i \frac{2\pi}{N}\sum_{j=1}^N \tilde{A}_y(k_{x0},k_{y0}+ \frac{2\pi j}{N},k_z) }\right]_{nm} \nonumber\\
\approx \langle \tilde{u}^n(k_{x0},k_{y0}+2\pi,k_z) |  \bigg[ \mathcal{P} \prod_{j=1}^N  \tilde{P}(k_{x0},k_{y0}+\frac{2\pi j}{N},k_z ) &  \left( 1  - \frac{2\pi }{N} \partial_{k_y}|_{(k_{x0},k_{y0}+ \frac{2\pi j}{N},k_z) }\right)   \tilde{P}(k_{x0},k_{y0}+\frac{2\pi j}{N},k_z) \bigg] |\tilde{u}^m(k_{x0},k_{y0},k_{z})\rangle \nonumber\\
&\approx \langle \tilde{u}^n(k_{x0},k_{y0},k_{z})|\tilde{V}(2\pi\hat{y})\tilde{\Pi}(k_{x0},k_{y0},k_{z} ) |\tilde{u}^m(k_{x0},k_{y0},k_{z}) \rangle, 
\label{eq:wilsondisc2}
\end{align}
where tildes indicate quantities obtained from the Wilson Hamiltonian (Eq.~(\ref{eq:WilsonHam})), $\tilde{P}(\mathbf{k})$ is the projector onto the occupied Wilson states (here the separated grouping of Wilson bands):
\begin{equation}
\tilde{P}(\mathbf{k}) = \sum_{n=1}^{n_{\rm occ}}|\tilde{u}^n(\mathbf{k})\rangle \langle \tilde{u}^n(\mathbf{k})|,
\label{eq:defproj2}
\end{equation}
and where in the last line of Eq.~(\ref{eq:wilsondisc2}), we have defined the ordered product of Wilson projectors,
\begin{equation} 
\tilde{\Pi}(k_{x0},k_{y0},k_{z}) \equiv \tilde{P}(k_{x0},k_{y0}  + 2\pi,k_z) \tilde{P}(k_{x0},k_{y0} +\frac{2\pi (N-1)}{N},k_z)\cdots \tilde{P}(k_{x0},k_{y0}+\frac{2\pi }{N},k_z).
\end{equation} 
The discretized nested loop in Eq.~(\ref{eq:wilsondisc2}) is also closed by a sewing matrix:
\begin{equation}
\left[\tilde{V}(2\pi\hat{y})\right]_{nm} =  |\tilde{u}^n(k_{x0},k_{y0},k_{z})\rangle\langle \tilde{u}^m(k_{x0},k_{y0}+2\pi,k_z)|,
\label{eq:nestedSew}
\end{equation}
that enforces the basepoint ($k_{y0}$) independence of Eq.~(\ref{eq:wilsondisc2})~\cite{multipole,WladTheory} in the same manner that Eq.~(\ref{eq:defsew}) does for Eq.~(\ref{eq:wilsondisc}).  The eigenvalues of $W_{2(k_{x0},k_{y0},k_{z})}$ are gauge-independent~\cite{multipole,WladTheory} and take the form of phases $\exp(i\theta_{2}(k_{z}))$.  We can thus define a Hermitian ``nested Wilson Hamiltonian,''
\begin{equation}
\left[H_{W_{2(k_{x0},k_{y0})}}(k_{z})\right]_{nm} = \left[-i\ln(W_{2(k_{x0},k_{y0},k_{z})})\right]_{nm} \equiv  H_{W_{2}}(k_{z})   = -i\ln(W_{2}(k_{z})),
\label{eq:WilsonHam2}
\end{equation}
where in the equivalence, we define the less formal expressions for the $y$-directed nested Wilson Hamiltonian and loop with suppressed Wilson band indices used throughout this work.  The eigenvalues of $H_{W_{2(k_{x0},k_{y0})}}(k_{z})$ take the form of real angles $\theta_{2}(k_{z})$ and form smooth and continuous ``nested Wilson bands''~\cite{WladTheory,HOTIBernevig}.  We refer to the values of $\theta_{2}(k_{z})$ as ``nested Wilson energies.''

\subsection{The Action of $\mathcal{I}$ Symmetry on the Nested Wilson Loop}
\label{sec:detW2Inv}

In this section, we derive the action of inversion symmetry ($\mathcal{I}$) on the nested Wilson loop $W_{2}(k_{z})$ as defined in~\ref{sec:W2}.  Specifically, we show that $\mathcal{I}$ acts to make the nested Wilson spectrum particle-hole symmetric in 3D, and to quantize the nested Berry phase in 2D.  We then show that this particle-hole symmetry preserves the robustness of odd winding in the nested Wilson loop $C_{\gamma_{2}}$, and thus indicates a phase with $\mathbb{Z}_{2}$-nontrivial magnetoelectric polarizability~\cite{VDBAxion,QHZ,AndreiInversion,AshvinAxion1,AshvinAxion2,WuAxionExp,VDBHOTI} $\theta=\pi$.  We begin by reproducing the result derived in Ref.~\onlinecite{ArisInversion} (and reproduced in Ref.~\onlinecite{DiracInsulator}) that $\mathcal{I}$ acts on $H_{W_{1}}(k_{y},k_{z})$ as a unitary particle-hole symmetry $\tilde{\chi}$ that flips the signs of $k_{y}$ and $k_{z}$:
\begin{align}
\mathcal{I}W_{1(k_{x0},k_\perp)}\mathcal{I}^{\dagger} & = \mathcal{I}V(2\pi\hat{x})\Pi(k_{x0},k_\perp )\mathcal{I}^{\dagger} \nonumber \\
&  = V^{\dagger}(2\pi\hat{x})\mathcal{I}\left[P(k_{x0}  + 2\pi,k_\perp)P(k_{x0} +\frac{2\pi (N-1)}{N},k_\perp)\cdots P(k_{x0}+\frac{2\pi }{N},k_\perp)\right]\mathcal{I}^{\dagger} \nonumber \\
& = V^{\dagger}(2\pi\hat{x})P(-k_{x0}  - 2\pi,-k_\perp)P(-k_{x0} -\frac{2\pi (N-1)}{N},-k_\perp)\cdots P(-k_{x0}-\frac{2\pi }{N},-k_\perp) \nonumber \\
& =  V^{\dagger}(2\pi\hat{x})V(2\pi\hat{x})\Pi^{\dagger}(-k_{x0},-k_\perp )V^{\dagger}(2\pi\hat{x}) \nonumber \\
& = W^{\dagger}_{1(-k_{x0},-k_\perp)},
\end{align}
in which we have used that $\mathcal{I}$ transforms Eq.~(\ref{eq:defsew}):
\begin{equation}
\mathcal{I}V(2\pi\hat{x})\mathcal{I}^{\dag} = V^{\dag}(2\pi\hat{x}).
\end{equation}
The Wilson Hamiltonian is therefore invariant under a unitary particle-hole symmetry, which we denote as $\tilde{\chi}$, that flips the signs of $k_{y}$ and $k_{z}$:
\begin{equation}
\tilde{\chi} H_{W_{1}}(k_{y},k_{z})\tilde{\chi}^{\dag} = -H_{W_{1}}(-k_{y},-k_{z}),
\label{eq:invWilson1}
\end{equation}
implying that for every Wilson eigenstate $|\tilde{u}^n(k_{x0},k_{y},k_{z})\rangle$ with eigenvalue $\theta_{1}(k_{y},k_{z})$, there is another eigenstate $\tilde{\chi}|\tilde{u}^n(k_{x0},k_{y},k_{z})\rangle$ with eigenvalue $-\theta_{1}(k_{y},k_{z})$.  We can therefore represent $\tilde{\chi}$ as:
\begin{equation}
\tilde{\chi}|\tilde{u}^n(k_{x0},k_{y},k_{z})\rangle = \tilde{U} |\tilde{u}^n(-k_{x0},-k_{y},-k_{z})\rangle,
\label{eq:chiOnN}
\end{equation}
where $\tilde{U}$ is a $\mathbf{k}$-independent unitary transformation that rotates the Wilson band index $n$.  

We can now determine the action of $\tilde{\chi}$ (and thus $\mathcal{I}$) on the nested Wilson loop.  It is important to note that if the nested Wilson loop is only calculated using nested Wilson projectors onto bands in half of the Wilson energy spectrum, as it originally was in Refs.~\onlinecite{multipole,WladTheory,HingeSM}, then $\tilde{\chi}$ is not a symmetry of the nested Wilson loop, as it takes Wilson bands from inside the nested Wilson projector to bands out of it.  Instead, as recently recognized in Refs.~\onlinecite{TMDHOTI,ZhidaBLG}, we will only choose nested Wilson projectors $\tilde{P}(\mathbf{k})$ onto particle-hole conjugate pairs of Wilson bands, such that:
\begin{equation}
\tilde{\chi}\tilde{P}(\mathbf{k})\tilde{\chi}^{\dag} = \tilde{P}(-\mathbf{k}),
\label{eq:invNestedProj}
\end{equation}
as $\tilde{P}(\mathbf{k})$ onto a block of particle-hole-symmetric Wilson bands is proportional to the identity in the basis of Wilson band indices, and the factors of $\tilde{U}$ and $\tilde{U}^{\dag}$ cancel out.  Eqs.~(\ref{eq:chiOnN}) and~(\ref{eq:invNestedProj}) also imply that $\tilde{\chi}$ transforms the nested sewing matrix (Eq.~(\ref{eq:nestedSew})):
\begin{align}
\tilde{\chi}\tilde{V}(2\pi\hat{y})\tilde{\chi}^{\dag} &= \tilde{\chi}\left[\tilde{P}(k_{x0},k_{y0},k_{z})\tilde{P}(k_{x0},k_{y0} + 2\pi,k_{z}) + (\mathds{1} - \tilde{P}(k_{x0},k_{y0},k_{z}))(\mathds{1} - \tilde{P}(k_{x0},k_{y0} + 2\pi,k_{z}))\right]\tilde{\chi}^{\dagger} \nonumber \\
&= \tilde{P}(-k_{x0},-k_{y0},-k_{z})\tilde{P}(-k_{x0},-k_{y0} -2\pi,-k_{z}) + (\mathds{1} - \tilde{P}(-k_{x0},-k_{y0},-k_{z}))(\mathds{1} - \tilde{P}(-k_{x0},-k_{y0} -2\pi, - k_{z})) \nonumber \\
&= \tilde{V}^{\dag}(2\pi\hat{y}).
\label{eq:invNestedV}
\end{align}

Using nested Wilson projectors that satisfy Eq.~(\ref{eq:invNestedProj}), we can finally deduce the action of $\tilde{\chi}$, and thus bulk $\mathcal{I}$, on the  $y$-directed nested Wilson loop:
\begin{align}
\tilde{\chi}W_{2(k_{x0},k_{y0},k_{z})}\tilde{\chi}^{\dag} & = \tilde{\chi}\tilde{V}(2\pi\hat{y})\tilde{\Pi}(k_{x0},k_{y0},k_{z} )\tilde{\chi}^{\dagger} \nonumber \\
&  = \tilde{V}^{\dagger}(2\pi\hat{y})\tilde{\chi}\left[\tilde{P} (k_{x0},k_{y0}  + 2\pi,k_z)\tilde{P}(k_{x0}, k_{y0} +\frac{2\pi (N-1)}{N},k_z)\cdots \tilde{P}(k_{x0},k_{y0}+\frac{2\pi }{N},k_z)\right]\tilde{\chi}^{\dagger} \nonumber \\
& = V^{\dagger}(2\pi\hat{y})\tilde{P}(-k_{x0},-k_{y0}  - 2\pi,-k_z)\tilde{P}(-k_{x0},-k_{y0} -\frac{2\pi (N-1)}{N},-k_z)\cdots \tilde{P}(-k_{x0},-k_{y0}-\frac{2\pi }{N},-k_z) \nonumber \\
& =  \tilde{V}^{\dagger}(2\pi\hat{y})\tilde{V}(2\pi\hat{y})\tilde{\Pi}^{\dagger}(-k_{x0},-k_{y0},-k_{z} )\tilde{V}^{\dagger}(2\pi\hat{y}) \nonumber \\
& = W^{\dagger}_{2(-k_{x0},-k_{y0},-k_{z})}.
\label{eq:W2InvEq}
\end{align}
The nested Wilson Hamiltonian is therefore invariant under a unitary particle-hole symmetry, which we denote as $\tilde{\tilde{\chi}}$, that flips the sign of $k_{z}$:
\begin{equation}
\tilde{\tilde{\chi}} H_{W_{2}}(k_{z})\tilde{\tilde{\chi}}^{\dag} = -H_{W_{2}}(-k_{z}),
\label{eq:invWilson2}
\end{equation}
implying that for every nested Wilson eigenstate $|\tilde{\tilde{u}}^n(k_{x0},k_{y0},k_{z})\rangle$ with eigenvalue $\theta_{2}(k_{z})$, there is another eigenstate $\tilde{\tilde{\chi}}|\tilde{\tilde{u}}^n(k_{x0},k_{y0},k_{z})\rangle$ with eigenvalue $-\theta_{2}(k_{z})$. 

\begin{figure}[h]
\centering
\includegraphics[width=0.9\textwidth]{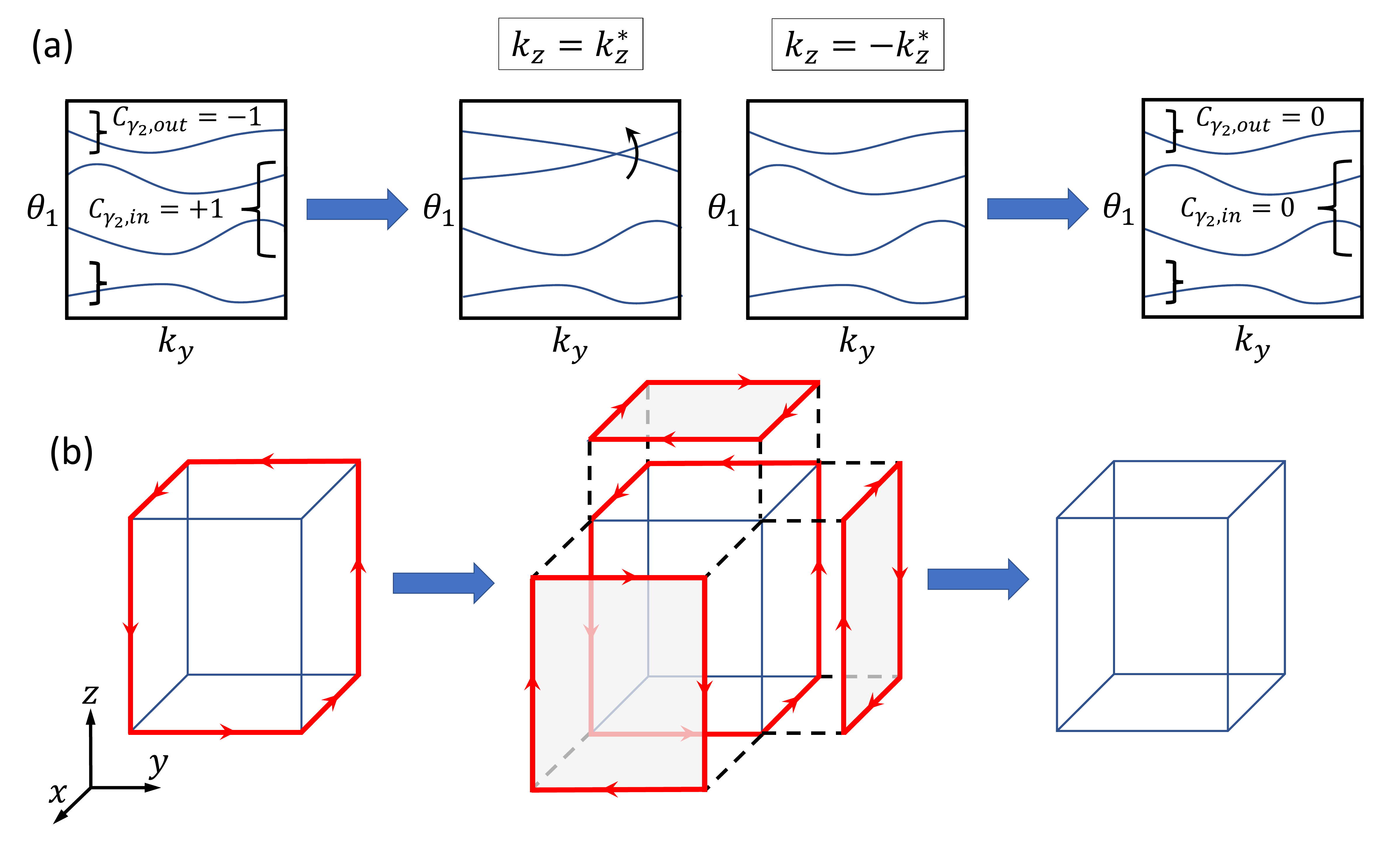}
\caption{(a) In a 3D magnetic insulator with no symmetries, gapless points can form between Wilson bands at any Wilson energy and crystal momentum, \emph{without corresponding gapless points in the energy spectrum}~\cite{Fidkowski2011,AndreiXiZ2,BarryFragile}.  These Wilson gap closings transfer Wilson Chern number between groupings of Wilson bands.  Therefore, even Wilson bands with nontrivial Wilson Chern numbers $C_{\gamma_{2}}$ are representative of a bulk trivial topology if individual (unpaired) Wilson crossings are allowed.  Specifically, if $\mathcal{I}$ (or $C_{2z}\times\mathcal{T}$ symmetry (\ref{sec:oneBandC2T} and~\ref{sec:twoBandsC2T})) is relaxed in an AXI with alternating winding between the inner and outer groupings of Wilson bands $C_{\gamma_{2},in/out} = \pm1$, then that winding can go to zero without closing an energy gap.  (b) In a finite-sized AXI, the Wilson gap closures in (a) are equivalent to removing the chiral hinge mode through an $\mathcal{I}$- (or $C_{2z}\times\mathcal{T}$-) asymmetric series of surface band inversions, or by gluing an $\mathcal{I}$- (or $C_{2z}\times\mathcal{T}$-) breaking arrangement of $|C|=1$ Chern insulators to crystal facets.}
\label{fig:noSyms}
\end{figure}

\begin{figure}[h]
\centering
\includegraphics[width=0.9\textwidth]{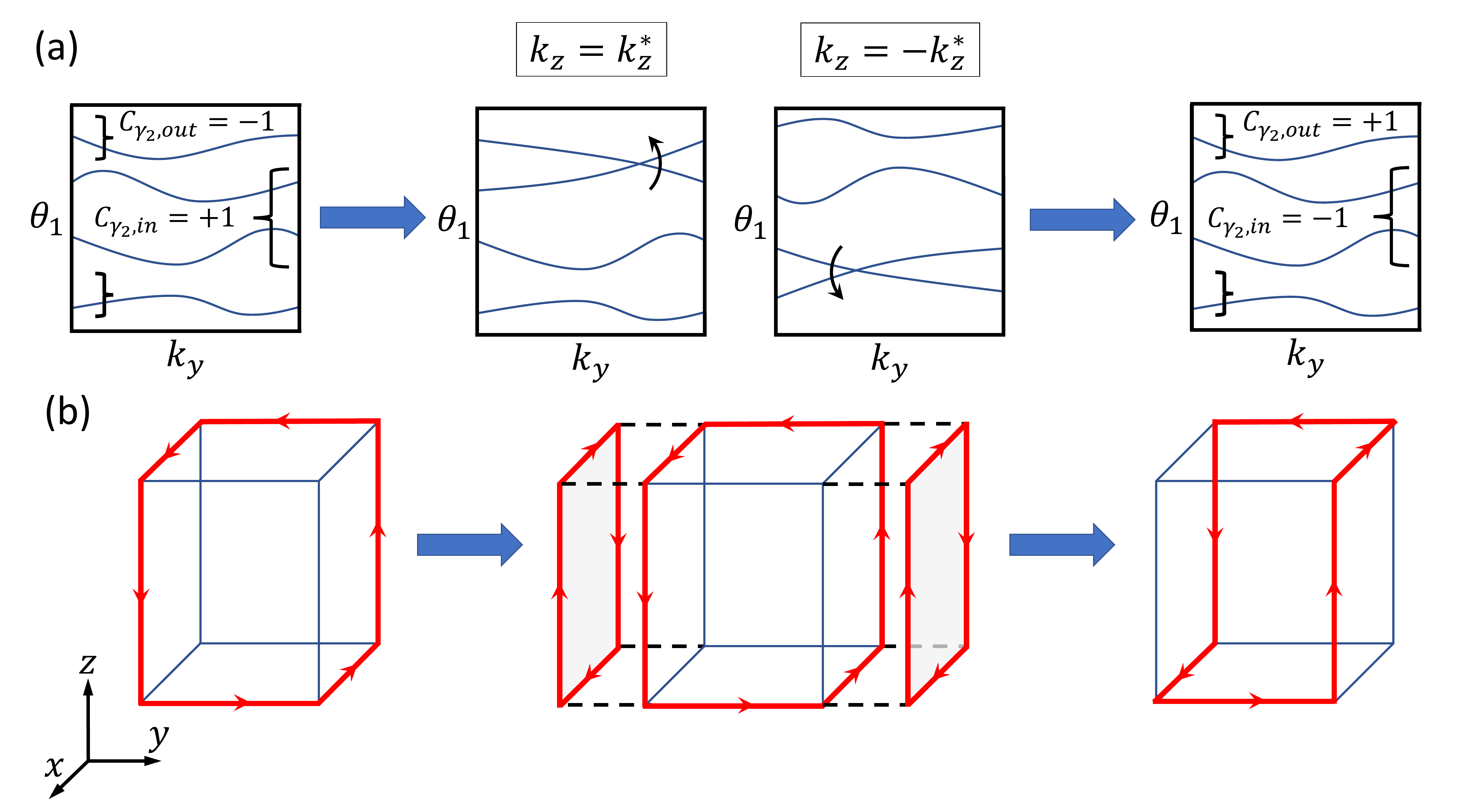}
\caption{(a)  When bulk $\mathcal{I}$ symmetry is present, then Wilson crossings in a 3D magnetic insulator can only form in pairs at $\theta_{1}(k_{y},k_{z})$ and $-\theta_{1}(-k_{y},-k_{z})$ (Eq.~(\ref{eq:invWilson1})).  Without further restrictions, these crossings can change the Wilson Chern numbers of particle-hole-symmetric groupings of Wilson bands by $\pm 2$.  As in Fig.~\ref{fig:noSyms}, these pairs of Wilson gapless points can form without closing a bulk gap.  Therefore if particle-hole-symmetric groupings of Wilson bands exhibit alternating odd values of $C_{\gamma_{2}}$, then their Wilson Chern numbers \emph{cannot} go to zero without closing a bulk gap, though they can change in value by $\pm 2$.  Thus, $C_{\gamma_{2}}\text{ mod }2$ represents a well-defined, bulk $\mathbb{Z}_{2}$ topological invariant (Eq.~(\ref{eq:thetaInv})).  (b) In a finite-sized $\mathcal{I}$-symmetric AXI, the Wilson gap closures in (a) are equivalent to moving the chiral hinge mode through an $\mathcal{I}$-symmetric series of surface band inversions, or by gluing an $\mathcal{I}$-preserving arrangement of $|C|=1$ Chern insulators to crystal facets.}
\label{fig:oddWindI}
\end{figure}

\begin{figure}[h]
\centering
\includegraphics[width=0.9\textwidth]{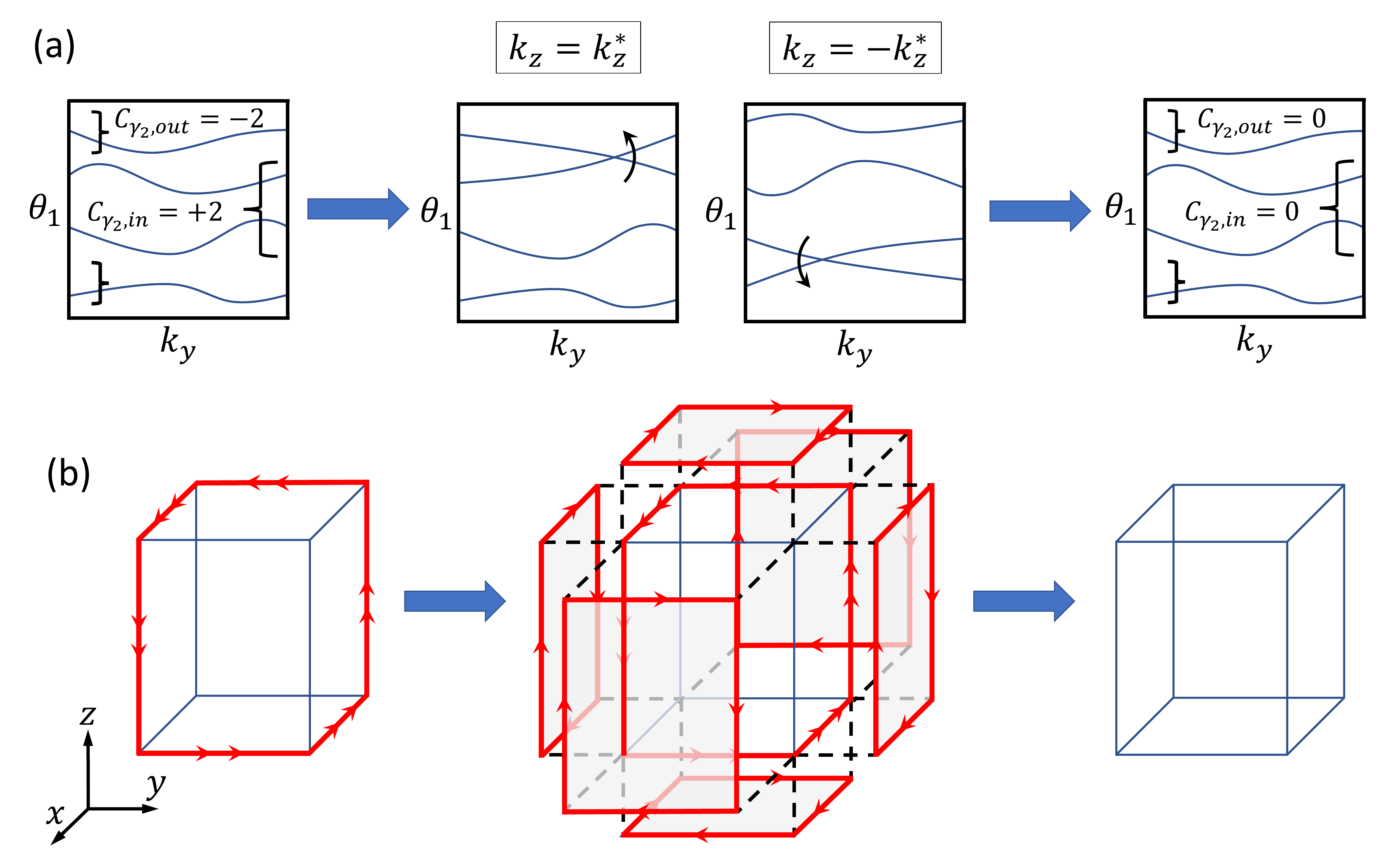}
\caption{(a)  When bulk $\mathcal{I}$ symmetry is present, then Wilson crossings in a 3D magnetic insulator can only form in pairs at $\theta_{1}(k_{y},k_{z})$ and $-\theta_{1}(-k_{y},-k_{z})$ (Eq.~(\ref{eq:invWilson1})).  Without further restrictions, these crossings can change the Wilson Chern numbers of particle-hole-symmetric groupings of Wilson bands by $\pm 2$.  As in Fig.~\ref{fig:noSyms}, these pairs of Wilson gapless points can form without closing a bulk gap.  Therefore if particle-hole-symmetric groupings of Wilson bands exhibit alternating even values of $C_{\gamma_{2}}$, then their Wilson Chern numbers \emph{can} go to zero without closing a bulk gap.  (b) In a finite-sized, $\mathcal{I}$-symmetric double AXI, such as the weak-SOC AXIs in Refs.~\onlinecite{YoungkukMonopole,TMDHOTI} taken with weak coupling between both spin sectors, the Wilson gap closures in (a) are equivalent to removing the two chiral hinge modes through an $\mathcal{I}$-symmetric series of surface band inversions, or by gluing an $\mathcal{I}$-preserving arrangement of $|C|=1$ Chern insulators to crystal facets.}
\label{fig:evenWindI}
\end{figure}

Furthermore, taking the determinant of both sides of Eq.~(\ref{eq:W2InvEq}), we determine that:
\begin{equation}
\det(W_{2}(k_{z})) = \left(\det(W_{2}(-k_{z}))\right)^{*},
\end{equation}
which, along with Eq.~(\ref{eq:gamma2}), implies that at $k_{z}=0,\pi$, $\det(W_{2}(k_{z}))$ is $\mathbb{Z}_{2}$ quantized:
\begin{equation}
\det(W_{2}(0)) = \pm 1,\ \det(W_{2}(0)) = \pm 1,
\end{equation}
and thus that the nested Berry phase $\gamma_{2}(k_{z})$ (Eq.~(\ref{eq:gamma2})) is also $\mathbb{Z}_{2}$ quantized:
\begin{equation}
\gamma_{2}(0) = 0,\pi,\ \gamma_{2}(\pi) = 0,\pi. 
\label{eq:QuantI}
\end{equation}

$\mathcal{I}$ symmetry also plays another, more subtle, role in enforcing topological winding in $W_{2}(k_{z})$.  In the case of Eq.~(\ref{eq:QuantI}) where $\gamma_{2}(0)\neq \gamma_{2}(\pi)$, $W_{2}(k_{z})$ is implied to wind an odd number of times.  In order to relate this winding to stable bulk topology, it must persist for all choices of parameters that do not close an energy gap.  However, Wilson bands are free to invert \emph{even if energy bands remain uninverted~\cite{Fidkowski2011,AndreiXiZ2,BarryFragile}}.  When bands within different $\tilde{\chi}$-symmetric groupings of Wilson bands touch, Wilson-band Chern number $C_{\gamma_{2}}$ can be passed between the groups of Wilson bands (Fig.~\ref{fig:noSyms}(a)).  Absent $\tilde{\chi}$ (and thus bulk $\mathcal{I}$) symmetry, these crossings can occur individually at any crystal momentum or Wilson energy, and thus the winding of $W_{2}(k_{z})$ can be removed \emph{without a bulk gap closure} (Fig.~\ref{fig:noSyms}).  Therefore, while it can \emph{appear} that an $\mathcal{I}$-broken AXI is not Wannierizable if it exhibits $C_{\gamma_{2}}\neq 0$, there is no obstruction to closing a Wilson gap and removing the winding of $W_{2}(k_{z})$.  This is in agreement with the recognition that, as $\theta$ transforms as the inner product of a vector and an axial vector~\cite{WilczekAxion,VDBAxion,QHZ,AndreiInversion,ChenGilbertChern,AshvinAxion1,AshvinAxion2,VDBHOTI}, \emph{i.e.} $\vec{E}\cdot\vec{B}$, then it is only quantized in the presence of bulk roto-inversion ($C_{ni}\times\mathcal{I}$, where $C_{ni}$ is an $n$-fold rotation about the $i$-axis) or roto-time-reversal ($C_{ni}\times\mathcal{T}$) symmetries.  Here, relaxing $\mathcal{I}$ allows $\theta$ to wind from the topological value ($\theta=\pi$) to the trivial value ($\theta=0$) without closing an energy gap.  Physically, this can be understood by drawing a connection between the Wilson and surface spectra~\cite{Fidkowski2011}.  In a finite-sized AXI, surface band inversions are permitted that flip the (half-integer) surface Chern number of a particular facet (Fig.~\ref{fig:slabInversion}); in an infinite AXI crystal, these band inversions only appear as gap closures in the Wilson spectrum.  It has been established in previous works that if $\mathcal{I}$-symmetry is relaxed in a finite-sized (magnetic) HOTI, then a series of \emph{surface-only} band inversions (Fig.~\ref{fig:noSyms}(b)) can be performed to remove all of the chiral hinge states and trivialize the bulk~\cite{HOTIBernevig,AshvinTCI,EslamInversion}.  We thus conclude that this process is also reflected in the bulk Wilson spectrum of an $\mathcal{I}$-broken AXI.

Conversely, when $\tilde{\chi}$ is a symmetry of $W_{1}(k_{y},k_{z})$, then we can show that odd values of $C_{\gamma_{2}}$ cannot be changed to even values without closing a bulk energy gap.  In a $\tilde{\chi}$-symmetric Wilson spectrum, a Wilson band inversion at $\theta_{1}(k_{y},k_{z})$ \emph{must} be accompanied by a second crossing at $-\theta_{1}(-k_{y},-k_{z})$ (Eq.~(\ref{eq:invWilson1})).  As each crossing transfers a unit of Wilson Chern number between $\tilde{\chi}$-symmetric groupings of Wilson bands, then this process of simultaneously inverting Wilson bands at $\tilde{\chi}$-related Wilson energies and crystal momenta either completely preserves $C_{\gamma_{2}}$ within each grouping of Wilson bands, or changes it by $\pm 2$ (Figs.~\ref{fig:oddWindI} and~\ref{fig:evenWindI}).  Thus, without closing an energy gap, $\tilde{\chi}$-symmetric groupings of Wilson bands with alternating \emph{even} values of $C_{\gamma_{2}}$ can be converted to groupings with $C_{\gamma_{2}}=0$ (Fig.~\ref{fig:evenWindI}), but groupings with alternating \emph{odd} values of $C_{\gamma_{2}}$ will still exhibit odd values $C_{\gamma_{2}}\pm 2$ (Fig.~\ref{fig:oddWindI}).  Therefore, we conclude that $C_{\gamma_{2}}$ within a $\tilde{\chi}$-symmetric grouping of Wilson bands is only a well-defined bulk topological invariant taken modulo $2$, and becomes ill-defined when $\mathcal{I}$ symmetry is relaxed.  As we have numerically confirmed in Figs.~\ref{fig:oneCoupled}(h-k) and Fig.~\ref{fig:twoCoupled}(h-k) that $C_{\gamma_{2}}=\pm 1$ characterizes an $\mathcal{I}$-symmetric AXI with $\theta=\pi$, then we conclude that, in the presence of bulk $\mathcal{I}$ symmetry, the $\mathbb{Z}_{2}$-valued magnetoelectric polarizability~\cite{WilczekAxion,VDBAxion,QHZ,AndreiInversion,ChenGilbertChern,AshvinAxion1,AshvinAxion2,VDBHOTI} is given by:
\begin{equation}
\frac{\theta}{\pi} = C_{\gamma_{2}}\text{ mod } 2.
\label{eq:thetaInv}
\end{equation}
$\theta$ can therefore only be changed while keeping $\mathcal{I}$ symmetry through an energy band inversion that changes $\gamma_{2}(k_{z})$ at $k_{z}=0,\pi$, and thus changes whether an odd or even number of BZ planes indexed by $k_{z}=0,\pi$ are equivalent to 2D insulators with anomalous corner charges (Fig.~\ref{fig:AXIcorners}).

\subsection{The Action of $C_{2z}\times\mathcal{T}$ Symmetry on the $z$-Directed Wilson Loop}
\label{sec:zW1C2T}

In this section, we derive the action of $C_{2z}\times\mathcal{T}$ symmetry on the $z$-directed Wilson loop $W_{1}(k_{x},k_{y})$ as defined in~\ref{sec:W2}.  Specifically, we show that at each perpendicular crystal momenta:
\begin{equation}
k_{\perp} = (k_{x},k_{y}),
\end{equation}
$C_{2z}\times\mathcal{T}$ acts an antiunitary time-reversal symmetry $\tilde{\Theta}$ that preserves the sign of $k_{\perp}$.  We show that this can preserve twofold-degenerate crossings at any point in $k_{\perp}$ and Wilson energy $\theta_{1}(k_{\perp})$, which correspond to twofold-degenerate linear degeneracies in the $z$-surface spectrum~\cite{YoungkukLineNode,FangFuNSandC2,ChenRotation,HarukiRotation,LiangTCI}. 

We derive this result by determining the action of $C_{2z}\times\mathcal{T}$ on Eq.~(\ref{eq:wilsondisc}).  As $(C_{2z}\times\mathcal{T})^{2}=+1$ whether $\mathcal{H}(k_{x},k_{y},k_{z})$ characterizes spinful or spinless electrons,  it acts on $P(\mathbf{k})$ Eq.~(\ref{eq:defproj}) as:
\begin{equation}
(C_{2z}\times\mathcal{T})P(k_{x},k_{y},k_{z})(C_{2z}\times\mathcal{T})^{-1} = P^{*}(k_{x},k_{y},-k_{z}).
\label{eq:repC2T}
\end{equation}
Therefore, for the $z$-directed Wilson loop:
\begin{align}
(C_{2z}\times\mathcal{T})W_{1(k_\perp,k_{z0})}(C_{2z}\times\mathcal{T})^{-1} & = (C_{2z}\times\mathcal{T})V(2\pi\hat{z})\Pi(k_\perp, k_{z0} )(C_{2z}\times\mathcal{T})^{-1} \nonumber \\
  = V^{T}(2\pi\hat{z})(C_{2z}\times\mathcal{T}) & \bigg[P(k_{\perp},k_{z0}  + 2\pi)  P(k_{\perp},k_{z0} +\frac{2\pi (N-1)}{N})\cdots P(k_{\perp},k_{z0}+\frac{2\pi }{N})\bigg](C_{2z}\times\mathcal{T})^{-1} \nonumber \\
 = V^{T} (2\pi\hat{z})& P^{*}(k_{\perp},-k_{z0}  - 2\pi)P^{*}(k_{\perp},-k_{z0} -\frac{2\pi (N-1)}{N})\cdots P^{*}(k_{\perp},-k_{z0}-\frac{2\pi }{N}) \nonumber \\
& =  V^{T}(2\pi\hat{z})V^{*}(2\pi\hat{z})\Pi^{T}(k_{\perp},-k_{x0})V^{T}(2\pi\hat{z}) \nonumber \\
& = W^{T}_{1(k_{\perp},-k_{z0})},
\label{eq:C2zTzWils}
\end{align}
in which we have used that $C_{2z}\times\mathcal{T}$ transforms Eq.~(\ref{eq:defsew}):
\begin{equation}
(C_{2z}\times\mathcal{T})V(2\pi\hat{z})(C_{2z}\times\mathcal{T})^{-1} = KV^{\dag}(2\pi\hat{z})K^{-1} = V^{T}(2\pi\hat{z}),
\end{equation}
where $K$ is complex conjugation.  Eq.~(\ref{eq:C2zTzWils}) implies that the Wilson Hamiltonian is invariant under an antiunitary time-reversal symmetry, which we denote as $\tilde{\Theta}$, that preserves the signs of $k_{x}$ and $k_{y}$:
\begin{equation}
\tilde{\Theta} H_{W_{1}}(k_{x},k_{y})\tilde{\Theta}^{-1} = H^{*}_{W_{1}}(k_{x},k_{y}),
\label{eq:C2TWilsonZ}
\end{equation}
implying that $H_{W_{1}}(k_{x},k_{y})$ is real.  For every Wilson eigenstate $|\tilde{u}^n(k_{x},k_{y},k_{z0})\rangle$ with eigenvalue $\theta_{1}(k_{x},k_{y})$, there is therefore an eigenstate $\tilde{\Theta}|\tilde{u}^n(k_{x},k_{y},k_{z0})\rangle$ (possibly the same state) with the same eigenvalue $\theta_{1}(k_{x},k_{y})$.  As shown in Refs.~\onlinecite{YoungkukLineNode,FangFuNSandC2,ChenRotation,HarukiRotation,LiangTCI}, any real 2D Hamiltonian, Wilson or energy, can host topological twofold linear degeneracies at generic crystal momenta, and thus bulk $C_{2z}\times\mathcal{T}$ protects robust twofold degeneracies in the $z$-directed Wilson loop at generic Wilson energies and crystal momenta. 

\subsection{The Action of $C_{2z}\times\mathcal{T}$ Symmetry on the $y$-Directed Nested Wilson Loop}
\label{sec:detW2C2T}

In this section, we derive the action of $C_{2z}\times\mathcal{T}$ symmetry on the $y$-directed nested Wilson loop $W_{2}(k_{z})$ as defined in~\ref{sec:W2}.  Specifically, we show that $C_{2z}\times\mathcal{T}$ acts to make the nested Wilson spectrum particle-hole symmetric in 3D (but with different crystal momenta flipped than previously in~\ref{sec:detW2Inv}), and to quantize the nested Berry phase in 2D.  We then show that this particle-hole symmetry preserves the robustness of odd winding in the nested Wilson loop $C_{\gamma_{2}}$, and thus indicates a phase with $\mathbb{Z}_{2}$-nontrivial magnetoelectric polarizability~\cite{VDBAxion,QHZ,AndreiInversion,AshvinAxion1,AshvinAxion2,WuAxionExp,VDBHOTI} $\theta=\pi$.  We begin by reproducing the result derived in Refs.~\onlinecite{ZhidaBLG,AdrianFragile,BarryFragile} that $C_{2z}\times\mathcal{T}$ acts on $H_{W_{1}}(k_{y},k_{z})$ as an antiunitary particle-hole symmetry $\tilde{\Xi}$ that flips the sign of $k_{z}$ while preserving the sign of $k_{y}$.  

For the $x$-directed Wilson loop, $C_{2z}\times\mathcal{T}$ does not change the direction of the product of projectors in Eq.~(\ref{eq:defproj}), as previously occurred for $\mathcal{I}$ in~\ref{sec:detW2Inv}, and so the action of $C_{2z}\times\mathcal{T}$ on $W_{1}(k_{y},k_{z})$ follows simply from Eqs.~(\ref{eq:defproj}),~(\ref{eq:defsew}), and~(\ref{eq:repC2T}):
\begin{align}
(C_{2z}\times\mathcal{T})W_{1(k_{x0},k_{y},k_{z})}(C_{2z}\times\mathcal{T})^{-1} & = (C_{2z}\times\mathcal{T})V(2\pi\hat{x})\Pi(k_{x0},k_{y},k_{z} )(C_{2z}\times\mathcal{T})^{-1}  \nonumber \\
&  = V^{*}(2\pi\hat{x})\Pi^{*}(k_{x0},k_{y},-k_{z}) \nonumber \\
& = W^{*}_{1(k_{x0},k_{y},-k_{z})}.
\end{align}
The Wilson Hamiltonian is therefore invariant under an antiunitary particle-hole symmetry, which we denote as $\tilde{\Xi}$, that flips the sign of $k_{z}$ while leaving $k_{y}$ invariant:
\begin{equation}
\tilde{\Xi} H_{W_{1}}(k_{y},k_{z})\tilde{\Xi}^{-1} = -H^{*}_{W_{1}}(k_{y},-k_{z}),
\label{eq:C2TWilson1}
\end{equation}
implying that for every Wilson eigenstate $|\tilde{u}^n(k_{x0},k_{y},k_{z})\rangle$ with eigenvalue $\theta_{1}(k_{y},k_{z})$, there is another eigenstate $\tilde{\Xi}|\tilde{u}^n(k_{x0},k_{y},k_{z})\rangle$ with eigenvalue $-\theta_{1}(k_{y},k_{z})$.  We can therefore represent $\tilde{\Xi}$ as:
\begin{equation}
\tilde{\Xi}|\tilde{u}^n(k_{x0},k_{y},k_{z})\rangle = \tilde{\mathcal{O}} (|\tilde{u}^n(k_{x0},k_{y},-k_{z})\rangle)^{*},
\label{eq:XiOnN}
\end{equation}
where $\tilde{\mathcal{O}}$ is a $\mathbf{k}$-independent orthogonal transformation that rotates the Wilson band index $n$.  We restrict that $\mathcal{O}$ be orthogonal, and not simply unitary, as previously occurred in Eq.~(\ref{eq:chiOnN}), as $\tilde{\Xi}$ is antiunitary and $\tilde{\Xi}^{2}=+1$.  

We can now determine the action of $\tilde{\Xi}$ (and thus $C_{2z}\times\mathcal{T}$) on the nested Wilson loop.  As previously in~\ref{sec:detW2Inv}, in order for $\tilde{\Xi}$ to be a symmetry of the nestesd Wilson loop, we must restrict ourselves to nested Wilson projectors $\tilde{P}(\mathbf{k})$ onto particle-hole conjugate pairs of Wilson bands~\cite{TMDHOTI,ZhidaBLG}, such that:
\begin{equation}
\tilde{\Xi}\tilde{P}(k_{x0},k_{y},k_{z})\tilde{\Xi}^{-1} = \tilde{P}^{*}(k_{x0},k_{y0},-k_{z}),
\label{eq:C2TNestedProj}
\end{equation}
as $\tilde{P}(\mathbf{k})$ onto a block of particle-hole-symmetric Wilson bands is proportional to the identity in the basis of Wilson band indices, and the factors of $\tilde{\mathcal{O}}$ and $\tilde{\mathcal{O}}^{-1}$ cancel out.   For the $y$-directed nested Wilson loop, $\tilde{\Xi}$ also does not change the direction of the product of projectors in Eq.~(\ref{eq:defproj2}), as previously occurred for $\tilde{\chi}$ in~\ref{sec:detW2Inv}, and so the action of $\tilde{\Xi}$ (and thus $C_{2z}\times\mathcal{T}$) on $W_{2}(k_{z})$ also follows simply from Eqs.~(\ref{eq:defproj2}),~(\ref{eq:nestedSew}), and~(\ref{eq:C2TNestedProj}):
\begin{align}
\tilde{\Xi}W_{2(k_{x0},k_{y0},k_{z})}\tilde{\Xi}^{-1} & = \tilde{\Xi}\tilde{V}(2\pi\hat{y})\tilde{\Pi}(k_{x0},k_{y0},k_{z} )\tilde{\Xi}^{-1}  \nonumber \\
&  = \tilde{V}^{*}(2\pi\hat{y})\tilde{\Pi}^{*}(k_{x0},k_{y0},-k_{z}) \nonumber \\
& = W^{*}_{2(k_{x0},k_{y0},-k_{z})}.
\label{eq:W2C2TEq}
\end{align}
The nested Wilson Hamiltonian is therefore invariant under an antiunitary particle-hole symmetry, which we denote as $\tilde{\tilde{\Xi}}$, that flips the sign of $k_{z}$:
\begin{equation}
\tilde{\tilde{\Xi}} H_{W_{2}}(k_{z})\tilde{\tilde{\Xi}}^{-1} = -H^{*}_{W_{2}}(-k_{z}),
\label{eq:C2TWilson2}
\end{equation}
implying that for every nested Wilson eigenstate $|\tilde{\tilde{u}}^n(k_{x0},k_{y0},k_{z})\rangle$ with eigenvalue $\theta_{2}(k_{z})$, there is another eigenstate $\tilde{\tilde{\Xi}}|\tilde{u}^n(k_{x0},k_{y0},k_{z})\rangle$ with eigenvalue $-\theta_{2}(k_{z})$.

\begin{figure}[h]
\centering
\includegraphics[width=0.9\textwidth]{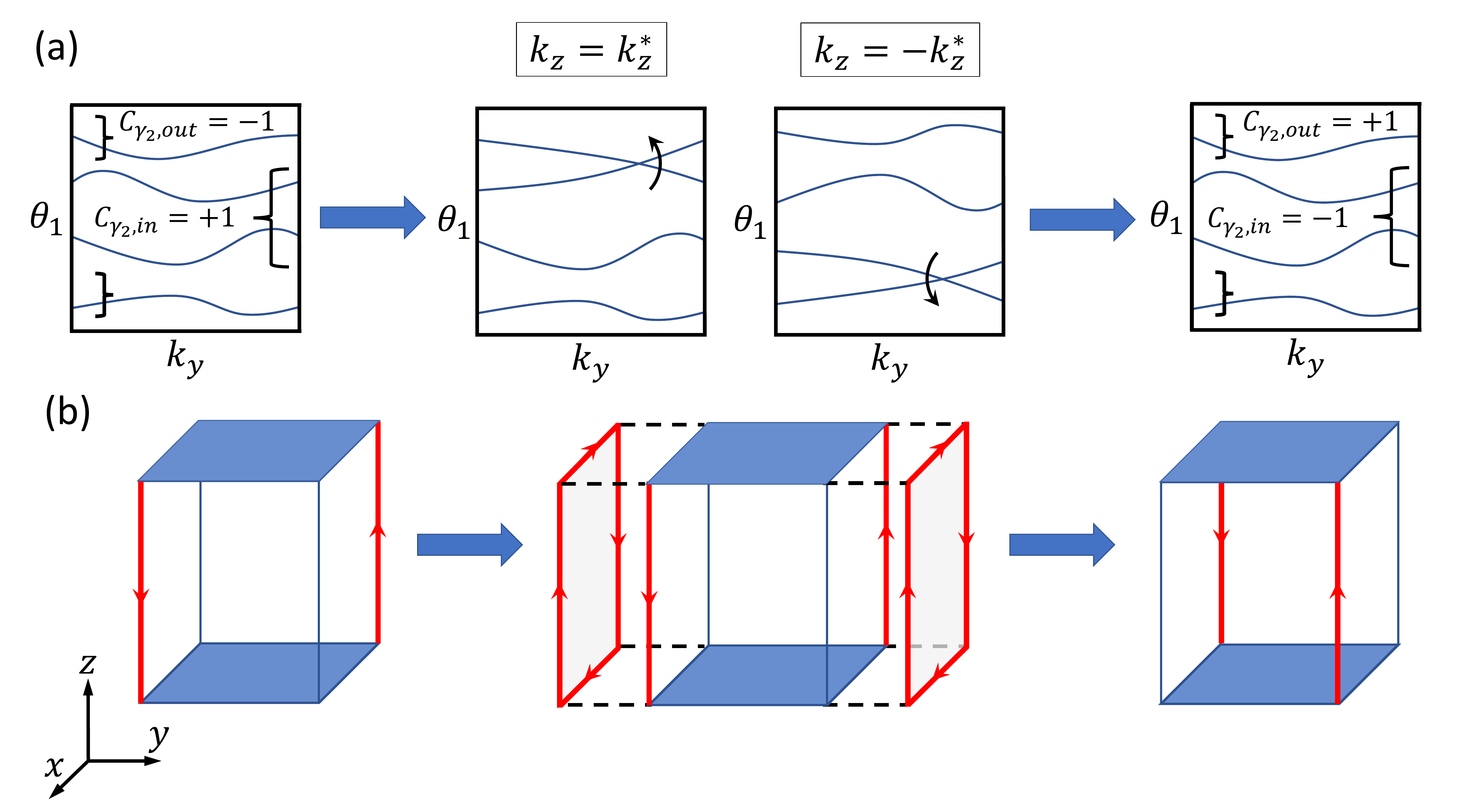}
\caption{(a)  When bulk $C_{2z}\times\mathcal{T}$ symmetry is present, then Wilson crossings in a 3D magnetic insulator can only form in pairs at $\theta_{1}(k_{y},k_{z})$ and $-\theta_{1}(k_{y},-k_{z})$ (Eq.~(\ref{eq:C2TWilson1})).  Without further restrictions, these crossings can change the Wilson Chern numbers of particle-hole-symmetric groupings of Wilson bands by $\pm 2$.  As in Fig.~\ref{fig:noSyms}, these pairs of Wilson gapless points can form without closing a bulk gap.  Therefore if particle-hole-symmetric groupings of Wilson bands exhibit alternating odd values of $C_{\gamma_{2}}$, then their Wilson Chern numbers \emph{cannot} go to zero without closing a bulk gap, though they can change in value by $\pm 2$.  Thus, $C_{\gamma_{2}}\text{ mod }2$ represents a well-defined, bulk $\mathbb{Z}_{2}$ topological invariant (Eq.~(\ref{eq:thetaC2T})).  (b) In a finite-sized $C_{2z}\times\mathcal{T}$-symmetric AXI, the Wilson gap closures in (a) are equivalent to moving the chiral hinge modes through a $C_{2z}\times\mathcal{T}$-symmetric series of band inversions on surfaces with normals in the $xy$-plane, or by gluing a $C_{2z}\times\mathcal{T}$-preserving arrangement of $|C|=1$ Chern insulators to crystal facets with normals in the $xy$-plane.}
\label{fig:oddWindC2T}
\end{figure}

\begin{figure}[h]
\centering
\includegraphics[width=0.9\textwidth]{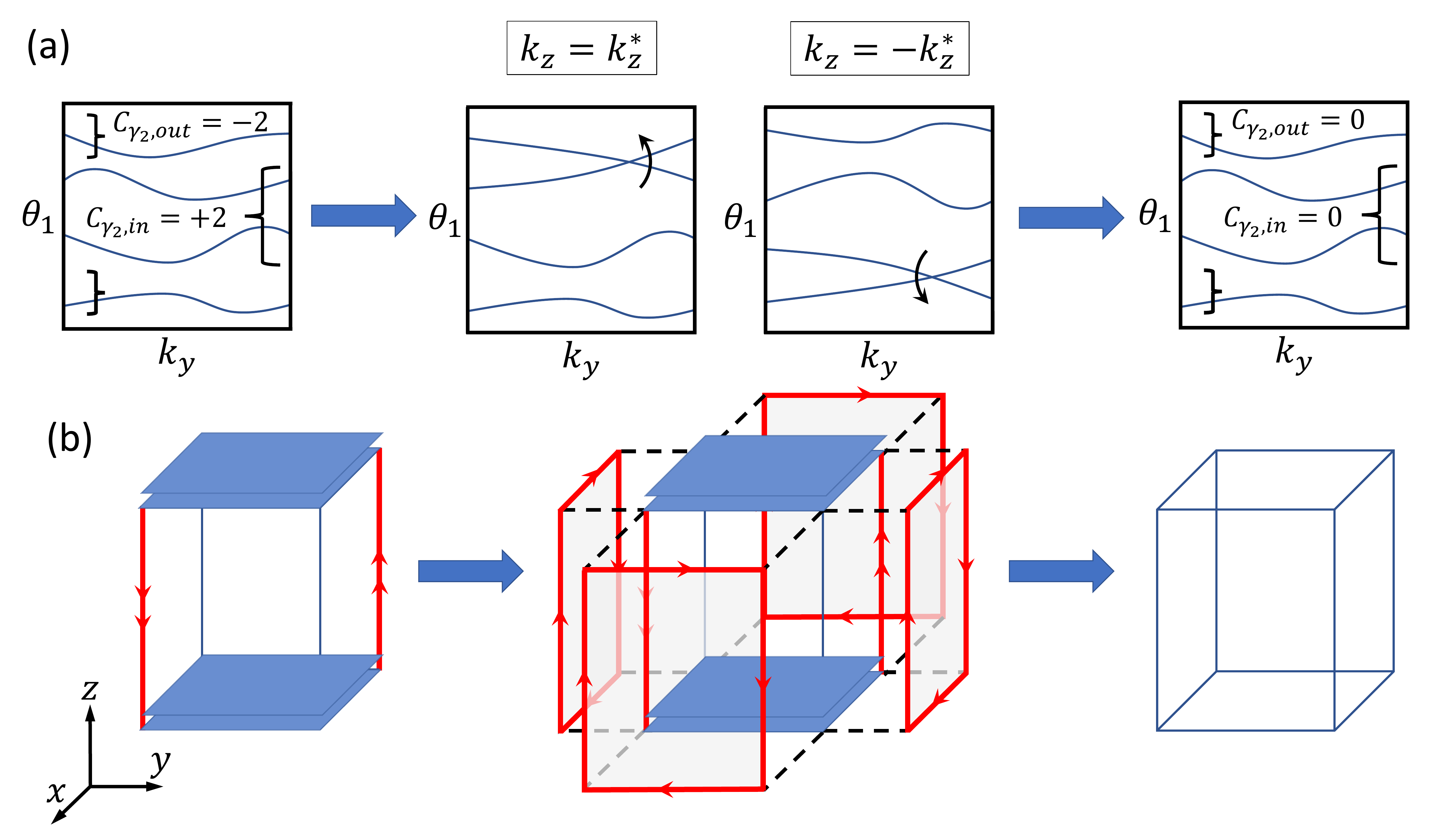}
\caption{(a)  When bulk $C_{2z}\times\mathcal{T}$ symmetry is present, then Wilson crossings in a 3D magnetic insulator can only form in pairs at $\theta_{1}(k_{y},k_{z})$ and $-\theta_{1}(k_{y},-k_{z})$ (Eq.~(\ref{eq:C2TWilson1})).  Without further restrictions, these crossings can change the Wilson Chern numbers of particle-hole-symmetric groupings of Wilson bands by $\pm 2$.  As in Fig.~\ref{fig:noSyms}, these pairs of Wilson gapless points can form without closing a bulk gap.  Therefore if particle-hole-symmetric groupings of Wilson bands exhibit alternating even values of $C_{\gamma_{2}}$, then their Wilson Chern numbers \emph{can} go to zero without closing a bulk gap.  (b) In a finite-sized, $C_{2z}\times\mathcal{T}$-symmetric double AXI, such as a rotation-non-anomalous crystalline insulator with two twofold cones on each $\pm z$-normal surface~\cite{FangFuNSandC2,ChenRotation,HarukiRotation,LiangTCI} (doubled blue planes), the Wilson gap closures in (a) are equivalent to removing the two chiral hinge modes through a $C_{2z}\times\mathcal{T}$-symmetric series of band inversions on surfaces with normals in the $xy$-plane, or by gluing a $C_{2z}\times\mathcal{T}$-preserving arrangement of $|C|=1$ Chern insulators to crystal facets with normals in the $xy$-plane.  When the two cones on each $\pm z$ normal surface are pairwise eliminated, this results in a 3D trivial insulator without surface or hinge states.}
\label{fig:evenWindC2T}
\end{figure}

Furthermore, taking the determinant of both sides of Eq.~(\ref{eq:W2C2TEq}), we determine that:
\begin{equation}
\det(W_{2}(k_{z})) = \left(\det(W_{2}(-k_{z}))\right)^{*},
\end{equation}
which, along with Eq.~(\ref{eq:gamma2}), implies that at $k_{z}=0,\pi$, $\det(W_{2}(k_{z}))$ is $\mathbb{Z}_{2}$ quantized:
\begin{equation}
\det(W_{2}(0)) = \pm 1,\ \det(W_{2}(0)) = \pm 1,
\end{equation}
and thus that the nested Berry phase $\gamma_{2}(k_{z})$ (Eq.~(\ref{eq:gamma2})) is also $\mathbb{Z}_{2}$ quantized:
\begin{equation}
\gamma_{2}(0) = 0,\pi,\ \gamma_{2}(\pi) = 0,\pi. 
\label{eq:QuantC2T}
\end{equation}
This quantization of $\gamma_{2}$ in $C_{2z}\times\mathcal{T}$-invariant planes ($k_{z}=0,\pi$) was previously explained in Ref.~\onlinecite{ZhidaBLG} using homotopy, and in Ref.~\onlinecite{TMDHOTI} (for $\mathcal{I}\times\mathcal{T}$, which for spinless electrons is equivalent to $C_{2z}\times\mathcal{T}$ in 2D~\cite{WiederLayers,SteveMagnet,BarryFragile,ZhidaBLG}) using the co-dimension arguments in Refs.~\onlinecite{TeoKaneDefect,YoungkukLineNode} applied to the  Altland-Zirnbauer classification~\cite{AZClass,KitaevClass}.  It was also related (again specifically for spinless $\mathcal{I}\times\mathcal{T}$) to the second Stiefel-Whitney invariant $w_{2}$ in Refs.~\onlinecite{YoungkukMonopole,KoreanFragile}.

As with $\mathcal{I}$ in~\ref{sec:detW2Inv}, $C_{2z}\times\mathcal{T}$ symmetry here also plays another, more subtle, role in enforcing topological winding in $W_{2}(k_{z})$.  In the case of Eq.~(\ref{eq:QuantC2T}) where $\gamma_{2}(0)\neq \gamma_{2}(\pi)$, $W_{2}(k_{z})$ is implied to wind an odd number of times.  In order to relate this winding to stable bulk topology, it must persist for all choices of parameters that do not close a bulk energy gap.  However, Wilson bands are free to invert \emph{even if energy bands remain uninverted~\cite{Fidkowski2011,AndreiXiZ2,BarryFragile}.}  When bands within different $\tilde{\Xi}$-symmetric groupings of Wilson bands touch, Wilson-band Chern number $C_{\gamma_{2}}$ can be passed between the groups of Wilson bands (Fig.~\ref{fig:noSyms}(a)).  Absent $\tilde{\Xi}$ (and thus bulk $C_{2z}\times\mathcal{T}$) symmetry, these crossings can occur individually at any crystal momentum or Wilson energy, and thus the winding of $W_{2}(k_{z})$ can be removed \emph{without a bulk gap closure} (Fig.~\ref{fig:noSyms}).  Therefore, as occurred previously in~\ref{sec:detW2Inv}, while it can \emph{appear} that a $C_{2z}\times\mathcal{T}$-broken AXI is not Wannierizable if it exhibits $C_{\gamma_{2}}\neq 0$, there is no obstruction to closing a Wilson gap and removing the winding of $W_{2}(k_{z})$. 

Conversely, when $\tilde{\Xi}$ is a symmetry of $W_{1}(k_{y},k_{z})$, then we can show that odd values of $C_{\gamma_{2}}$ cannot be changed to even values without closing an energy gap.  In a $\tilde{\Xi}$-symmetric Wilson spectrum, a Wilson band inversion at $\theta_{1}(k_{y},k_{z})$ \emph{must} be accompanied by a second crossing at $-\theta_{1}(k_{y},-k_{z})$ (Eq.~(\ref{eq:C2TWilson1})).  As each crossing transfers a unit of Wilson Chern number between $\tilde{\Xi}$-symmetric groupings of Wilson bands, then this process of simultaneously inverting Wilson bands at $\tilde{\Xi}$-related Wilson energies and crystal momenta either completely preserves $C_{\gamma_{2}}$ within each grouping of Wilson bands, or changes it by $\pm 2$ (Fig.~\ref{fig:oddWindC2T} and Fig.~\ref{fig:evenWindC2T}).  Thus, without closing an energy gap, $\tilde{\Xi}$-symmetric groupings of Wilson bands with alternating \emph{even} values of $C_{\gamma_{2}}$ can be converted to groupings with $C_{\gamma_{2}}=0$ (Fig.~\ref{fig:evenWindC2T}), but groupings with alternating \emph{odd} values of $C_{\gamma_{2}}$ will still exhibit odd values $C_{\gamma_{2}}\pm 2$ (Fig.~\ref{fig:oddWindC2T}).  Therefore, we conclude that $C_{\gamma_{2}}$ within a $\tilde{\Xi}$-symmetric grouping of Wilson bands is only a well-defined bulk topological invariant taken modulo $2$, and becomes ill-defined when $C_{2z}\times\mathcal{T}$ symmetry is relaxed.  As we have numerically confirmed in Figs.~\ref{fig:oneCoupledC2T}(h-k) and Fig.~\ref{fig:twoCoupledC2T}(e-h) that $C_{\gamma_{2}}=\pm 1$ characterizes a $C_{2z}\times\mathcal{T}$-symmetric crystalline AXI with $\theta=\pi$ inherited from its parent $C_{2z}$- and $\mathcal{T}$-symmetric TI phase (Eq.~(\ref{eq:C2andT})), then we conclude that, in the presence of bulk $C_{2z}\times\mathcal{T}$ symmetry, the $\mathbb{Z}_{2}$-valued magnetoelectric polarizability~\cite{WilczekAxion,VDBAxion,QHZ,AndreiInversion,ChenGilbertChern,AshvinAxion1,AshvinAxion2,VDBHOTI} is given by:
\begin{equation}
\frac{\theta}{\pi} = C_{\gamma_{2}}\text{ mod } 2.
\label{eq:thetaC2T}
\end{equation}
$\theta$ can therefore only be changed while keeping $C_{2z}\times\mathcal{T}$ symmetry through an energy band inversion that changes $\gamma_{2}(k_{z})$ at $k_{z}=0,\pi$, and thus changes whether an odd or even number of BZ planes indexed by $k_{z}=0,\pi$ are equivalent to 2D insulators with anomalous corner charges (Fig.~\ref{fig:C2Tcorners}).

\end{appendix}
\bibliography{axionFragileArx}
\end{document}